\documentclass[preprint] {article}
\usepackage{amsmath,amssymb,amsfonts}
\usepackage{geometry}
\usepackage{graphicx}
\usepackage{subcaption}
\usepackage{braket}
\usepackage{float}
\allowdisplaybreaks
\newcommand {\bp}{\begin{pmatrix}}
\newcommand {\ep}{\end{pmatrix}}
\newcommand{\be}{\begin{equation}} \newcommand{\ee}{\end{equation}}
\newcommand{\bea}{\begin{eqnarray}}\newcommand{\eea}{\end{eqnarray}}

\DeclareMathOperator{\sgn}{sgn}
\begin{document}
\title{ Complex dynamical properties of coupled Van der Pol-Duffing oscillators
with balanced loss and gain}

\author{ Puspendu Roy\footnote {{\bf email}:puspenduroy716@gmail.com}
\ and \ Pijush K. Ghosh\footnote{{\bf email:} 
pijushkanti.ghosh@visva-bharati.ac.in}
\footnote{Corresponding Author} 
}
\date{Department of Physics, Siksha-Bhavana, \\ 
Visva-Bharati University, \\
Santiniketan, PIN 731 235, India.}

\maketitle
\begin{abstract}

We consider a Hamiltonian system of coupled Van der Pol-Duffing(VdPD) oscillators with balanced
loss and gain. The system is analyzed perturbatively by using Renormalization
Group(RG) techniques as well as Multiple Scale Analysis(MSA). Both the methods
produce identical results in the leading order of the perturbation. The RG flow equation
is exactly solvable and the slow variation of amplitudes and phases in time 
can be computed analytically.  The system is analyzed numerically and shown
to admit periodic solutions in regions of parameter-space, confirming the results of the
linear stability analysis and perturbation methods. The complex dynamical behavior of  the system
is studied in detail by using time-series, Poincar$\acute{e}$-sections, power-spectra,
auto-correlation function and bifurcation diagrams. The Lyapunov exponents are computed
numerically. The numerical analysis reveals chaotic behaviour in the system beyond a
critical value of the parameter that couples the two VdPD oscillators through linear coupling,
thereby providing yet another example of Hamiltonian chaos in a system with balanced
loss and gain. Further, we modify the nonlinear terms of the model to make it a
non-Hamiltonian system of coupled VdPD oscillators with balanced loss
and gain. The non-Hamiltonian system is analyzed perturbativly as well as numerically
and shown to posses regular periodic as well as chaotic solutions. It is seen that
the ${\cal{PT}}$-symmetry is not an essential requirement for the existence of regular
periodic solutions in both the Hamiltonian as well as non-Hamiltonian systems. 

\end{abstract}
\vspace{0.1in}
\newpage

\section{Introduction}

Systems with balanced loss and gain have received considerable interests over the last few
years\cite{pkg-review}. One of the characteristic features of such a system is that the flow
preserves the volume in the position-velocity state space, although the individual
degrees of freedom are subjected to loss or gain\cite{pkg-review}. The system is
non-dissipative and may admit periodic solutions within some regions in the parameter-space.
The Hamiltonian formulation is possible for suitable choice of the potential and the
corresponding quantum theory is well defined within appropriate Stoke wedges\cite{ben}.
The mathematical model\cite{ben} for an experimentally realized ${\cal{PT}}$-symmetric
coupled resonators\cite{bpeng} provides a prototype for a system with balanced
loss and gain that incorporates all these features. The periodic solutions
and the quantum bound states exist only in the ${\cal{PT}}$-symmetric phase. The
phase-transition from ${\cal{PT}}$-symmetric to ${\cal{PT}}$-broken phase is accompanied
by a change in the nature of the solution, namely, the bounded solution becomes
unbounded. This observation has generated some interest and lead to introduction of many
${\cal{PT}}$-symmetric Hamiltonian systems with balanced loss and
gain\cite{ben1,ivb, ds-pkg,khare, pkg-ds, ds-pkg1,p6-deb,pkg-1}. The examples of
Hamiltonian system with balanced loss and gain include systems with nonlinear
interaction\cite{ivb,ds-pkg, khare,pkg-ds, ds-pkg1,p6-deb}, many-particle
systems\cite{ben1, pkg-ds, ds-pkg1,p6-deb}, systems with space-dependent loss-gain
terms\cite{ds-pkg1}, systems with Lorentz interaction\cite{pkg-1} etc..

It has been shown recently that non-${\cal{PT}}$ symmetric systems with balanced loss and gain
may also admit periodic solutions \textemdash the requirement of ${\cal{PT}}$-symmetry is not
necessary\cite{pkg-pr}. The most general form of parity transformation that is
embedded in the generic Lorentz transformation is considered to check the symmetry of these
non-${\cal{PT}}$ symmetric systems. This is consistent with the present understanding that
the ${\cal{CPT}}$-symmetry of ${\cal{PT}}$-symmetric quantum systems is expected to correspond
to the ${\cal{CPT}}$-theorem\cite{cpt} of a local Lorentz-invariant field theory at a more
fundamental level. It may be noted that the CPT theorem, which was originally derived for a
hermitian Hamiltonian, has now been extended to the case of non-hermitian Hamiltonian\cite{nh-cpt}.
Further, there is no signature of violation of Lorentz invariance in nature so far and the
CPT is considered to be a fundamental symmetry of nature. This justifies the consideration
of only linear transformation to describe parity operation for a classical system. This is also not
in contradiction with the corresponding quantum theory, since the parity and time-reversal operators
are not unique for a quantum system\cite{pkg-review}. 

The example of a non-${\cal{PT}}$ symmetric mechanical system with balanced loss-gain that admits
periodic solution is described by a Duffing oscillator coupled to an anti-damped oscillator with
a variable angular frequency\cite{pkg-pr}. The Hamiltonian describes a multi-stable system  with
very rich dynamical behaviour. It exhibits regular periodic solutions within regions of the
parameter-space. The chaotic behaviour is observed in the system beyond a critical value of
the parameter that couples the Duffing oscillator to the anti-damped harmonic oscillator,
thereby providing the first example of Hamiltonian chaos in a system with balanced loss and
gain. The system contains positional non-conservative forces\cite{dissi} or curl-forces\cite{berry}
which are known to admit chaotic behaviour.
 The standard Duffing oscillator exhibits chaotic behaviour only if an external forcing
term is added to the system. Further, the system is non-Hamiltonian. On the other hand, there
is no external forcing term in the coupled Duffing oscillator model with balanced loss and gain.
The term that couples to the anti-damped oscillator may be interpreted as the forcing term. 

The Van der Pol oscillator is a prototype for systems with self-excited limit cycle oscillations. The
equation has been used to model oscillations in vacuum tube triode circuit\cite{triode},
biology\cite{bio-1,bio-2}, phonation\cite{vocal}, seismology\cite{sesimo}, quantum
synchronisation\cite{qs}, quantum criticality\cite{qc} etc. The Van der Pol
equation is nonlinear due to the space-dependent gain/loss coefficient which makes it 
damped in one region of space and anti-damped in the remaining part of the line. If a
nonlinear restoring force with cubic nonlinearity is added to the system in addition to
the linear restoring force, a double-well potential is obtained and the system is known
as VdPD oscillator\cite{vdpd-1,vdpd-2,vdpd-3,vdpd-4,vdpd-5,vdpd-6}. The dynamical
behaviour of the system is quite rich depending on different phases of the double-well
potential\cite{vdpd-4}. The VdPD has been studied extensively in the context of
convective instability of a binary fluid mixture in a porous medium\cite{vdpd-1}, hydrodynamics\cite{vdpd-2},
analog electronic system\cite{vdpd-3}, bifurcation and chaos\cite{vdpd-4}, integrability\cite{vdpd-5,vdpd-6}.
Various models of coupled Van der Pol oscillators\cite{cvdp-1,cvdp-2} and coupled VdPD oscillators\cite{cvdp-3}
have also been considered.

The purpose of this article is to study systems of coupled VdPD oscillators with balanced loss and gain. 
We present two different non-${\cal{PT}}$-symmetric models of coupled VdPD oscillators of which one is
a Hamiltonian system and the other one non-Hamiltonian. The Hamiltonian system is a generalization
of the coupled Duffing oscillator model\cite{pkg-pr} by making the balanced loss-gain terms space-dependent.
The two oscillators are coupled via loss-gain terms in addition to coupling through the potential terms.
The space-dependence of the loss-gain terms is such that one of the oscillators is dissipative within
an elliptic region and anti-damped in the remaining part of the configuration space. The requirement
of balanced loss-gain terms necessitates exactly opposite behaviour for the second oscillator.
The non-Hamiltonian system describes two coupled VdPD oscillators with the same space-dependent loss-gain
terms. This model is a generalization of non-${\cal{PT}}$-symmetric non-Hamiltonian coupled Duffing oscillators
with constant balanced loss-gain\cite{pkg-review,khare-0}.

We analyze the Hamiltonian system perturbatively by using MSA\cite{msa} and RG\cite{rg-1} methods.
The results in the leading order of the perturbation are identical for both the methods. The RG
flow equation defines a dimer model in which the two modes are coupled nonlinearly through the balanced
loss-gain terms along with coupling via nonlinear interaction. The RG flow equation is exactly solvable
and the slow variation of the amplitudes and the phases are determined analytically.
The system admits regular periodic solutions within regions of parameter-space which are
confirmed by numerical solutions. Thus, the list of non-${\cal{PT}}$-symmetric
systems with balanced loss and gain admitting regular periodic solution is enlarged by the inclusion of this
model. The numerical investigations reveal chaotic behaviour in these systems beyond some critical value of the linear
coupling term. The chaotic behaviour is seen in the standard VdPD oscillator in presence of a driving term.
Although the nonlinear oscillator considered in this article is undriven, the chaotic behaviour may
be attributed to the coupling between the two degrees of freedom of a given model. The complex dynamical
behavior of  the systems is investigated numerically by computing time-series, Poincar$\acute{e}$-sections,
power-spectra, auto-correlation function, bifurcation diagrams and the Lyapunov exponents. 
We get yet another example of Hamiltonian chaos for systems with balanced loss-gain.

We analyze a second system which differs from the first model in the velocity-independent nonlinear
terms. This is a generalization of the non-Hamiltonian system of two coupled Duffing oscillators
introduced in \cite{pkg-review,khare-0} by allowing the loss-gain terms to be space-dependent. The
model considered in Ref. \cite{khare-0} is ${\cal{PT}}$-symmetric and existence of regular periodic
solution is attributed to unbroken ${\cal{PT}}$-symmetry. The system considered in Ref. \cite{pkg-review}
admits regular periodic solutions in ${\cal{PT}}$-symmetric as well as non-${\cal{PT}}$-symmetric regimes and
reduces to the model of Ref. \cite{khare-0} in a particular limit.
The second model is non-Hamiltonian and analyzed perturbatively as well as numerically. The RG flow
equation differs from the first model only in the equation for the phase of one of the modes and is exactly
solvable. The system admits regular periodic solution in the ${\cal{PT}}$-symmetric as well as
non-${\cal{PT}}$-symmetric regimes. The systems admits chaotic behaviour which is investigated numerically
by computing time-series, Poincar$\acute{e}$-sections, power-spectra, auto-correlation function, bifurcation
diagrams and the Lyapunov exponents.

The plan of the article is the following. We introduce the Hamiltonian system describing coupled
VdPD oscillators in the next section. The general formalism along with ${\cal{PT}}$-symmetry and
linear stability analysis are presented at the beginning. The equations of motion are analyzed by using MSA
and RG techniques in Secs. 2.2.1 and 2.2.2, respectively. The solutions obtained by these two
methods are shown to be identical in the leading order of the perturbation. The RG flow equation is
solved in Sec. 2.2.1. The results of numerical investigations
are presented in Sec. 2.3 \textemdash regular solutions are described in Sec. 2.3.1, while
the analysis for chaotic dynamics given in Sec. 2.3.1. The non-Hamiltonian system describing coupled
VdPD is introduced in Sec. 3 and the details of numerical investigations are presented in Sec. 3.1.
Finally, the results are summarized in Sec. 4. 

\section{Hamiltonian system}

The Hamiltonian system is described by the equations of motion,
\bea
&& \ddot{x} + 2\gamma \left(1-r x^{2}-s y^{2}\right)\dot{x}+\omega^{2}x+\beta_{1}y+ g x^{3}=0,\nonumber \\
&& \ddot{y} - 2\gamma \left(1-r x^{2}-s y^{2}\right)\dot{y}+\omega^{2}y+\beta_{2}x+3 g x^{2}y=0,
\label{vdpd-eqn}
\eea
\noindent where $\omega$ is the angular frequency corresponding to the linear restoring force, $\beta_1,\beta_2$ are the
strengths of the linear coupling between the two modes and $g$ is the strength of the nonlinear interaction. The parameter
$\gamma$ is the strength of the constant loss-gain, while $r$ and $s$ are strengths of the space-dependent
loss-gain terms. The equation $r x^{2}+s y^{2} = 1$ corresponds to vanishing loss-gain and
separates two regions in the configuration space corresponding to dissipative and anti-damped regions. 
For $\gamma > 0$, the first equation in Eq. (\ref{vdpd-eqn}) is dissipative within the
elliptic region $r x^{2}+s y^{2} < 1$, while it is anti-damped for $r x^{2}+s y^{2} > 1$.
The Eq. (\ref{vdpd-eqn}) defines a system of balanced loss and gain in the sense that the
flow in the position-velocity state space preserves the volume, although individual degrees
of freedom are subjected to gain or loss.
This is manifested in the second equation which has exactly opposite behaviour in the respective regions of
the configuration space in respect to the first equation due to the requirement of balanced loss-gain\cite{pkg-1}.

There are several interesting limits of the system defined by Eq. (\ref{vdpd-eqn}).
The mathematical model\cite{ben} describing an experimentally realized ${\cal{PT}}$-symmetric coupled
resonators\cite{bpeng} is obtained for $r=s=g=0, \beta_1 =\beta_2$. The coupled Duffing oscillators with
balanced loss-gain\cite{pkg-pr} is obtained in the limit $r=s=0$. The $x$-degree of freedom decouples
completely for $s=\beta_1=0$, while the $y$-degree of freedom is unidirectionally coupled to it. This
limit corresponds to Hamiltonian formulation of the standard VdPD oscillator with the $y$ degree
of freedom treated as auxiliary variable. The standard techniques for analyzing a Hamiltonian system like canonical
perturbation theory, quantizaion, KAM theory, integrability etc. may be used to study VdPD oscillator. 
The canonical perturbation theory has been used successfully\cite{sagar,sagar-1} for the standard Van der Pol oscillator,
i.e, $s=\beta_1=g=0$.

The equations of motion (\ref{vdpd-eqn}) may be obtained from the Lagrangian,
\bea
{\cal{L}} & = & \dot{x}\dot{y} + \gamma \left [ \left( x-\frac{r x^{3}}{3} -s y^{2} x\right) \dot{y}-\left (y-r x^{2} y-
\frac{s y^{3}}{3}\right) \dot{x} \right ]  - \omega^{2} x y - \frac{1}{2} \left ( \beta_2 x^2 +
\beta_1 y^2 \right ) - g x^3 y.\nonumber
\label{lag}
\eea
\noindent The canonical momenta are,
\bea
P_x=\dot{y} - \gamma \left (y-r x^{2} y-\frac{s y^{3}}{3}\right), \ \ P_y=\dot{x} + \gamma \left( x-\frac{r x^{3}}{3} -s y^{2} x\right),
\eea
\noindent and the Hamiltonian is evaluated as,
\bea
H & = & P_x P_y + \gamma \left [ \left (y-r x^{2} y-\frac{s y^{3}}{3}\right) P_y -
\left( x-\frac{r x^{3}}{3} -s y^{2} x\right) P_x \right ]\nonumber \\
& - & \gamma^{2} \left (y-r x^{2} y-\frac{s y^{3}}{3}\right) \left( x-\frac{r x^{3}}{3} -
s y^{2} x\right)+ \omega^{2} x y + \frac{1}{2} \left ( \beta_2 x^2 + \beta_1 y^2 \right ) + g x^3 y.
\label{hami}
\eea
\noindent The Hamiltonian formulation for generic systems with balanced loss-gain is given in Refs. \cite{pkg-ds,p6-deb}
that is used to obtain ${\cal{L}}$ and $H$. The Hamiltonian for the specific case of coupled Van der Pol oscillators,
i.e. $\beta_1=\beta_2=g=0$ was obtained in Ref. \cite{sagar}. We define the generalized momenta as,
\bea
\Pi_x= P_x + \gamma A_x, \ \Pi_y = P_y + \gamma A_y, \ A_x(x,y) \equiv y-r x^{2} y-\frac{s y^{3}}{3}, \
A_y(x,y) \equiv  - \left ( x-\frac{r x^{3}}{3} -s y^{2} x \right ),
\eea
\noindent where $A_x$ and $A_y$ may be interpreted as ``fictitious gauge potential"\cite{p6-deb}. The
magnetic field corresponding to this fictitious gauge potential is equivalent to the space-dependence
of gain-loss terms\cite{p6-deb}. If the system is considered in the background of realistic magnetic
field, there will be additional contribution to the gauge potentials\cite{pkg-1}
The Hamiltonian can be rewritten as,
\bea
H= \Pi_x \Pi_y + V(x,y), \ V(x,y) \equiv \omega^{2} x y + \frac{1}{2} \left ( \beta_2 x^2 + \beta_1 y^2 \right ) + g x^3 y.
\eea
\noindent  In general, the Hamiltonian is not positive-definite.

The independent scales in the system may be fixed by employing the following transformations,
\bea
t \rightarrow \omega^{-1} t, \ x \rightarrow {\vert \beta_2 
\vert}^{-\frac{1}{2}} x, \
y \rightarrow {\vert \beta_1 \vert}^{-\frac{1}{2}} y, \ \beta_1 \neq 0,  \beta_2 \neq 0.
\label{scale}
\eea
\noindent The model can be described in terms of five independent parameters $\Gamma, \beta$,
$\alpha$, $a$ and $b$ defined as,
\bea
\Gamma = \frac{\gamma}{\omega}, \ \beta= \frac{\sqrt{{\vert {\beta_1} \vert}
{\vert {\beta_2}\vert}}}{\omega^2}, \ 
\alpha= \frac{g}{{\vert \beta_2 \vert} \omega^2}, \ a=\frac{r}{{\vert \beta_2 \vert}}, \ b=\frac{s}{{\vert \beta_1 \vert}}.
\label{scale-1}
\eea
\noindent The total number of independent parameters is reduced from seven to five and convenient for analyzing the system.
The equations of motion have the following expressions:
\bea
&& \ddot{x} + 2 \Gamma(1-a x^{2}-b y^{2}) \dot{x}+ x + \sgn(\beta_1) \ \beta y + \alpha x^{3}=0, \nonumber \\
&& \ddot{y} - 2 \Gamma(1-a x^{2}-b y^{2}) \dot{y}+ y + \sgn(\beta_2) \ \beta x+ 3 \alpha x^{2}y=0,
\label{vdpd-eqn-1}
\eea 
\noindent where $\sgn(x)$ is the signum function. The scale transformation (\ref{scale}) implies,
\bea
&& P_x \rightarrow \frac{\omega}{\sqrt{\beta_1}} \tilde{P}_x, \ \ P_y \rightarrow
\frac{\omega}{\sqrt{\beta_2}} \tilde{P}_y, \ \ H \rightarrow \beta^{-1} \tilde{H},\nonumber \\
&& \tilde{P}_x \equiv \dot{y} - \Gamma y \left (1-a x^{2} -\frac{b y^{2}}{3}\right),
\tilde{P}_y \equiv \dot{x} + \Gamma x \left( 1-\frac{a x^{2}}{3} -b y^{2} \right).
\eea
\noindent Defining generalized momenta $\tilde{\Pi}_x=\tilde{P}_x + \Gamma y \left (1-a x^{2} -\frac{b y^{2}}{3}\right), \ 
\tilde{\Pi}_y=\tilde{P}_y - \Gamma x \left( 1-\frac{a x^{2}}{3} -b y^{2} \right)$, the Hamiltonian $\tilde{H}$
can be rewritten as,
\bea
\tilde{H}=\tilde{\Pi}_x \tilde{\Pi}_y + V(x,y), \ \
V(x,y)= xy + \frac{\beta}{2} \left [ \sgn(\beta_2) \ x^2 + \sgn(\beta_1) \ y^2 \right ] + \alpha x^3 y.
\eea
\noindent The Hamiltonian $\tilde{H}$ and the equations of motion in (\ref{vdpd-eqn-1}) will be
considered for further analysis and the results in terms of the original variables may be obtained
by inverse scale transformations. The Hamiltonian $\tilde{H}$ or equivalently the energy
$E=\dot{x} \dot{y} + V(x,y)$ is a constant of motion, but neither semi-positive definite nor bounded
from below. The energy may be bounded from below for specific orbits in the phase-space to be
determined from the equations of motion.

The parity ${\cal{P}}$ and time-reversal symmetry ${\cal{T}}$ in $2+1$ space-time dimensions may be defined
as follows:
\bea
&& {\cal{T}}: t \rightarrow -t, \ \tilde{P}_x \rightarrow -\tilde{P}_x, \ \tilde{P}_y \rightarrow -\tilde{P}_y\nonumber \\
&& {\cal{P}}: \left (x, y\right ) \rightarrow \left ( y, x \right ), \ \left  ( \tilde{P}_x,  \tilde{P}_x \right )
\rightarrow \left  ( \tilde{P}_x,  \tilde{P}_x \right ).
\label{dis-s}
\eea
\noindent The system defined by Eq. (\ref{vdpd-eqn-1}) is ${\cal{PT}}$ symmetric for $a=b, \ \sgn(\beta_1) = \sgn(\beta_2),
\ \alpha=0$ and becomes non-${\cal{PT}}$-symmetric if one or more of these conditions are violated.
The parity ${\cal{P}}$ in Eq. (\ref{dis-s}) belongs to $O(2)$ transformation and is consistent with the general expectation
that ${\cal{CPT}}$-symmetry of ${\cal{PT}}$-symmetric quantum systems will correspond to ${\cal{CPT}}$ theorem at
a more fundamental level\cite{pkg-review}. Any possibility of non-linear transformations with the properties
of parity operator for which Eq. (\ref{vdpd-eqn-1}) may be ${\cal{PT}}$-symmetric for $a \neq b$ or $\alpha \neq 0$ or
$\sgn(\beta)_1 \neq \sgn(\beta)_2$ is
discarded for the same reason, since the ${\cal{CPT}}$-theorem dwells on Lorentz transformation which is linear.
The linear stability analysis along with numerical investigations show that the system with $a=b=0$ does not admit any
periodic solution for  $\sgn(\beta_1) = - \sgn(\beta_2)$ \cite{pkg-pr}. We choose $\sgn(\beta_1) = \sgn(\beta_2)$ in this
article for simplicity and consider the case $\beta_1 > 0, \beta_2 >0$.  The results for $\beta_1<0, \beta_2<0$ may be
reproduced by simply taking $\beta \rightarrow -\beta$. We choose $\alpha \neq 0$ with (i) $a=b\neq 0$ and (ii) $a \neq b$ for which
the system is necessarily non-${\cal{PT}}$-symmetric and will be shown to admit regular periodic solution as well chaotic behaviour.

\subsection{Linear Stability Analysis}

The Hamilton's equations of motion read,
\bea
&& \dot{x}  = \tilde{\Pi}_y, \ 
\dot{y}= \tilde{\Pi}_x, \ \
\dot{\tilde{P_x}}= \Gamma\left(1-ax^{2}-by^{2}\right)
\tilde{\Pi}_x -2axy \Gamma \tilde{\Pi}_y
 -\beta x-y-3 \alpha x^2 y\nonumber \\
&& \dot{\tilde{P_y}}= -\Gamma\left(1-ax^{2}-by^{2}\right) \tilde{\Pi}_y
-2bxy\Gamma \tilde{\Pi}_x
 -\beta y-x- \alpha x^3.
\label{hami-eqm}
\eea
\noindent The equilibrium points and their stability may be analyzed by employing standard techniques.
In particular, the equilibrium points are determined by the solutions of the algebraic equations
obtained by putting the right hand side of Eq. ( \ref{hami-eqm}) equal to zero. It should be noted that
a Hamiltonian system admits either center i.e. closed orbit in the phase-space surrounding an
equilibrium point or hyperbolic point signalling instability\cite{center}. The stable equilibrium point
of a Hamiltonian system is necessarily a center and is not asymptotically stable. The system admits
five equilibrium points $P_0, P_1^{\pm}, P_2^{\pm}$ in the phase-space $(x,y,\tilde{P}_{x},\tilde{P}_{y})$
of the system,
\bea
&& P_0=(0,0,0,0), P_1^{\pm}= \left (\pm \delta_+, \pm\eta_+, \mp\Gamma \eta_+ \left(1-a {\delta_+}^{2} -\frac{b{\eta_+}^{2}}{3}\right),
\pm \Gamma \delta_+ \left( 1- \frac{a{\delta_+}^{2}}{3} - b {\eta_+}^{2}\right)\right),\nonumber \\
&& P_2^{\pm}=\left (\pm \delta_-, \pm \eta_-, \mp \Gamma \eta_- \left(1-a {\delta_-}^{2} - \frac{b{\eta_-}^{2}}{3}\right),
\pm \Gamma \delta_- \left( 1-\frac{a{\delta_-}^{2}}{3} - b {\eta_-}^{2}\right)\right),
\label{eqpoint0}
\eea
\noindent where $\delta_{\pm}$ and $\eta_{\pm}$ are defined as follows: 
\bea
\delta_{\pm}=\frac{1}{\sqrt{3 \alpha}} \left [ -2  \pm \sqrt{ 1 + 3 \beta^2} 
\right]^{\frac{1}{2}}, \ \
\eta_{\pm}=-\frac{\delta_{\pm}}{3 \beta} \left [ 1 \pm \sqrt{1 + 3 \beta^2}
\right ].
\label{eqpoint1}
\eea
\noindent We denote the coordinates of a generic equilibrium point in the phase-space
as $Z_0 \equiv (x_0, y_0, \tilde{P}_{x_0}, \tilde{P}_{y_0})$. The expressions for the coordinates
of a given equilibrium point  may be obtained from Eqs. (\ref{eqpoint0},\ref{eqpoint1}).
The critical points of the Hamiltonian $\tilde{H}$ are also equilibrium points for the
Eq. (\ref{hami-eqm}), since both are determined from the equation
$\tilde{H}_Z \equiv \frac{\partial \tilde{H}}{\partial Z}=0, Z \equiv (x, y, \tilde{P}_x, \tilde{P}_y)$.
The Dirichlet theorem\cite{dirichlet} states that  a critical point is stable if the Hessian
$\tilde{H}_{ZZ}$ at that point is definite \textemdash no conclusion can be drawn for an indefinite
Hessian. It has been shown in Ref. \cite{pkg-pr} that $\tilde{H}_{ZZ}$
for $a=b=0$ is not definite at $P_0$ and the decision for its stability is inconclusive. 
This results holds for $a \neq 0 \neq b$, since $\tilde{H}_{ZZ}$ evaluated at $P_0$ is same
for both the cases. Thus, the eigenvalues of $\tilde{H}_{ZZ}$ do not depend on $a$ and $b$.
A similar analysis for the points $P_1^{\pm}$ and $P_2^{\pm}$ are cumbersome and not attempted
in this article. We employ linear stability analysis and numerical investigations to explore
the stability of these points. The equilibrium points of a Hamiltonian system are either
center or hyperbolic points.

The linear stability analysis of Eq. (\ref{hami-eqm}) for $a=0=b$ has been analyzed
in detail in Ref. \cite{pkg-pr}. We observe that the effect of non-zero $a, b$ is absent
for the Point $P_0$. The effect on the remaining points $P_1^{\pm}, P_2^{\pm}$
is merely to shift the coordinates $\tilde{P}_{x_0}, \tilde{P}_{y_0}$ by keeping the coordinates
$x_0, y_0$ unchanged. The criteria for linear stability of an equilibrium point do not depend
on the non-linear terms, i.e. the terms with the coefficients $a, b, \alpha$.
This may be seen by considering small fluctuations around an equilibrium point  in the phase
space as $(x = x_0 + \xi_1, y = y_0 + \xi_2 , \tilde{P}_x = \tilde{P}_{x0} + \xi_3,\
\tilde{P}_y = \tilde{P}_ {y_0} + \xi_4)$. We obtain an equation of the form $\dot{\xi}=M \xi$ by 
keeping only the terms linear in $\xi_i$ in Eq. (\ref{hami-eqm}), where $\xi$ is a vector
with four components $\xi_i$ and $M$ is a $4 \times 4$ constant matrix whose eigenvalues determine the stability
of small fluctuations around an equilibrium point. The matrix $M$ is independent of $P_{x_0}$ and
$P_{y_0}$ and so is its eigenvalues\cite{pkg-pr}. Thus, the stability criteria are independent of whether
$a, b$ are vanishing or not. We state the results from Ref. \cite{pkg-pr} for completeness.\\
{\bf Point $P_0$}: The point is an equilibrium point for any $\alpha$ and $\beta \neq 0$. The solutions are
stable in a region of the parameter-space defined by the conditions,
\bea
-\frac{1}{\sqrt{2}} < \Gamma < \frac{1}{\sqrt{2}}, \ \
4 \Gamma^2 \left ( 1 - \Gamma^2 \right ) < \beta^2 < 1.
\label{stab-con}
\eea
{\bf Point $P_1^{\pm}$}: The equilibrium points exist for $\alpha > 0, \beta^2 > 1$ and $\alpha < 0, 0 < \beta^2 < 1$.
The expressions for the co-ordinates in the phase-space becomes complex if the above conditions are not satisfied.
The stable solutions are obtained in the region $\alpha >0$ for $\beta^2 > 1, \Gamma^2 \leq \frac{\sqrt{2}-1}{\sqrt{2}}$.
The point is unstable $\alpha < 0, 0 < \beta^2 < 1$.\\
{\bf Point $P_2^{\pm}$}: The equilibrium points exist for $\alpha < 0, 0 < \beta^2 < 1, \beta^2 > 1$; otherwise,
the co-ordinates become complex. None of the points are stable.\\
This is a multi-stable system \textemdash three out of five equilibrium points are stable.

\subsection{Perturbative Solution}

In this section, the  Hamiltonian of two coupled VdPD oscillators with balanced loss and
gain is analyzed by perturbative methods. The standard perturbation technique fails due to the
appearance of secular terms. The Poincar$\acute{e}$-Lindstedt method is also problematic
for systems with multiple time-scales. The MSA perturbation  method can be applied successfully
to a host of nonlinear equations except for systems with hidden time-scales varying with
non-integer powers of the perturbation-parameter \textemdash the Mathieu equation being one
such example\cite{rg-1}. The RG technique is free from this problem, since no time-scale is assumed
a priory and it is determined by the flow equations. We analyze the system perturbatively
by using RG techniques as well as MSA method. We find that the results match exactly
in the leading order of the perturbation.

We introduce the column matrices $X$, $V$ and a function $Q(x,y)$ as follows,
\bea
X=\begin{pmatrix}
x \\ y
\end{pmatrix}, \ \
{V}(x,y) = \begin{pmatrix} x^3\\ 3 x^2 y \end{pmatrix}, Q(x,y)=1-a x^{2}-b y^{2}.
\eea
\noindent The Eq. (\ref{vdpd-eqn-1}) can be expressed in a compact form,
\bea
\ddot{X} + \left ( I + \beta \sigma_1 \right ) X + 2 \Gamma \sigma_3 Q(x,y) \dot{X} + \alpha V(x,y)=0
\label{mat-eqn}
\eea 
\noindent where $\sigma_a, a=1, 2, 3$ are the Pauli matrices. The perturbation analysis can be
implemented in various ways by identifying one or more small parameters depending on the
physical situations. As discussed in Ref. \cite{pkg-pr} and apparent from linear stability analysis,
the periodic solutions are obtained for small gain-loss parameter $\Gamma$. Either the parameter
$\beta$ or $\alpha$ or both of them can be chosen to be small parameters in order to implement
perturbation scheme. In this article, we consider $\Gamma$, $\beta$ and $\alpha$ as small parameters by choosing  $\Gamma= 
\epsilon \Gamma_0, \alpha=\epsilon \alpha_0, \beta=\epsilon \beta_0 $, where the parameter
$\epsilon$ is taken to be small, i.e. $\epsilon << 1$.
The Eqn. (\ref{mat-eqn}) can be expressed as,
\bea
\ddot{X} +X + \epsilon \left [ 2 \Gamma_0 \sigma_3 Q(x,y) \dot{X}+ \beta_0 \sigma_1 X
+ \alpha_0 \tilde{V}(x) \right ] =0.
\label{vdpd-eqn-2}
\eea
\noindent The coordinates are expressed in powers of the small parameter $\epsilon$
\bea
 X =\sum_{n=0}^{\infty} \epsilon^n X^{(n)}, \ \ X^{(n)} \equiv \bp x_n\\y_n\ep.
\label{var-scale-1}
\eea
\noindent The equations at different orders of the small parameter $\epsilon$ are obtained
by substituting Eq. ({\ref{var-scale-1}}) in Eq. (\ref{vdpd-eqn-2}) and equating the terms
with the same coefficient $\epsilon^n$ to zero. The equations at the zeroth and the first
orders are determined as,
\bea
\label{zero}
&& {\cal{O}}(\epsilon^0): \frac{d^2 X^{(0)}}{d t^2} + X^{(0)}=0,\\
&& {\cal{O}}(\epsilon): \frac{d^2 X^{(1)}}{d t^2} +
 X^{(1)} = - 2 \Gamma_{0} \sigma_3 Q_0 \frac{d X^{(0)}}{d t} - \beta_0 \sigma_1 X^{(0)} - \alpha_0 
\begin{pmatrix}
x_0^3\\
3 x_0^2 y_0
\end{pmatrix},
\label{order1}
\eea
\noindent where we have defined $Q_0 \equiv Q(x_0,y_0)=1-a x_0^2-b x_0^2$.
The unperturbed solution of Eq.  (\ref{zero}) has the form,
\bea
X^{(0)}= {\cal{A}}  \exp(it) + c.c.
\label{sol_0}
\eea
\noindent where c.c. denotes complex conjugate and the transpose of the constant column matrix
${\cal{A}}$ is denoted as ${\cal{A}}^T \equiv (A, B )$.
The solution $X^{(0)}$ of the zeroth order equation is substituted in 
Eq. (\ref{order1}) resulting in the following inhomogeneous equation,
\bea
&& \frac{d^2 X^{(1)}}{d t^2} +  X^{(1)}  = D\nonumber \\
&& D \equiv  -e^{i t} \left [ \beta_0 \sigma_1 {\cal{A}}
+ 2 i \Gamma_0 \sigma_3 \bp A (1-a {\vert A \vert}^2 - 2 b {\vert B \vert}^2 ) + b B^{2}A^{*}\\
B (1-b {\vert B \vert}^{2} - 2 a {\vert A \vert}^2 ) +a A^{2} B^{*} \ep \right . \nonumber \\
& + & \left . 3 \alpha_0 A \bp {\vert A \vert}^2\\ AB^* +2 A^*B \ep\ \right ]
 +  e^{3 i t} \bp -\alpha_{0}A^{3}+2i\Gamma_{0} A \left ( a A^{2}+ b B^{2} \right )\\-3\alpha_{0}A^2 B-
2 i \Gamma_{0} B \left ( a  A^2 + b B^2 \right ) \ep + c.c.
\label{polar-form_0}
\eea
\noindent The general solution up to the first order in $\epsilon$ is obtained as,
\bea
X & = & {\cal{A}} e^{i t}
+ \frac{i \epsilon t}{2} e^{i t} \left [  \beta_0 \sigma_1 {\cal{A}}
+ 2 i \Gamma_0 \sigma_3 \bp A (1-a {\vert A \vert}^2 - 2 b {\vert B \vert}^2 ) + b B^{2}A^{*}\\
B (1-b {\vert B \vert}^{2} - 2 a {\vert A \vert}^2 ) +a A^{2} B^{*} \ep
\; \; \; \right . \nonumber \\
& + & \left . 3 \alpha_0 \bp {\vert A \vert}^2 A\\ A^2B^* +2 {\vert A \vert}^2B \ep\ \right ]
 +  \frac{\epsilon e^{3 i t}}{8} \bp \alpha_{0}A^{3}-2i\Gamma_{0} A \left ( a A^{2}
+ b B^{2} \right )\\3\alpha_{0}A^2B+ 2 i \Gamma_{0} B \left ( a  A^2 + b B^2 \right ) \ep
 + c.c. + {\cal{O}}(\epsilon^2),
\label{sec-sol}
\eea
\noindent Note the appearance of secular terms which make
the series divergent for large $t > \frac{1}{\epsilon}$. There are different regularization schemes.
The Poincar$\acute{e}$-Lindstedt method is not suitable for systems having multiple
time-scales.  We use the MSA method\cite{msa} in the next section. We employ RG method\cite{rg-1} in Sec. \ref{RGT-head}
to regularize the system and show that the results from both the methods agree at the first order of perturbation.

\subsubsection{Method of multiple time-scales}

We solve Eq. (\ref{vdpd-eqn-2}) perturbatively by using MSA methods.
The unperturbed part of the system is described by two decoupled harmonic oscillators satisfying the equation
$\ddot{X} + X=0$. The terms with the coefficient $\epsilon$ in Eq.({\ref{vdpd-eqn-2}}) is
treated as perturbation, which contain the effect of the loss-gain, coupling and nonlinear terms.
The coordinates are expressed in powers of the small parameter $\epsilon$ and multiple time-scales are introduced
as follows,
\bea
T_n= \epsilon^n t, \ \
X = \sum_{n=0}^{\infty} \epsilon^n X^{(n)}(T_0, T_1, \dots).
\label{msa-1}
\eea
\noindent Substituting Eq.(\ref{msa-1}) in Eq.({\ref{vdpd-eqn-2}}) and equating
the terms with the same coefficient $\epsilon^n$ to zero,
the following equations up to ${\cal{O}}(\epsilon)$ are obtained,
\bea
\label{zero.1}
{\cal{O}}(\epsilon^0): && \frac{\partial^2 X^{(0)}}{\partial T_0^2} + X^{(0)}=0,\\
{\cal{O}}(\epsilon): && 
 \frac{\partial^2 X^{(1)}}{\partial {T_{0}}^2} + X^{(1)} =
- \left [2\frac{\partial^2 X^{(0)}}{\partial T_{0}\partial T_{1}} +2 \Gamma_0 \sigma_3 Q_0 \frac{\partial
X^{(0)}}{\partial T_{0}}  +\beta_{0} \sigma_1 X^{(0)} + \alpha_0 \begin{pmatrix} x_0^3\\
3 x_0^2 y_0
\end{pmatrix} \right ].
\label{order1.1}
\eea
\noindent These equations are to be solved consistently to get the perturbative results.
Eq.({\ref{zero.1}}) is solved as,
\bea
X^{(0)}=\bp A(T_{1},T_{2}..)\\ B(T_{1},T_{2}..) \ep  \exp(iT_{0})+c.c.
\label{sol_1}
\eea
\noindent The integration constants $A(T_{1},T_{2}, \dots)$
and $B(T_{1},T_{2}, \dots)$ are independent of $T_0$ and depend on slower time-scales.
Eq. (\ref{order1.1}) is inhomogeneous, since the right hand side is independent of $X^{(1)}$.
We define a column matrix ${\cal{B}}$,
\bea
{\cal{B}} & = & -e^{i T_0} \left [ 2i \frac{\partial{\cal{A}}}{\partial T_1} + \beta_0 \sigma_1 {\cal{A}}
+ 2 i \Gamma_0 \sigma_3 \bp A (1-a {\vert A \vert}^2 - 2 b {\vert B \vert}^2 ) + b B^{2}A^{*}\\
B (1-b {\vert B \vert}^{2} - 2 a {\vert A \vert}^2 ) +a A^{2} B^{*} \ep \right . \nonumber \\
& + & \left . 3 \alpha_0 A \bp {\vert A \vert}^2\\ AB^* +2 A^*B \ep\ \right ]
 +  e^{3 i T_0} \bp -\alpha_{0}A^{3}+2i\Gamma_{0} A \left ( a A^{2}+ b B^{2} \right )\\-3\alpha_{0}A^2B-
2 i \Gamma_{0} B \left ( a  A^2 + b B^2 \right ) \ep + c.c.
\label{polar-form1}
\eea
\noindent Substituting Eq.(\ref{sol_1}) in Eq.(\ref{order1.1}), we obtain
$\frac{\partial^2 X^{(1)}}{\partial t^2} +  X^{(1)}  = {\cal{B}}$.
The term with the factors $e^{\pm i T_0}$  in ${\cal{B}}$ produce secular terms in $X^{(1)}$ which must
be eliminated for a uniform expansion. This is accomplished by setting the coefficients of $e^{\pm i  T_{0}}$
equal to zero:
\bea
2i \frac{\partial{\cal{A}}}{\partial T_1} + \beta_0 \sigma_1 {\cal{A}}
+ 2 i \Gamma_0 \sigma_3 \bp A (1-a {\vert A \vert}^2 - 2 b {\vert B \vert}^2 ) + b B^{2}A^{*}\\
B (1-b {\vert B \vert}^{2} - 2 a {\vert A \vert}^2 ) +a A^{2} B^{*} \ep 
+  3 \alpha_0 A \bp {\vert A \vert}^2\\ AB^* +2 A^*B \ep = 0 .
\label{polar-form2}
\eea
\noindent The above equation governing the slow variation of the amplitudes results in a dimer model with the loss-gain
terms depending linearly as well as nonlinearly on the dependent variables. The two modes of the dimer are coupled
via loss-gain terms in addition to coupling through interaction terms. The VdPD oscillator
in Eq. (\ref{vdpd-eqn-1}) reduces to the coupled Duffing oscillator system\cite{pkg-pr} with balanced loss and gain for
$a=b=0$. The dimer model corresponding to the Duffing oscillator system\cite{pkg-pr} is reproduced from Eq. (\ref{polar-form2})
for $a=b=0$, which  has been shown to be  exactly solvable admitting regular periodic solutions. 

We introduce the Stokes variables as follows,
\bea
Z_a = 2 {\cal{A}}^{\dagger} \sigma_a {\cal{A}}, \ \
R = 2 {\cal{A}}^{\dagger} {\cal{A}} = \sqrt{Z_1^2+Z_2^2+Z_3^2},
\eea
\noindent which reduces Eq. (\ref{polar-form2}) into four equations in terms $Z_a$ and $R$:
\bea
&& \frac{\partial Z_1}{\partial T_1}=0, \ \ Z_1(0) \equiv C_1\nonumber \\
&&\frac{\partial Z_2}{\partial T_1}= \left(\beta_0 +\frac{3 \alpha C_1}{4}\right) Z_3 + \frac{3 \alpha_0 C_1}{4} R +\frac{\Gamma_0 Z_2}{2} \left [
(b-a)R -(a+b) Z_3 \right ]\nonumber \\
&& \frac{\partial Z_3}{\partial T_1} = \frac{\Gamma_0}{2} \left [
(a+b) Z_2^2 - R (4-R(a+b)+Z_3(b-a) )\right ] -\left(\beta_0 +\frac{3 \alpha_0 C_1}{4}\right) Z_2\nonumber \\
&& \frac{\partial R}{\partial T_1} = \frac{\Gamma_0}{2} \left [
\left(b-a\right) Z_2^2 - Z_3 (4-R(a+b) + Z_3(a-b) \right ]
+\frac{3 \alpha_0}{4} C_1 Z_2
\label{stoke-eq}
\eea
\noindent The quantity $Z_1$ is a constant of motion and it's value at $t=0$ is denoted as the constant $C_1$.
It seems that the system of coupled nonlinear equations defined by Eq. (\ref{stoke-eq}) is not amenable to analytical
solutions in its full generality. The linear stability analysis around the equilibrium point $Z_2=Z_3=R=0$
gives a set of coupled linear equations which have the form given by Eq. (\ref{stoke-eq}) with $a=b=0$.
It may be noted that Eq. (\ref{stoke-eq}) is exactly solvable in the limit $a=b=0$ and has been studied earlier
in the context of coupled Duffing-oscillators system with BLG\cite{pkg-pr}. The general solution of Eq. (\ref{stoke-eq})
for small fluctuations around the equilibrium point may be obtained from Ref.  \cite{pkg-pr}
and one particular solution is reproduced\footnote{The corresponding solution
in Ref. \cite{pkg-pr} contains $\epsilon^2 t$ instead of $\epsilon t$ as in Eq. (\ref{amp-msa}). This is because
$T_1=\epsilon^2 t$ in Ref. \cite{pkg-pr}, while $T_1=\epsilon t$ in the present article.}  below for the sake of completeness:
\bea
A(t) & = & \sqrt{\frac{C_1 \beta_0}{\eta}} \
e^{i \frac{2 \eta^2 + 3 C_1 \beta_0}{4 \eta} \epsilon t},\nonumber \\
B(t) & = & \sqrt{\frac{C_1}{2 \eta} (2 \beta_0+3 \alpha C_1)}
\ \ e^{i \left [ \frac{ 2 \eta^2 +
3 C_1 \beta_0}{4 \eta} - \tan^{-1} \left ( \frac{2 \Gamma_0}{\eta} \right )
\right ] \epsilon t}.
\label{amp-msa}
\eea
\noindent The stationary modes are obtained in regions of the parameter-space determined by the condition:
\bea
&&\eta^2 \equiv \beta_0^2 + \frac{3}{2} \alpha \beta_0 C_1 - 4 \Gamma_0^2 \geq 0,\nonumber \\
&& \textrm{For} \ \ \alpha \geq 0: C_1 \beta_0 > 0,\ \
\textrm{For} \ \ \alpha < 0:  0 < C_1 \beta_0 < \frac{2 \beta_0^2}{3 {\vert \alpha \vert}}
\label{condi-msa}
\eea
\noindent The expressions for $(A, B)$ in Eq. (\ref{amp-msa}) along with the condition (\ref{condi-msa}) define
approximate slow variation of the amplitudes to the leading order in the perturbation. 

It appears that Eq. (\ref{polar-form2}) is not analytically solvable with its full generality. The effect of
space-dependence of the BLG terms is not seen in the approximate expressions (\ref{amp-msa}) obtained by
linearizing the dimer system (\ref{stoke-eq}) which do not contain
the parameters $a, b$. We find exact solutions of Eq. (\ref{polar-form2}) for $a \neq 0 \neq b$ with specific
initial conditions and the effects of space-dependent BLG terms can be observed.  We use polar decomposition of ${\cal{A}}$,
\bea
A= A_0(T_1, T_2, \dots)  e^{i \theta_{A}(T_1, T_2, \dots)},
B= B_0(T_1,T_2,\dots)  e^{i \theta_{B}(T_1, T_2, \dots)}.
\label{polar-form}
\eea 
\noindent where the radial variables $ A_0, B_0$ and the phases $\theta_A, \theta_B$ depend on slower time scales.
The following equations are obtained after separating the real and imaginary parts of Eq.  (\ref{polar-form2}),
\bea
&& \frac{\partial A_0}{\partial T_1} + \Gamma_0 A_0 \left [ 1-a A_0^2 - 2 b B_0^2 + b B_0^2 
\cos\left(2\left(\theta_B-\theta_A\right)\right) \right ]
+\frac{1}{2} \sin\left(\theta_B-\theta_A \right ) \beta_0 B_0 = 0,\nonumber \\
&& \frac{\partial B_0}{\partial T_1} - \Gamma_0 B_0 \left [ 1-2 a A_0^2 - b B_0^2 + a A_0^2 
\cos\left(2\left(\theta_B-\theta_A\right)\right) \right ]
-\frac{1}{2} \sin\left(\theta_B-\theta_A \right ) \beta_0 A_0\nonumber \\
&& + \frac{3 \alpha_0}{2} B_0 A_0^2 \sin\left ( 2 \left (
\theta_A -\theta_B \right ) \right ) = 0,\nonumber \\
&& \frac{\partial \theta_B}{\partial T_1} - \frac{3 \alpha_0}{2} A_0^2 \left [2+
\cos\left ( 2 \left (\theta_B-\theta_A\right ) \right)\right ]
 - \frac{\beta_0 A_0}{2 B_0} \cos(\theta_B-\theta_A)
+ \Gamma_0 a A_0^2 \sin \left(2\left(\theta_B-\theta_A\right ) \right )=0,\nonumber \\
&& \frac{\partial \theta_A}{\partial T_1} + \frac{3 \alpha_0}{2} A_0^2 - \frac{\beta_0 B_0}{2 A_0} \cos(\theta_B-\theta_A)
+ \Gamma_0 b B_0^2 \sin \left(2\left(\theta_B-\theta_A\right ) \right )=0.
\label{p-scale}
\eea
\noindent We take the limit of small $\theta_A, \theta_B$ in the above equations and obtain,
\bea
\label{flow-1}
&& \frac{\partial A_0}{\partial t}=-\Gamma A_0 \left [ 1-(a A_0^{2}+b B_0^{2}) \right ]  \label{A_0},\\
\label{flow-2}
&&\frac{\partial B_0}{\partial t}=\Gamma B_0 \left [ 1-(a A_0^{2}+bB_0^{2}) \right ] \label{B_0}\\
&&
\frac{\partial \theta_{A}}{\partial t}=\frac{\beta B_0}{2 A_0}-\frac{3\alpha A_0^{2}}{2}, \ \
\frac{\partial \theta_{B}}{\partial t}=\frac{\beta A_0}{2 B_0}+\frac{9\alpha A_0^{2}}{2}
\label{flow-3}
\eea
\noindent which are exactly solvable. The limit of small $\theta_A, \theta_B$ is equivalent to
choosing the initial conditions $X(0)={\cal{A}}$ and $\dot{X}(0)=0$. This leads the conditions
on the zeroth order solutions $X^{(0)}(0)={\cal{A}}, \dot{X}^{(0)}(0)=0$ and
$\theta_A^{(0)}=0=\theta_B^{(0)} \ \forall \ t$. The phases  $\theta_A, \theta_B$
receive non-vanishing contribution only at the ${\cal{O}}(\epsilon)$ and higher orders.
The simplest case $a=b=0$ corresponds to coupled Duffing
oscillators model of Ref. \cite{pkg-pr} and there are no periodic solutions for small $\theta_A,
\theta_B$ or equivalently with the initial conditions $X^{(0)}(0)={\cal{A}}, \dot{X}^{(0)}(0)=0$.
The amplitude $A_0$ decays with time for $\Gamma >0$ and becomes unbounded for
$\Gamma < 0$. Similarly, the amplitude $B_0$ grows exponentially with time for $\Gamma >0$
and decays for $\Gamma < 0$. The periodic solutions are obtained with different initial
conditions\cite{pkg-pr}. However, periodic solutions are obtained in this limit for
$a \neq 0 \neq b$. The effect of the space-dependent loss-gain terms is to change
the nature of the solutions. Multiplying Eq. (\ref{flow-1}) by $B_0$, Eq.
(\ref{flow-2}) by $A_0$ and adding the resulting equations determine $B_0 = \frac{K}{A_0}$,
where $K$ is an integration constant. Substituting $B$ in Eqs. (\ref{flow-1},\ref{flow-3}) and
denoting $p=A_0^2$, we obtain the following equations:
\bea
\label{flow-A}
&& \frac{\partial p}{\partial t} = {2 \Gamma} \left ( a p^2 -  p + b K^2 \right )\\
&& \theta_A = \int dt \left [ \frac{\beta K}{2 p} + \frac{3 \alpha}{2} p \right ] + \theta_{A0}, \ \
\theta_B= \int dt \left ( \frac{\beta}{2 K} + \frac{9 \alpha}{2} \right ) p + \theta_{B0},
\label{flow-T}
\eea
\noindent where $\theta_{A0}$ and $\theta_{B0}$ are integration constants. The primary task is
to find the solution for $p$ so  that $B_0=\pm \frac{K}{\sqrt{p}}$ is determined directly and
$\theta_A, \theta_B$ are obtained after performing the integration. For $\Gamma=0$, $p$ is constant
and $\theta_A, \theta_B$ vary linearly with $t$. We consider $\Gamma \neq 0$ for which Eq. (\ref{flow-A}) has three
distinct solutions in three regions of parameter-space characterized by
$\bigtriangleup \equiv 1 - 4 ab K^2$, where $ \bigtriangleup$
is the discriminant of the quadratic polynomial $a p^2- p  + bK^2$.
The solution for $\bigtriangleup < 0$ is singular at finite time and discarded. However, the solutions
for $\bigtriangleup =0$ and $\bigtriangleup > 0$ can be chosen to be finite for all time $t \geq 0$ with
appropriate choice of initial conditions. The solutions for $A_0, B_0, \theta_A, \theta_B$ in these
two regions for $\Gamma \neq 0$ are described below.

The solution for $\bigtriangleup = 0, i.e. K=\frac{1}{2 \sqrt{ab}}$
has the form,
\bea
A_0(t) & = & \sqrt{\frac{1}{2a}} \left ( 1 - \frac{1}{\Gamma t + c_1} \right )^{\frac{1}{2}}, \ \
B_0(t) = \sqrt{\frac{1}{2b}} \left ( 1 - \frac{1}{\Gamma t + c_1} \right )^{-\frac{1}{2}},\nonumber \\
\theta_{A}(t) & = & \left ( \frac{\beta}{2} \sqrt{\frac{a}{b}} + \frac{3 \alpha}{4a} \right ) t
+\frac{1}{\Gamma} ln \left [  \frac{\left ( \Gamma t +c_1-1 \right )^{\frac{\beta}{2} { \sqrt\frac{a}{b}}}}
{\left ( \Gamma t + c_1 \right )^{\frac{3 \alpha}{4a}} }\right ]
+ \frac{\beta(c_1-1)}{2\Gamma} \sqrt{\frac{a}{b}} + \theta_{A0}\nonumber \\
\theta_{B}(t) & = & \left ( \frac{\beta}{2} \sqrt{\frac{b}{a}} +\frac{9\alpha}{4a} \right )
\left  ( t -\frac{1}{\Gamma} \ln({\Gamma t}+c_1) \right )+\theta_{B0}
\label{delgeq0}
\eea
\noindent where $c_1$ is an integration constant and we have chosen a positive $K$ without loss of
any generality.  The factor $\sqrt{\frac{a}{b}}$ is real either
for $(a, b) >0$ or $(a,b) <0$. We choose $c_1 > 1 $ for $\Gamma >0$ so that singularity 
at finite $t$ within the range $0 \leq t < \infty$ is avoided and $A_0, B_0$ are real.
We choose $c_1 <0$ for $\Gamma <0$ and the imaginary terms appearing in $\theta_A$ and $\theta_B$
from the negative of the logarithm may be absorbed in the integration constants $\theta_{A0}$
and $\theta_{B0}$, respectively. Alternatively, the parameters may be chosen such that the quantities
within the  logarithm are always positive. The solution for $\bigtriangleup >0$,
\bea
A_0(t) & = & \sqrt{\frac{1}{2a}} \left [ 1- \frac{\sqrt{\bigtriangleup}}{2}
\coth\left(\frac{\sqrt{\bigtriangleup}}{2} \left ( \Gamma t + c_2\right) \right ) \right ]^{\frac{1}{2}}, \
B_0(t)=\frac{K \sqrt{{2a}}}{ \left [ 1- \frac{\sqrt{\bigtriangleup}}{2} \coth\left(\frac{\sqrt{\bigtriangleup}}{2}
\left ( \Gamma t + c_2\right) \right ) \right ]^{\frac{1}{2}}}\nonumber \\
\theta_A (t)& = & \left ({3 \alpha} + \frac{\beta}{4 b K} \right ) t + \frac{\beta c_2}{4 b K\Gamma} +
\left (\frac{\beta}{4b K} -{3 \alpha} \right ) \frac{1}{\Gamma}  ln \left \{ \sinh \left
( \frac{\sqrt{\bigtriangleup}}{2} \left ( \Gamma t + c_2 \right )\right) \right \}\nonumber \\
& + & \frac{\beta}{4b K \Gamma} ln \left \{ 2 \coth\left ( \frac{\sqrt{\bigtriangleup}}{2} \left ( \Gamma t +
c_2 \right )\right) - \sqrt{\bigtriangleup} \right \} + \theta_{A0}\nonumber \\
\theta_B(t) & = & \frac{1}{4a} \left (\frac{\beta}{K} +{9 \alpha} \right ) \left [ t - \frac{1}{\Gamma}
ln \left \{ \sinh \left ( \frac{\sqrt{\bigtriangleup}}{2} \left ( \Gamma t + c_2 \right )\right) \right \} \right ]
+ \theta_{B0}
\label{delge0}
\eea
\noindent where the integration constant $c_2$ satisfies the condition $\coth^{-1}
\left ( \frac{\sqrt{\bigtriangleup}}{2} \right ) < c_2 < \coth^{-1} \left ( \frac{2}{\sqrt{\bigtriangleup}} \right )$
for $\Gamma > 0$ which implies $\bigtriangleup > 4$. The condition $\bigtriangleup > 4$ can be implemented in various ways:
(i) $ a >0, b> 0, K<0$, (ii) $a<0, b<0, K <0$, (iii) $a >0, b <0, K>0$, (iv) $a<0, b >0, K > 0$.
We choose the cases (i) and (iii) so that $A_0$ and $B_0$ are real. The constant $c_2$ may be chosen as  $c_2 <0$ for
$\Gamma <0$ and the imaginary terms appearing in $\theta_A$ and $\theta_B$
from the negative of the logarithm may be absorbed in the integration constants $\theta_{A0}$
and $\theta_{B0}$, respectively. Alternatively, the parameters may be chosen such that the quantities
within the the logarithm are always positive.  
The constants $\theta_{A0}$ and $\theta_{B0}$ may be chosen appropriately for both $\bigtriangleup=0$
and $\bigtriangleup > 0$ so that $\theta_A(0)=0=\theta_B(0)$. The integration constants $c_1$ and $c_2$ may
be chosen appropriately to fix the values of $A_0(0)$ and $B_0(0)$ depending on the physical
requirement.

\subsubsection{RG Techniques\label{RGT-head}}

The RG technique involves introducing an intermediate time $\tau$ within the time-interval
$(t_0,t)$ to split it into two sub-intervals as $t-t_0= (t-\tau) + (\tau-t_0)$, where 
the initial time $t_0=0$ for the present case. The objective is to cancel the secular terms
at the moment $\tau$ and construct the slow time-dependence of the integration constant
${\cal{A}}$ of the unperturbed problem. We denote ${\cal{A}}={\cal{A}}(0)$ to distinguish it
from the amplitude ${\cal{A}}(t)$ where slow time-variation has been incorporated. The
${\cal{A}}{(0)}$ is expanded as follows,
\bea
{\cal{A}}(0)= {\cal{A}}(\tau) +\sum_{n=1}^{\infty} \epsilon^n {\cal{A}}^T(\tau) {\cal{A}}^{(n)}, \
{\cal{A}}^{(n)} \equiv \bp a^{(n)}\\ b^{(n)} \ep
\label{rg-con}
\eea
\noindent where $a^{(n)}, b^{(n)}$ are
constants.  We rewrite the overall multiplicative factor $t$ in the second term of Eq. (\ref{sec-sol})
as $t=(t-\tau)+\tau$, substitute the expressions for ${\cal{A}}(0)$
by using Eq. (\ref{rg-con}) and keep the terms up to the order of $\epsilon$. The coefficients
$a^{(1)}, b^{(1)}$ are chosen such that the terms proportional
to $\tau$ are cancelled by keeping the terms proportional to $t-\tau$:
\bea
{\cal{A}}^{(1)} = - \frac{i \tau}{2} \left [  \beta_0 \sigma_1 {\cal{A}}
+ 2 i \Gamma_0 \sigma_3 \bp A (1-a {\vert A \vert}^2 - 2 b {\vert B \vert}^2 ) + b B^{2}A^{*}\\
B (1-b {\vert B \vert}^{2} - 2 a {\vert A \vert}^2 ) +a A^{2} B^{*} \ep
+ 3 \alpha_0 \bp {\vert A \vert}^2 A\\ A^2B^* +2 {\vert A \vert}^2B \ep\ \right ]
\eea
\noindent The resulting expression for $X$ depends on $\tau$ explicitly:
\bea
X & = & e^{it} \left [ {\cal{A}} 
+ \frac{i \epsilon (t-\tau)}{2} \left \{  \beta_0 \sigma_1 {\cal{A}}
+ 2 i \Gamma_0 \sigma_3\bp A (1-a {\vert A \vert}^2 - 2 b {\vert B \vert}^2 ) + b B^{2}A^{*}\\
B (1-b {\vert B \vert}^{2} - 2 a {\vert A \vert}^2 ) +a A^{2} B^{*} \ep
\; \; \; \right . \right . \nonumber \\
& + & \left . \left . 3 \alpha_0 \bp {\vert A \vert}^2 A\\ A^2B^* +2 {\vert A \vert}^2B \ep\ \right \} \right ]
 +  \frac{\epsilon e^{3 i t}}{8} \bp \alpha_{0}A^{3}-2i\Gamma_{0} A \left ( a A^{2}
+ b B^{2} \right )\\3\alpha_{0}A^2B+ 2 i \Gamma_{0} B \left ( a  A^2 + b B^2 \right ) \ep
 + c.c. + {\cal{O}}(\epsilon^2)\nonumber
\label{sec-sol-1}
\eea
\noindent The parameter $\tau$ does not appear in the original problem
and should not be present in the final solution. The $\tau$-dependence of $X$ is removed by imposing
$\frac{\partial X}{\partial \tau}|_{\tau=t}=0, \forall \ t $ leading to the RG flow equation which is
same as Eq. (\ref{polar-form2}) with the substitution of $T_1=\epsilon t, \alpha=\epsilon \alpha_0, \beta=\epsilon
\beta_0, \Gamma=\epsilon \Gamma_0$. The removal of $\tau$ requires setting $\tau=t$ for which the secular
terms vanish and the expression of $X$ is given as,
\bea
X  = e^{it} {\cal{A}} +  \frac{\epsilon e^{3 i t}}{8} \bp \alpha_{0}A^{3}-2i\Gamma_{0} A \left ( a A^{2}
+ b B^{2} \right )\\3\alpha_{0}A^2B+ 2 i \Gamma_{0} B \left ( a  A^2 + b B^2 \right ) \ep
 + c.c. + {\cal{O}}(\epsilon^2),
\eea
\noindent where the modulus and phases of $A(t), B(t)$, i.e. $(A_0, \theta_A), (B_0, \theta_B)$ are
given by Eq. (\ref{delgeq0}) for $\bigtriangleup=0$ and from Eq. (\ref{delge0}) for $\bigtriangleup > 0$.

\subsection{Numerical Solution}

\begin{figure}[htbp]
\centering
\begin{subfigure}{.5\textwidth}
\centering
\includegraphics[width=0.8\linewidth]{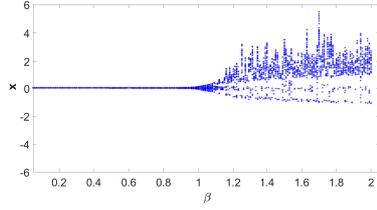}
\caption{ $\alpha=0.5, \Gamma=0.01, a=1.0, b=1.0$}
\label{bifurcation-xvsbeta-h}
\end{subfigure}%
\begin{subfigure}{.5\textwidth}
\centering
\includegraphics[width=0.8\linewidth]{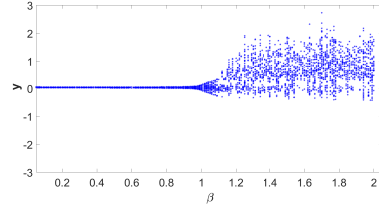}
\caption{ $\alpha=0.5, \Gamma=0.01, a=1.0, b=1.0$}
\label{bifurcation-yvsbeta-h}
\end{subfigure}%
\caption{(Color online) Bifurcation diagrams for $\beta$ 
with the initial conditions $x(0)=0.01$, $y(0)=0.02$, $\dot{x}(0)=0.03$, $\dot{y}(0)=0.04$}
\label{fig-bifr}
\end{figure}

The Hamiltonian system is described in terms of five independent parameters $\Gamma$, $\beta, \alpha, a$ and $b$.
We have numerically studied the bifurcation diagram by varying one of these parameters and keeping the remaining
four  parameters as fixed. The bifurcation diagrams for varying $\beta$ is  shown in Fig. \ref{fig-bifr}
for $\Gamma=0.01$ and $\alpha=0.5$ with the initial values of the dynamical variables chosen around the point $P_0$.
\noindent  The time-series has different behaviours depending on the values of $\beta$. 
The regular periodic solutions are obtained in the range $0 \leq \beta < 1.05$ and the chaotic behavior
onsets from $\beta \geq \beta_c =1.05$. The bifurcation diagram is symmetric with respect to the point $\beta=0$ and the
results are shown only for $\beta \geq 0$. The qualitative feature of the bifurcation diagram is similar to
the  case with $a=b=0$. The crossover from regular to chaotic dynamics as $\beta$ is varied through $\beta_c$ may
be understood by interpreting the $x$ degree of freedom as describing a forced VdPD oscillator with the identification
of $\beta y$ as the forcing term. Unlike the standard forced  VdPD oscillator, the forcing is determined in a nontrivial
way from the solution of the system. The chaotic behaviour of $y$ degree of freedom is induced via its coupling to
the $x$ degree of freedom. The two oscillators are also coupled via the space-dependent loss-gain terms. However,
the bifurcation diagram for varying $b$ and constant $\alpha, \beta, \Gamma, a$ does not show any chaotic behaviour
under the similar initial conditions for $\beta =0$ as well as $\beta \neq 0$. 

\subsubsection{Regular Dynamics}

\begin{figure}[ht!]
\begin{subfigure}{.5\textwidth}
\centering
\includegraphics[width=.8\linewidth]{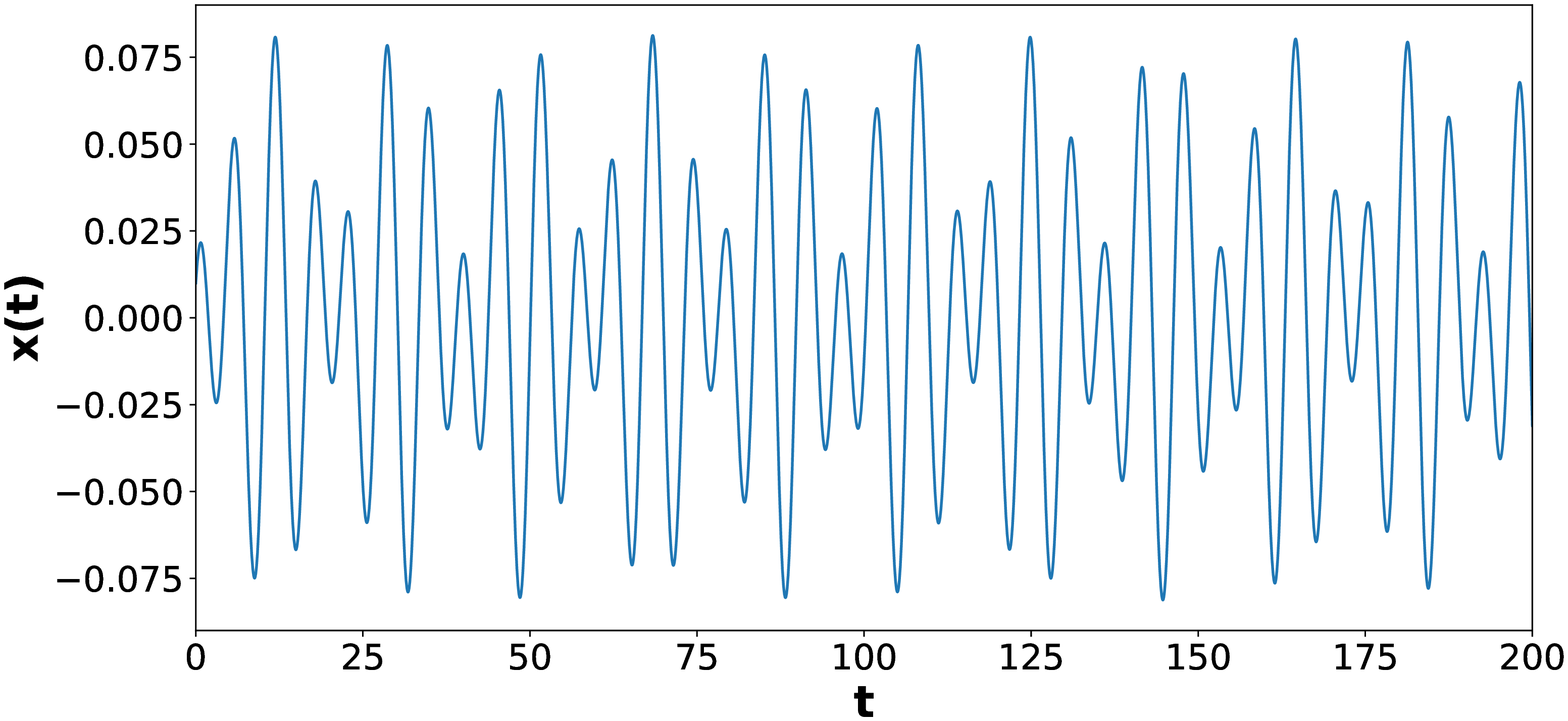} 
\caption{ $\alpha=1.0, \beta=0.5,\Gamma=0.2,a=1.0,b=1.0$}
\label{xtimeseries1}
\end{subfigure}%
\begin{subfigure}{.5\textwidth}
\centering
\includegraphics[width=.8\linewidth]{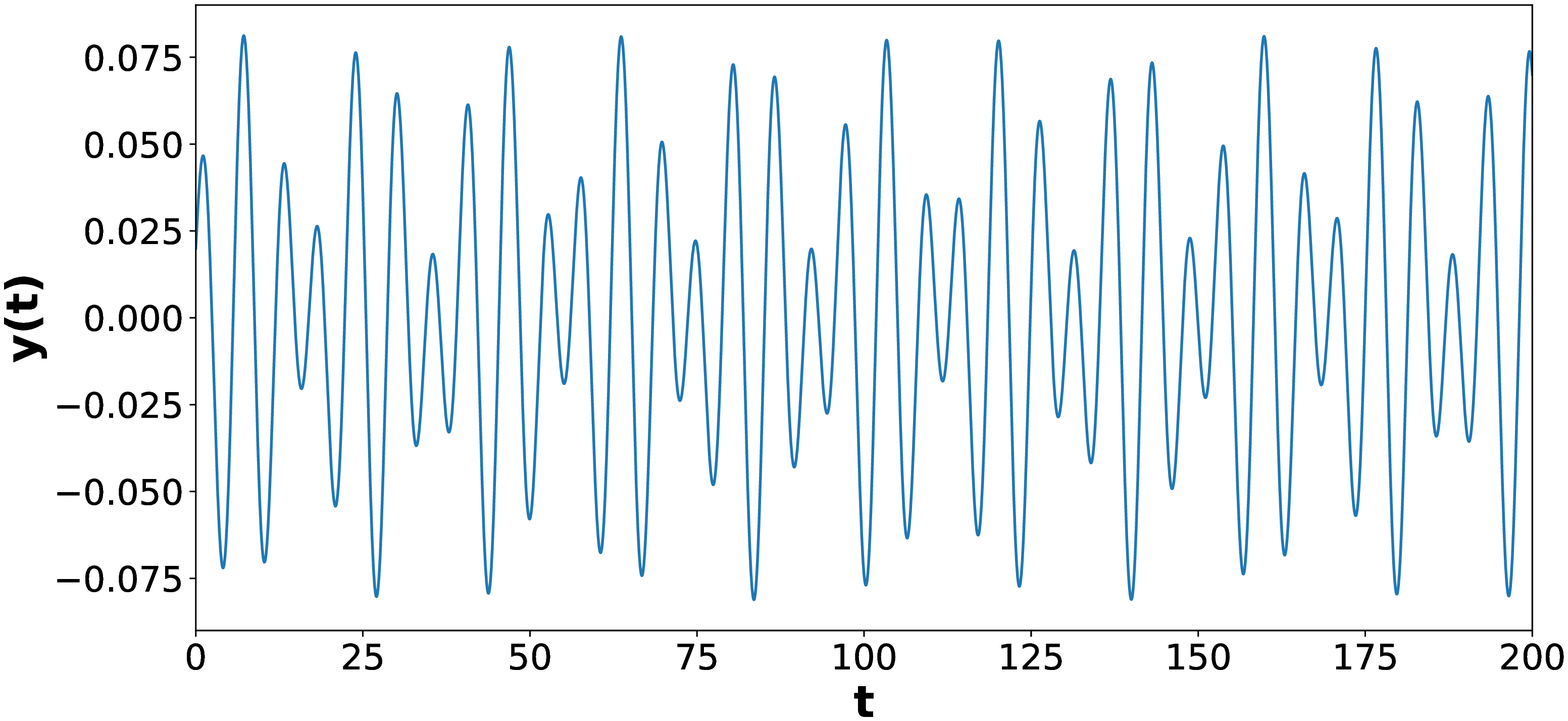}  
\caption{$\alpha=1.0, \beta=0.5,\Gamma=0.2,a=1.0,b=1.0$}
\label{ytimeseries1}
\end{subfigure}%
\newline
\begin{subfigure}{.5\textwidth}
\centering
\includegraphics[width=.8\linewidth]{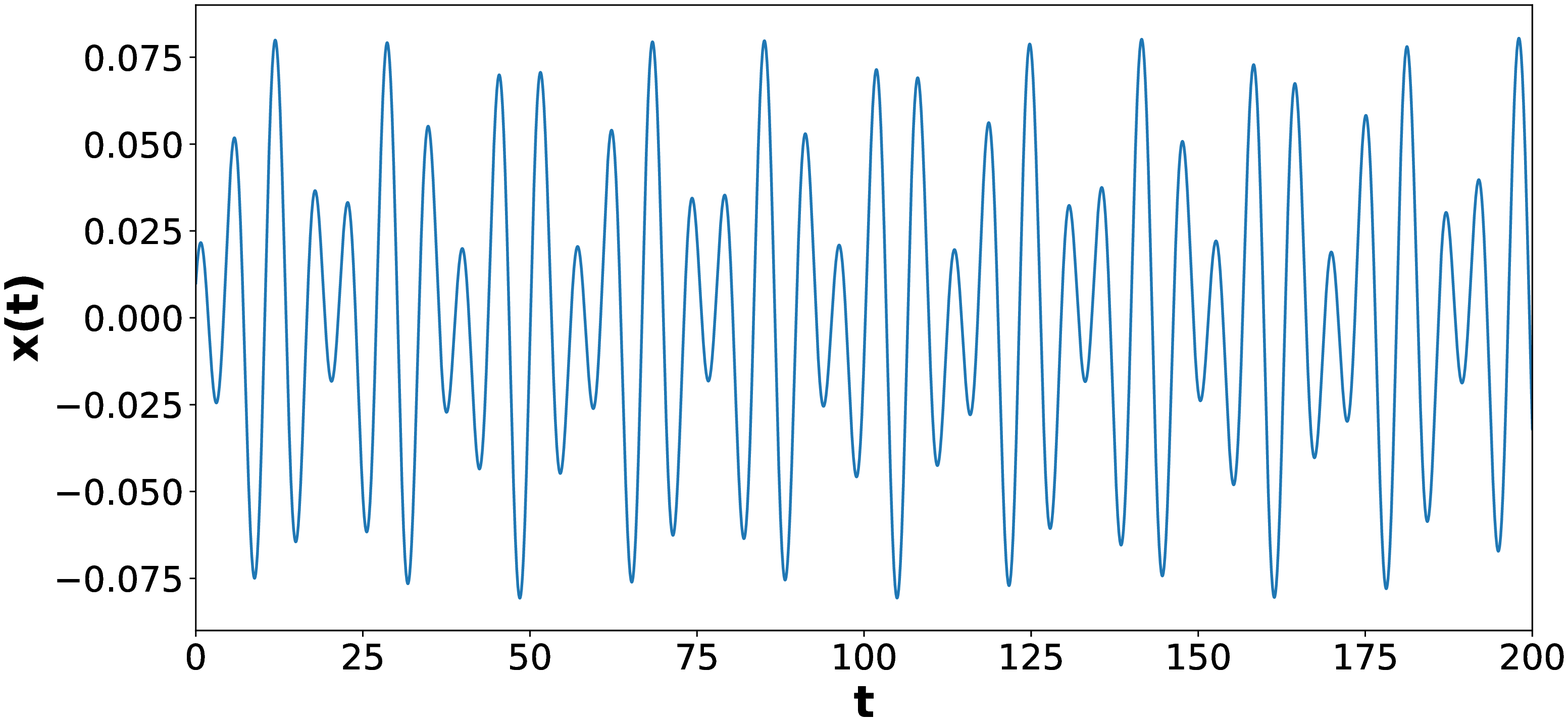}
\caption{$\alpha=1.0, \beta=0.5,\Gamma=0.2,a=1.0,b=5.0$}
\label{xtimeseries2}
\end{subfigure}%
\begin{subfigure}{.5\textwidth}
\centering
\includegraphics[width=.8\linewidth]{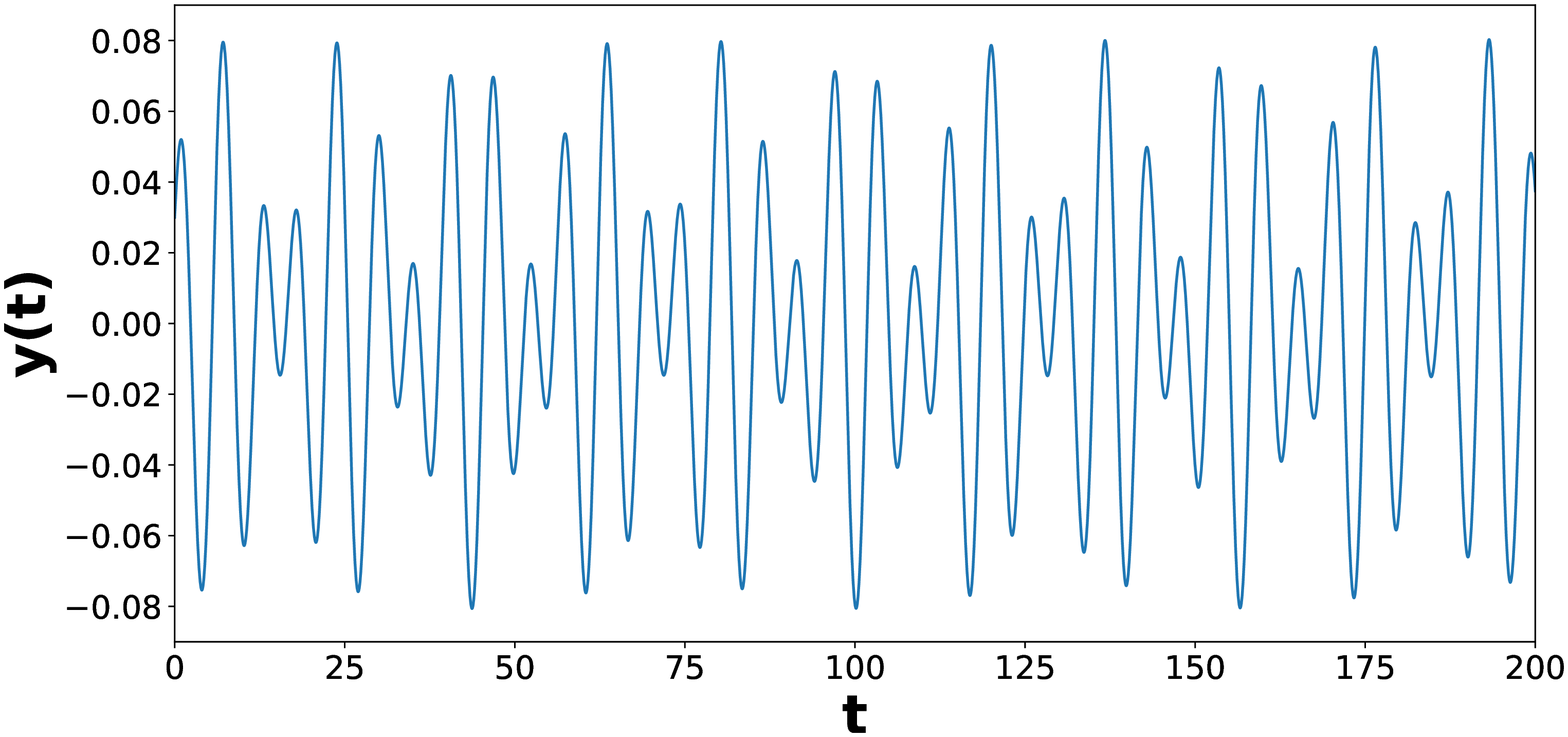}
\caption{$\alpha=1.0, \beta=0.5,\Gamma=0.2,a=1.0,b=5.0$}
\label{ytimeseries2}
\end{subfigure}%
\caption{(Color online) \ Regular solutions of Eq. (\ref{vdpd-eqn-1}) in the vicinity of the point
$P_0$ with the initial conditions $x(0)=0.01, y(0)=0.02, \dot{x}(0)=0.03$ and $\dot{y}(0)=0.04$.}
\label{time-series-P0}
\end{figure}

The time-series of the dynamical variables  in the vicinity of the point $P_0$ is shown in
Fig. \ref{time-series-P0} for $\Gamma=0.2, \beta=0.5$,  $\alpha= 1.0$, $a =b $  and $a \neq b$.
Periodic solutions in Figs. \ref{xtimeseries1} and \ref{ytimeseries1} correspond to $a=1.0,b=1.0$,
while Figs.  (\ref{xtimeseries2}) and (\ref{ytimeseries2}) correspond to $a=1.0,b=5.0$.
\noindent It may be noted that the time evolution of the dynamical variables
with the same initial conditions and fixed $\Gamma, \beta, a, b$ show similar
oscillatory behaviour for positive as well as negative $\alpha$. There are minute changes in
amplitudes and phases and that too in the limit of large $t$. The Lyapunov exponents and
the autocorrelation functions for the time series representing the periodic solutions in
Fig. \ref{time-series-P0} have been calculated to confirm that these solutions are indeed regular. 
The numerical results are consistent with perturbative analysis.

\begin{figure}[ht!]
\begin{subfigure}{.5\textwidth}
\centering
\includegraphics[width=.8\linewidth]{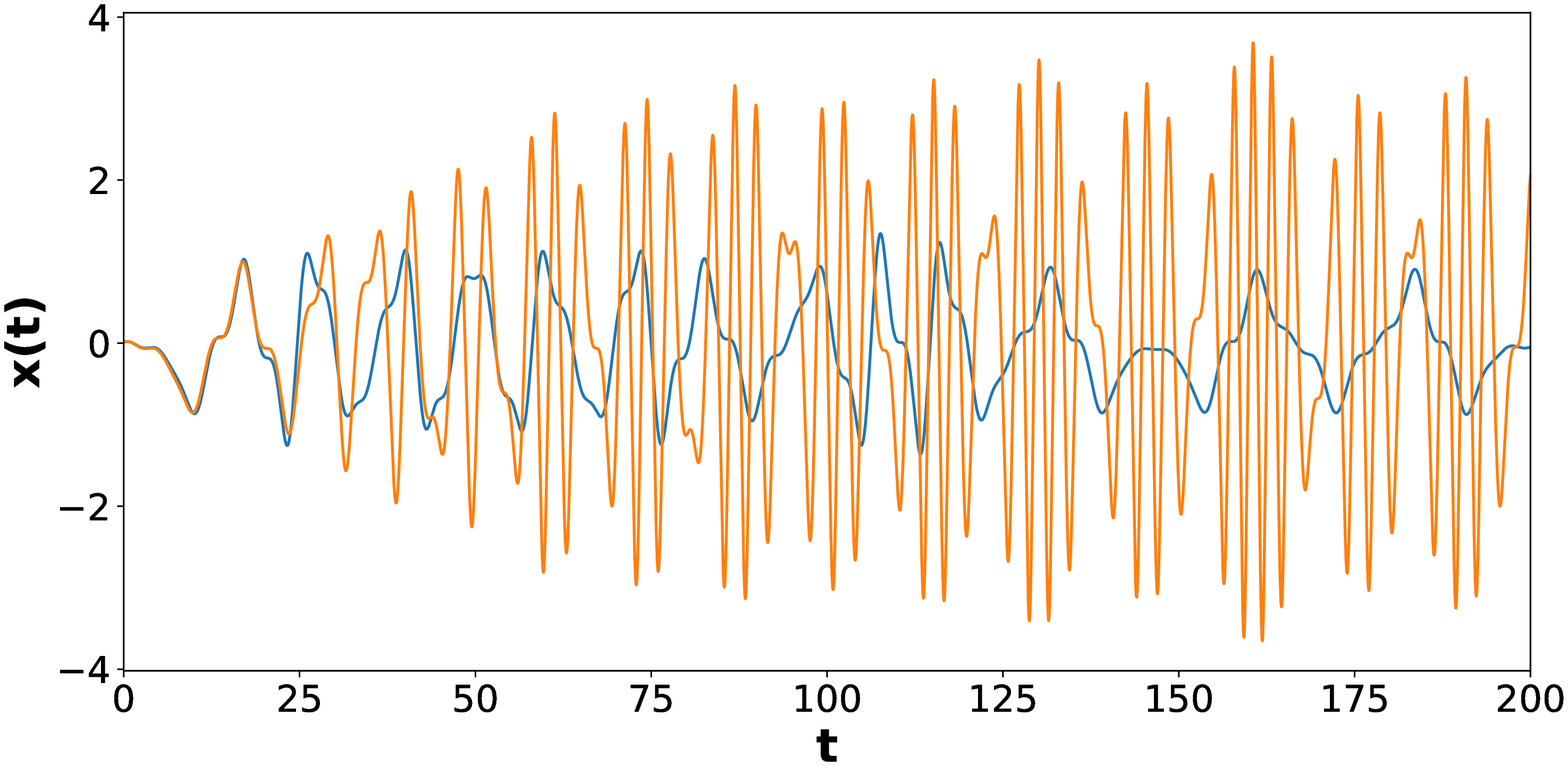} 
\caption{ $\Gamma=0.01, \beta=1.3,\alpha=0.5, a=1.0, b=1.0$}
\label{chaotic1}
\end{subfigure}%
\begin{subfigure}{.5\textwidth}
\centering
\includegraphics[width=.8\linewidth]{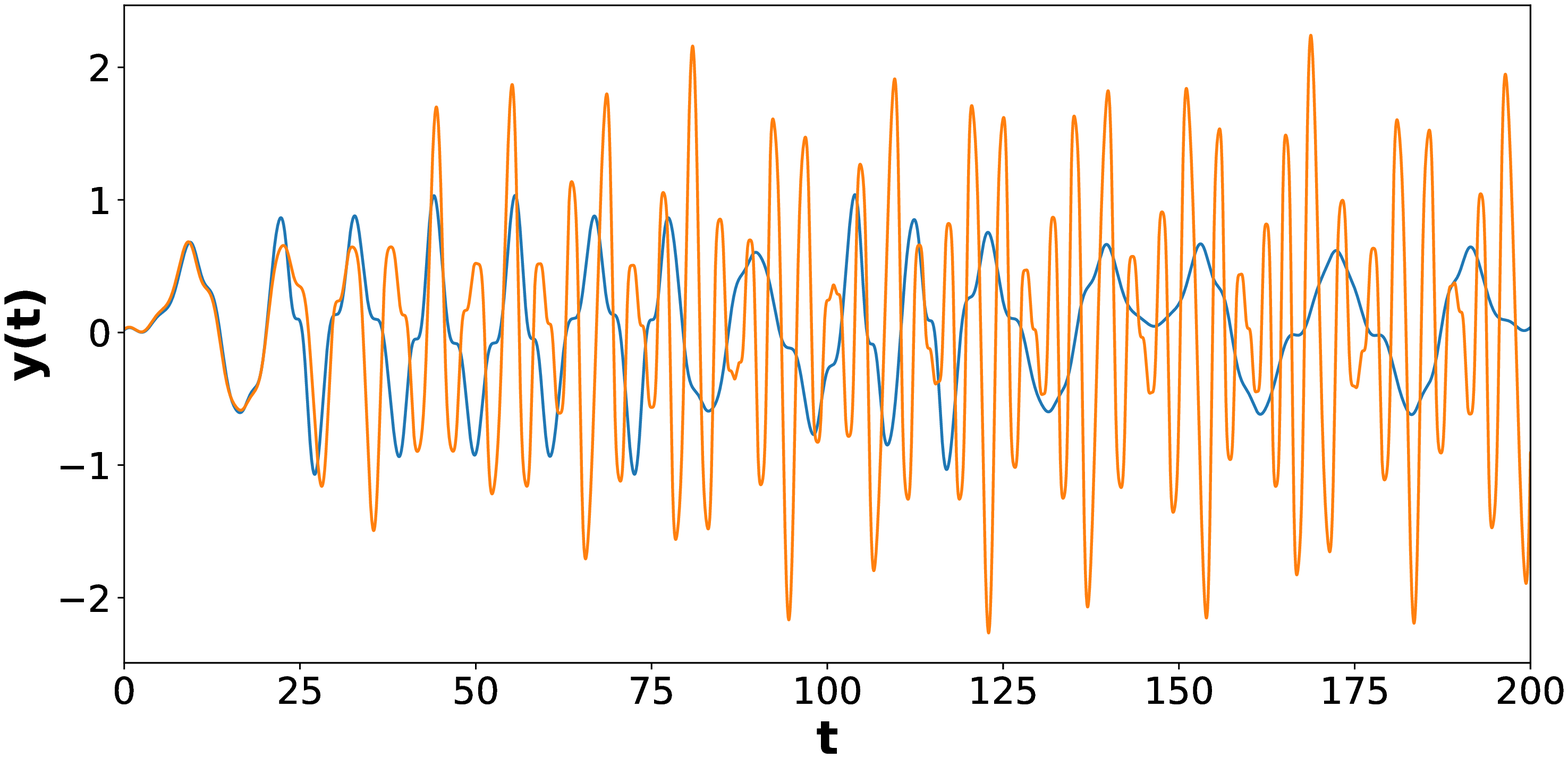}  
\caption{$\Gamma=0.01, \beta=1.3,\alpha=0.5, a=1.0, b=1.0$}
\label{chaotic2}
\end{subfigure}%

\caption{(Color online) Chaotic solutions of Eq. (\ref{vdpd-eqn-1}) with two sets of initial
conditions (a) $x(0)=0.01,y(0)=0.02,\dot{x}(0)=0.03,\dot{y}(0)=0.04$ (blue color) and 
(b) $x(0)=0.01,y(0)=0.025,\dot{x}(0)=0.03,\dot{y}(0)=0.04$ (orange color).
} 
\label{chaotic}
\end{figure}

\begin{figure}[ht!]
\begin{subfigure}{.5\textwidth}
\centering
\includegraphics[width=.8\linewidth]{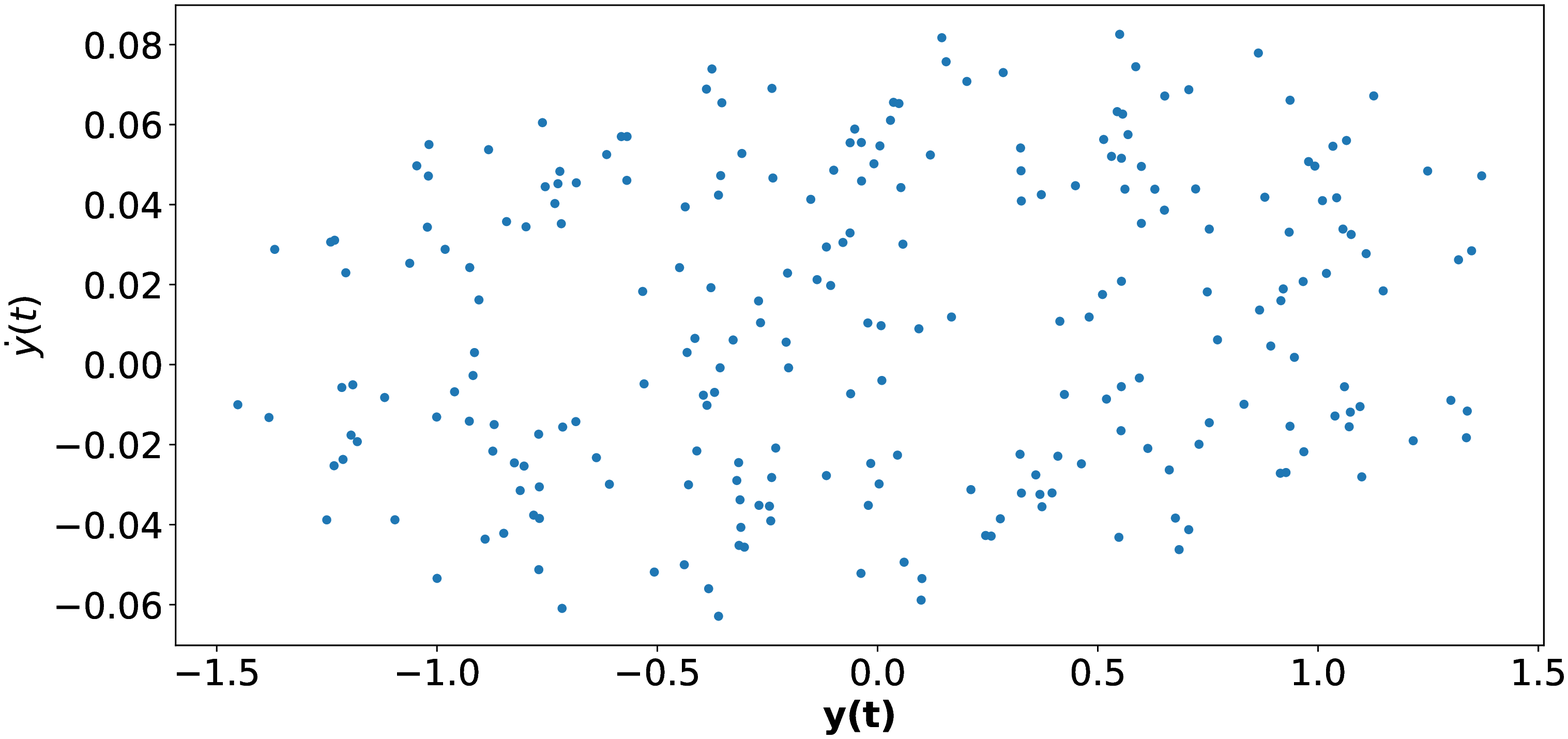}
\caption{Poincar$\acute{e}$ section: $\dot{y}(t)$ VS. $y(t)$ plot}
\label{multi-poincare}
\end{subfigure}%
\begin{subfigure}{.5\textwidth}
\centering
\includegraphics[width=.8\linewidth]{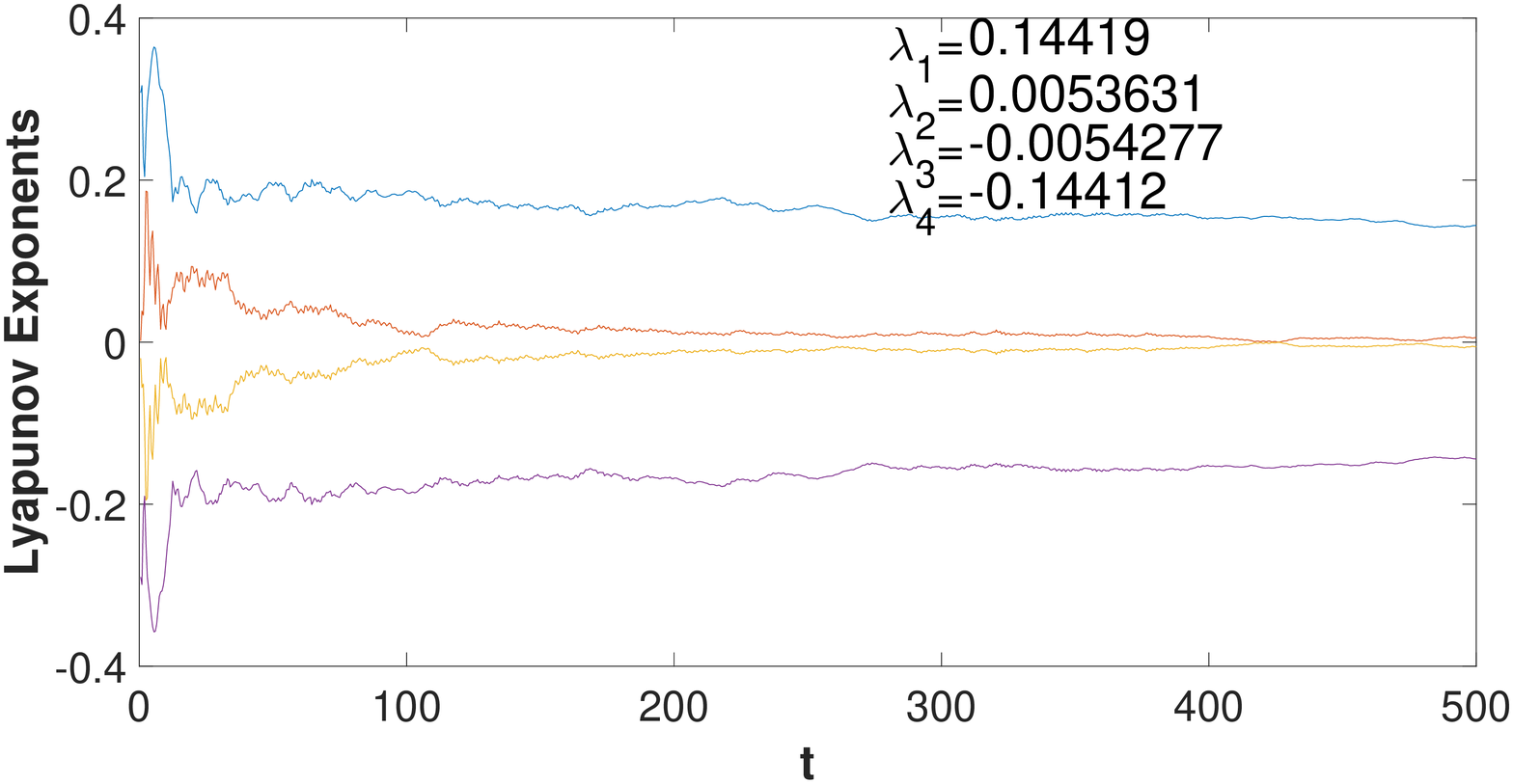}
\caption{Lyapunov exponents}
\label{multi-lyap}
\end{subfigure}%
\newline
\begin{subfigure}{.5\textwidth}
\centering
\includegraphics[width=.8\linewidth]{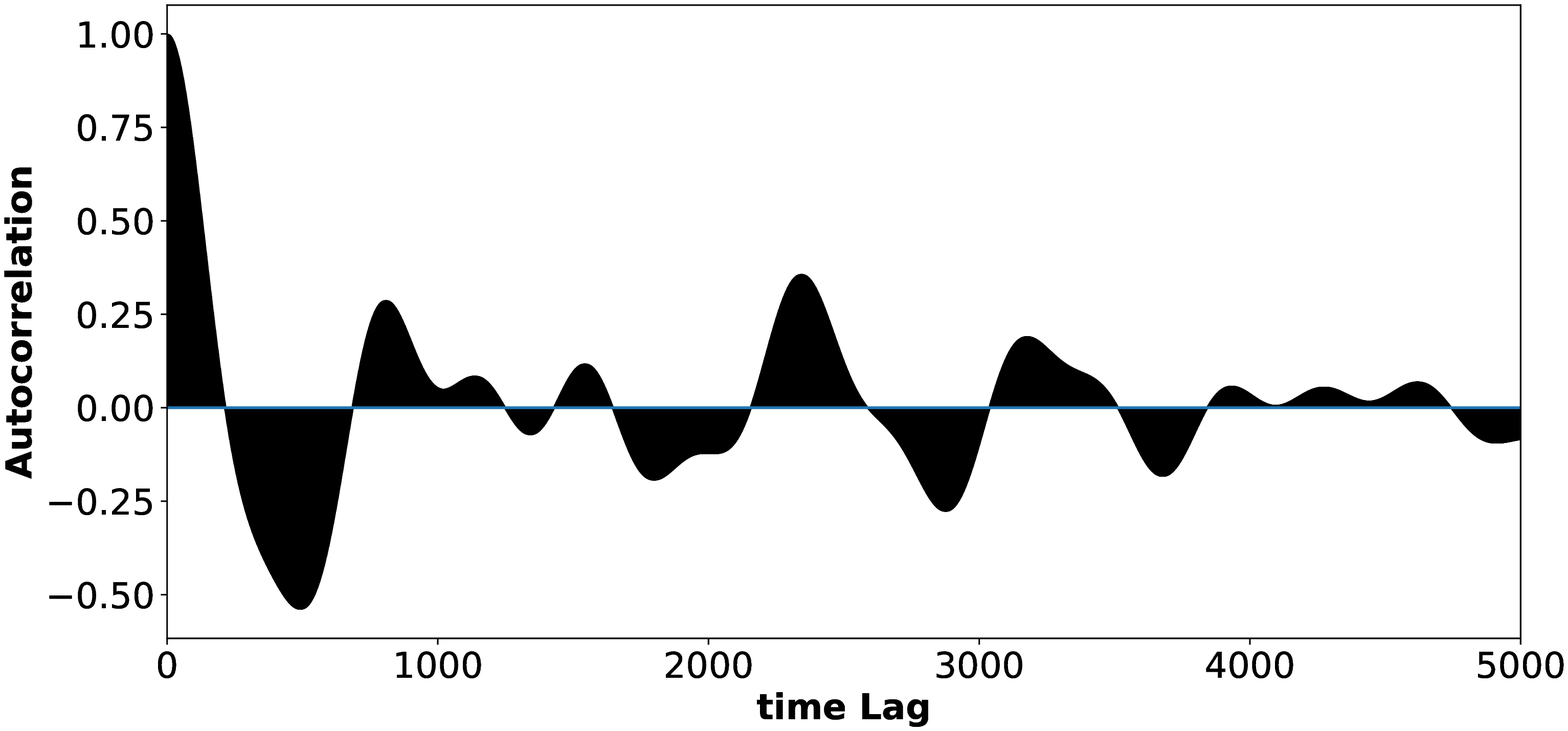}
\caption{Autocorrelation function of $x(t)$}
\label{multi-xauto}
\end{subfigure}%
\begin{subfigure}{.5\textwidth}
\centering
\includegraphics[width=.8\linewidth]{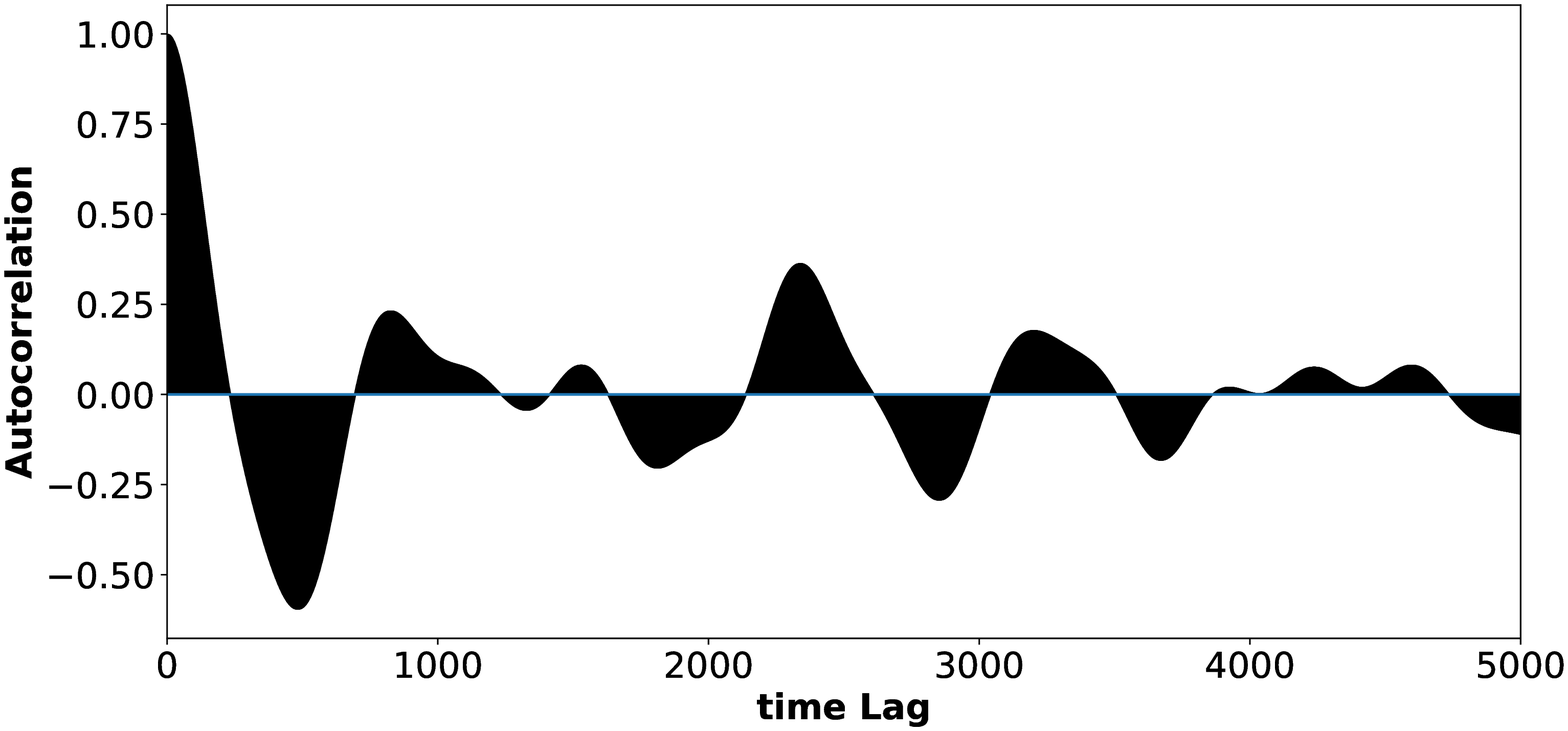}
\caption{Autocorrelation function of $y(t)$}
\label{multi-yaoto}
\end{subfigure}%
\newline
\begin{subfigure}{.5\textwidth}
\centering
\includegraphics[width=.8\linewidth]{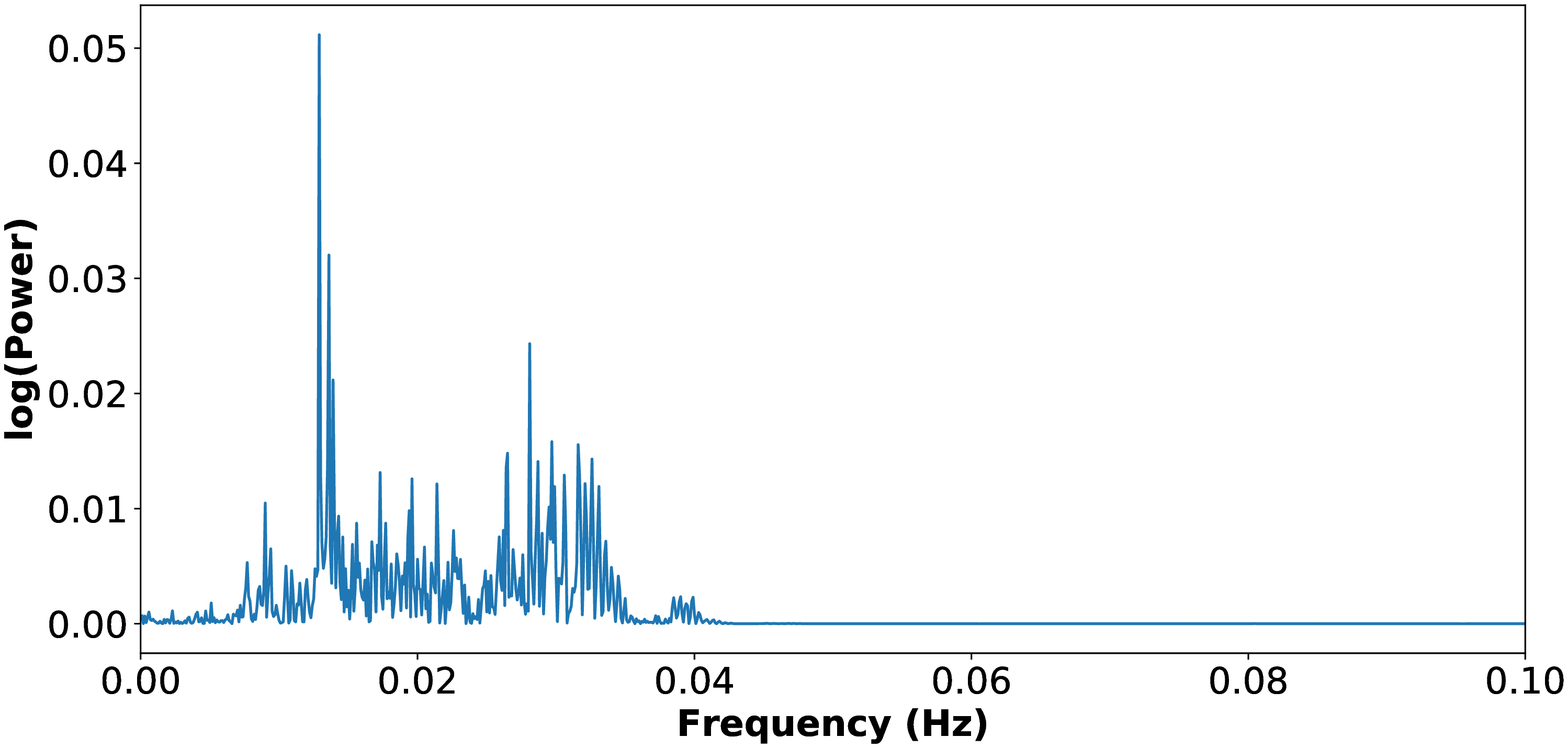}
\caption{Powerspectra of $x(t)$}
\label{multi-xpower}
\end{subfigure}%
\begin{subfigure}{.5\textwidth}
\centering
\includegraphics[width=.8\linewidth]{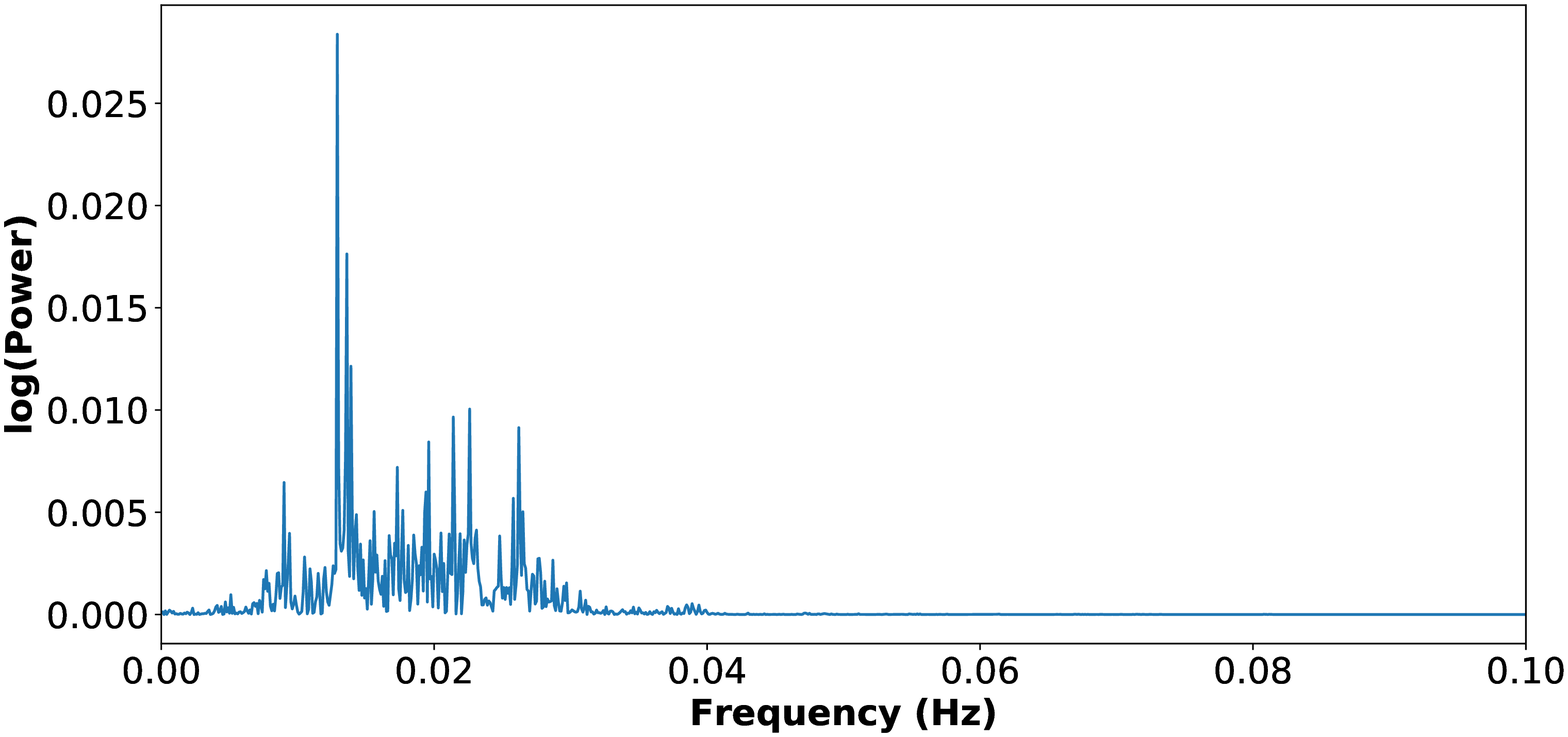}
\caption{Powerspectra of $y(t)$}
\label{multi-ypower}
\end{subfigure}%
\caption{(Color online)  Poincar$\acute{e}$ section, Lyapunov exponents, autocorrelation function
and power spectra for $\Gamma=0.01,\beta=1.3,\alpha=0.5,a=1.0,b=1.0$ with the initial condition
$x(0)=0.01,y(0)=0.02,\dot{x}(0)=0.03,\dot{y}(0)=0.04$. The plots corresponding to the
largest ($\lambda_1=0.14419$) and the smallest ($\lambda_2=-0.14412$)  Lyapunov exponents
are denoted by blue and violet colors, respectively.} 
\label{multi}
\end{figure}

\subsubsection{Chaotic Dynamics}

The bifurcation diagram is given in Fig. \ref{fig-bifr}, where $x$ and $y$ are plotted as a function of $\beta$
for fixed values of $\alpha=0.5, \Gamma=0.01, a=1.0, b=1.0$. It is seen that the system is chaotic for
$\beta > \beta_c=1.05$. The study of sensitivity of the dynamical variables to the initial conditions is one of
the important methods to check whether the system is chaotic or not. The two sets of initial conditions:
(a) $x(0)=0.01,y(0)=0.02,\dot{x}(0)=0.03,\dot{y}(0)=0.04$ and (b) $x(0)=0.01,y(0)=0.02,\dot{x}(0)=0.03,\dot{y}(0)=0.025$ are considered in order to study the sensitivity of the dynamical variables to the initial
conditions in different regions of the parameters. It may be noted that these two initial conditions are identical
except for the values of $\dot{y}(0)$ which differ by $0.015$. The Fig. \ref{chaotic} represents the time
series of the dynamical variables in the chaotic regime for $\beta=1.3$. A few other independent methods are
also employed to confirm the chaotic behaviour in the model. In this regard,
the auto-correlation function, Lyapunov exponent, Poincar$\acute{e}$ section and power spectra are
plotted in Fig. \ref{multi}. The Lyapunov exponents are computed up to seven decimal places
$(0.14419, 0.0063631, -0.0054277, -0.14412)$.
It is known that the sum of the Lyapunov exponents are zero for a Hamiltonian system,
which is valid for the present case if values up to the 2nd decimal places are considered
with an error of the order of $10^{-3}$.

\begin{figure}[ht!]
\begin{subfigure}{.5\textwidth}
\centering
\includegraphics[width=.8\linewidth]{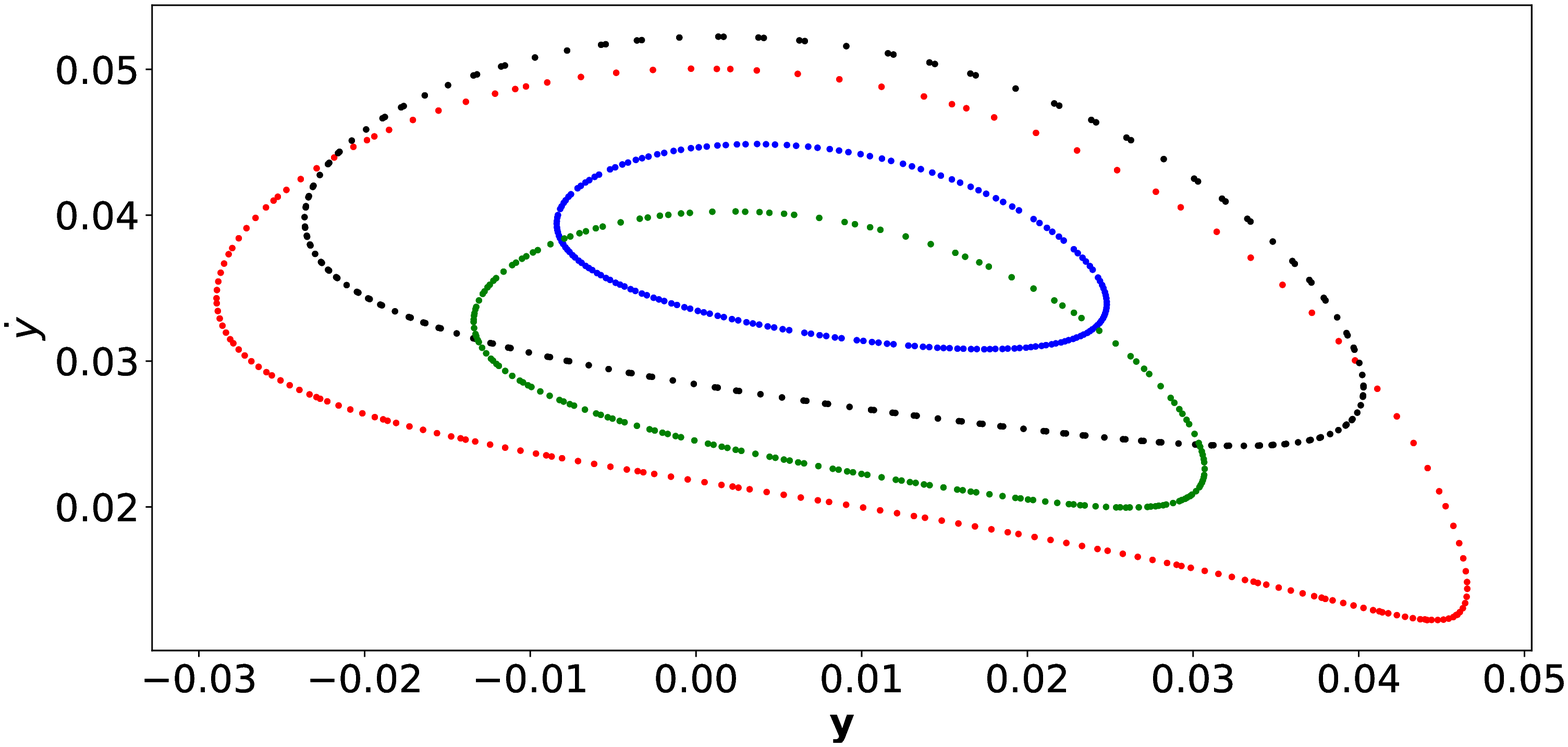} 
\caption{ $\alpha=0.5,\beta=0.4,\Gamma=0.01,a=1.0,b=1.0$}
\label{poincare1}
\end{subfigure}%
\begin{subfigure}{.5\textwidth}
\centering
\includegraphics[width=.8\linewidth]{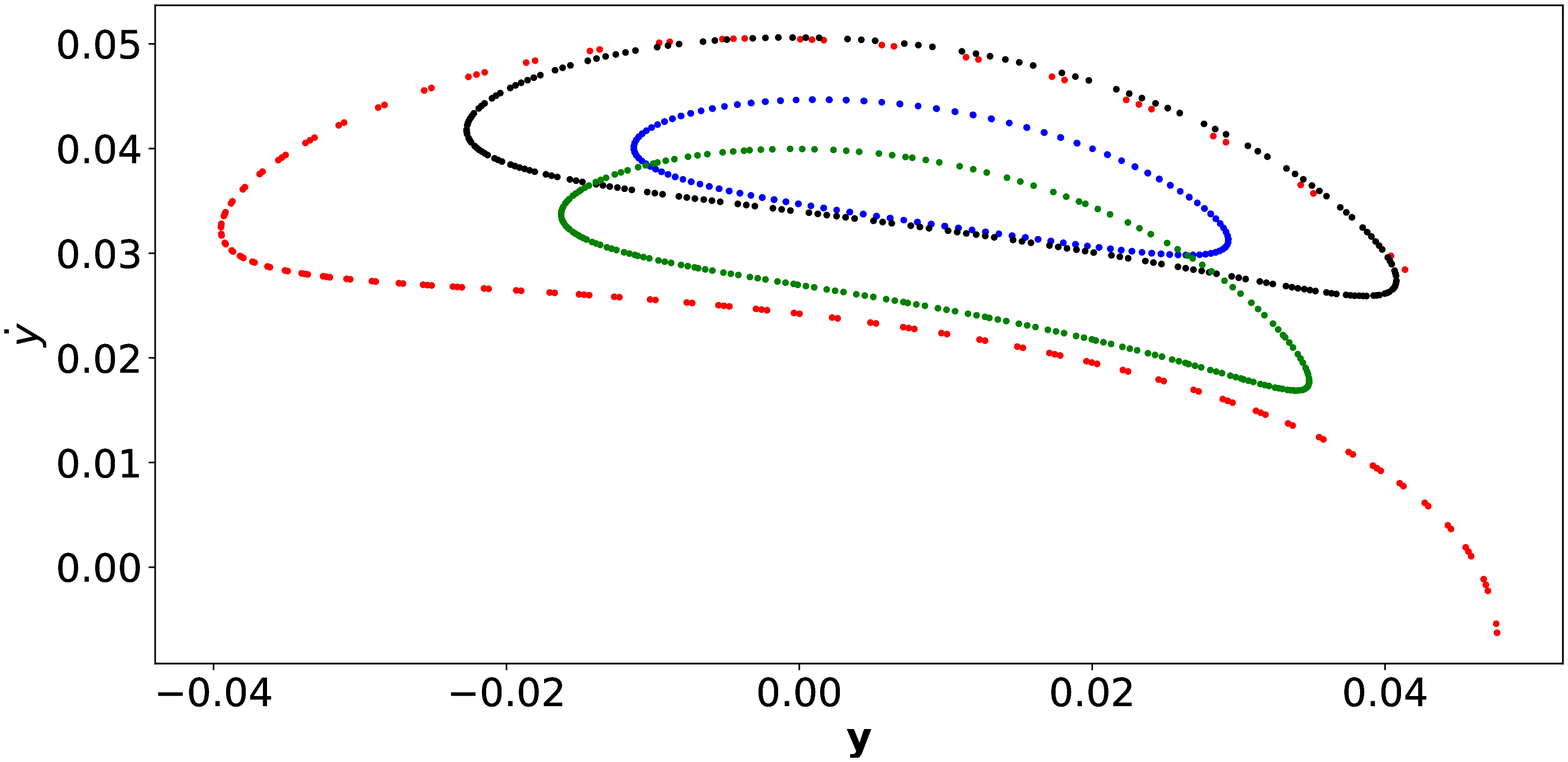}  
\caption{$\alpha=0.5,\beta=0.7,\Gamma=0.01,a=1.0,b=1.0$}
\label{poincare2}
\end{subfigure}%
\newline
\begin{subfigure}{.5\textwidth}
\centering
\includegraphics[width=.8\linewidth]{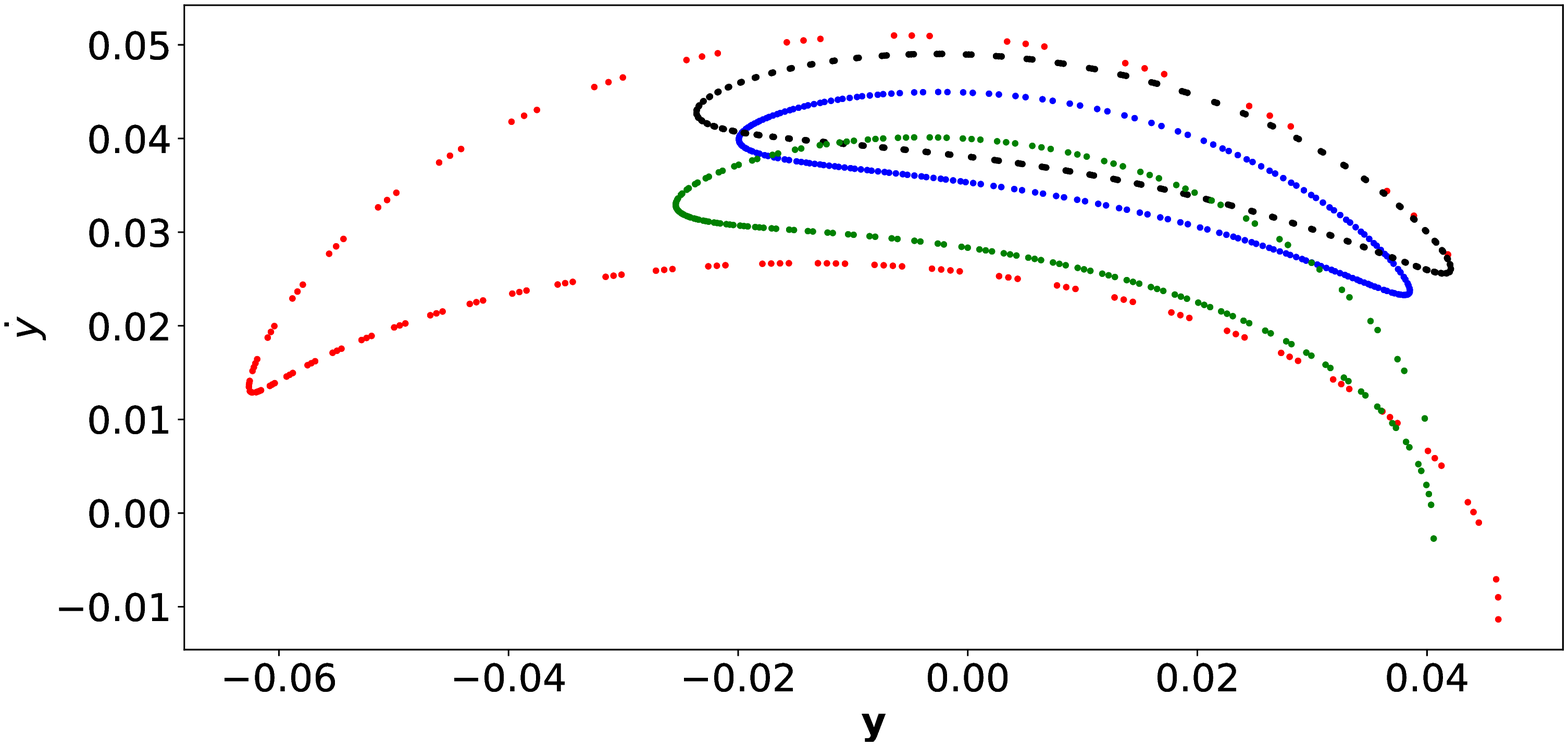}
\caption{$\alpha=0.5,\beta=0.88,\Gamma=0.01,a=1.0,b=1.0$}
\label{poincare3}
\end{subfigure}%
\begin{subfigure}{.5\textwidth}
\centering
\includegraphics[width=.8\linewidth]{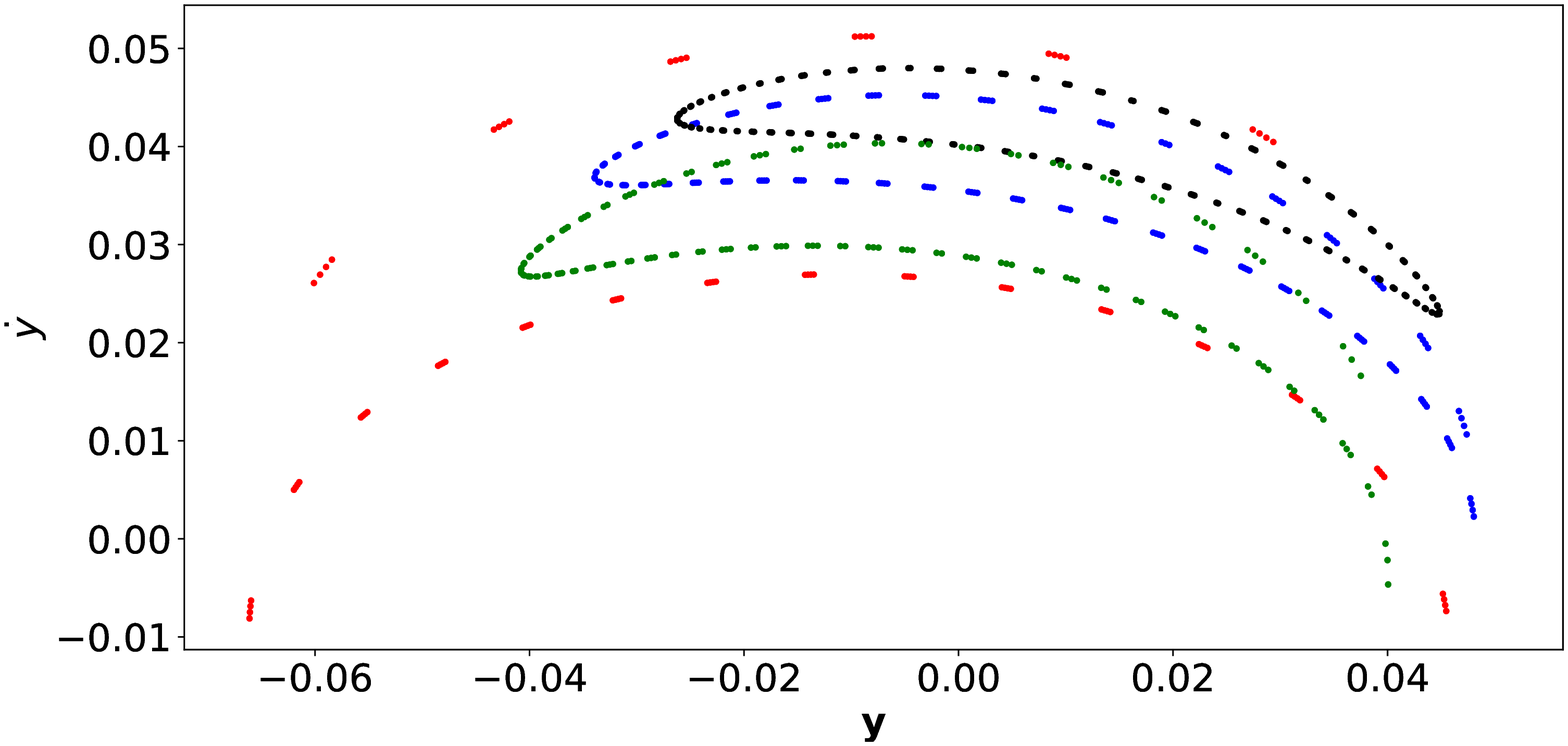}
\caption{$\alpha=0.5,\beta=0.95,\Gamma=0.01,a=1.0,b=1.0$}
\label{poincare4}
\end{subfigure}%
\newline
\begin{subfigure}{.5\textwidth}
\centering
\includegraphics[width=.8\linewidth]{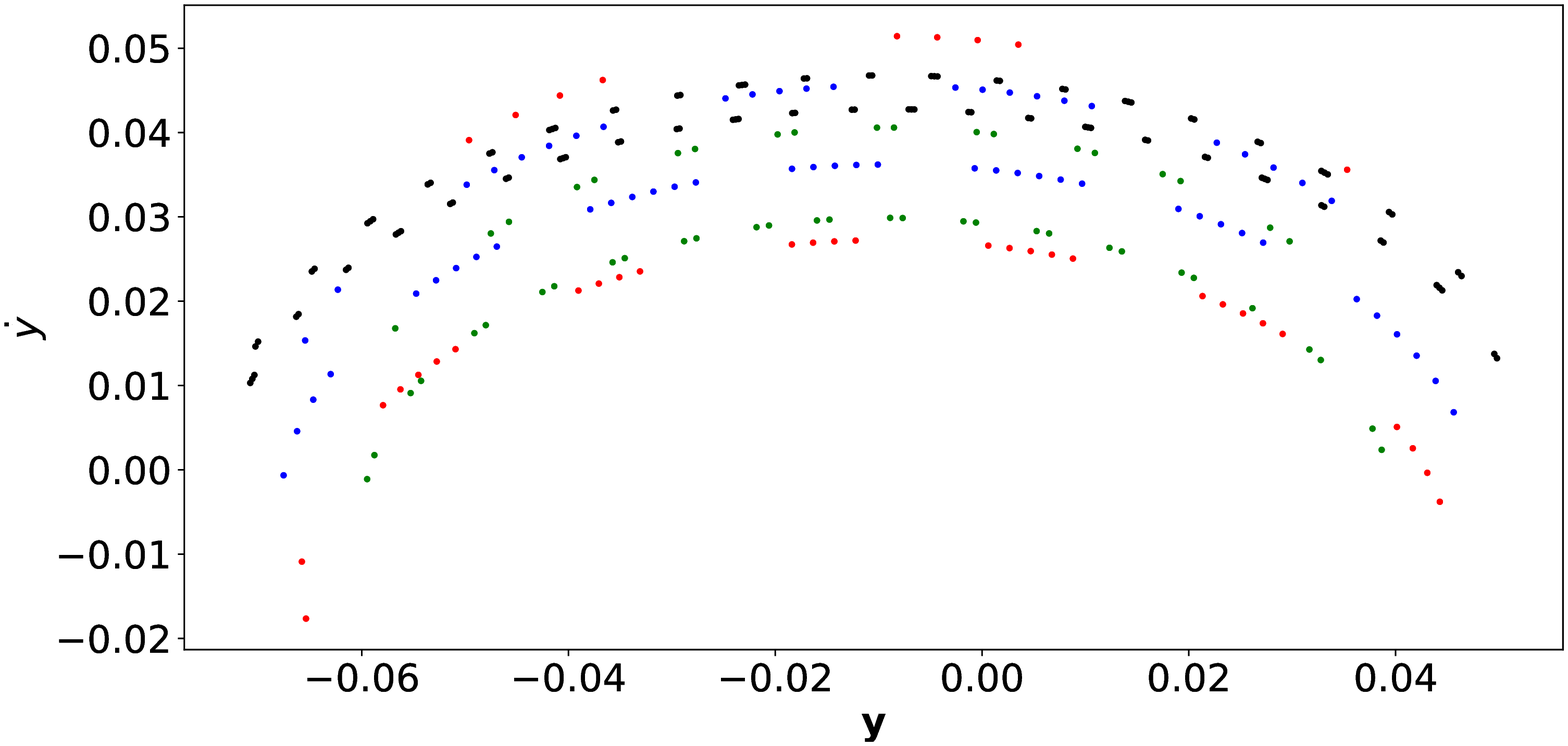}
\caption{$\alpha=0.5,\beta=1.0,\Gamma=0.01,a=1.0,b=1.0$}
\label{poincare5}
\end{subfigure}%
\begin{subfigure}{.5\textwidth}
\centering
\includegraphics[width=.8\linewidth]{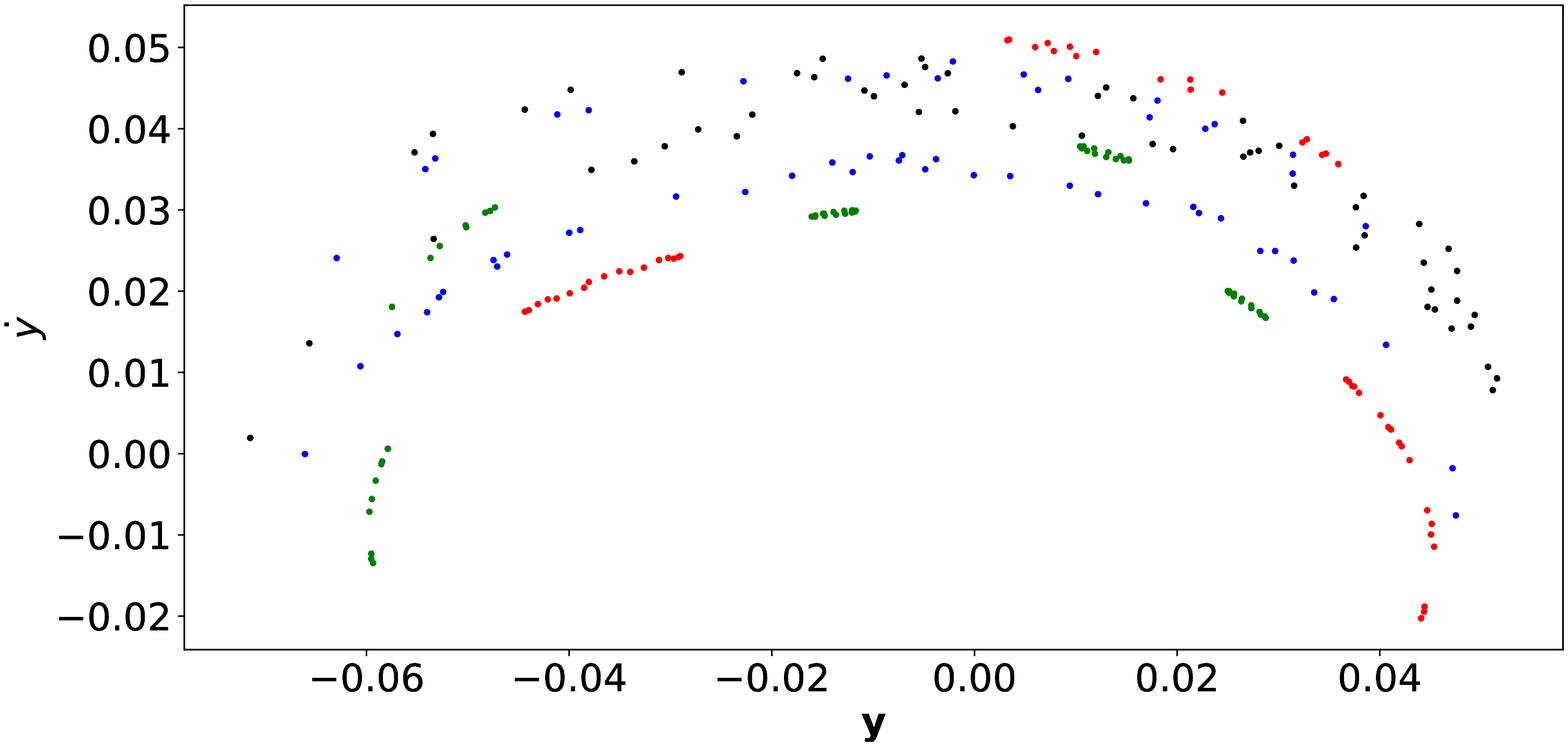}
\caption{$\alpha=0.5,\beta=1.04,\Gamma=0.01,a=1.0,b=1.0$}
\label{poincare6}
\end{subfigure}%
\caption{(Color online) Poincar$\acute{e}$ sections of Eq. ({\ref{vdpd-eqn-1}}) with four sets of initial
conditions: (a) $x(0)=0.01,y(0)=0.02,\dot{x}(0)=0.03,\dot{y}(0)=0.04$ (blue color) ,
(b) $x(0)=0.01,y(0)=0.03,\dot{x}(0)=0.015,\dot{y}(0)=0.04$ (red color),(c) $x(0)=0.01,y(0)=0.04,\dot{x}(0)=0.025,\dot{y}(0)=0.03$ (black color) and (d) $x(0)=0.01,y(0)=0.025,\dot{x}(0)=0.03,\dot{y}(0)=0.02$ (green color)
}
\label{poincareh}
\end{figure}

The route to chaos may be studied in terms of the qualitative changes in a
set of Poincar$\acute{e}$ sections of the system as the parameter $\beta$ is
varied from the non-chaotic to the chaotic region. The set of
Poincar$\acute{e}$ sections corresponds to independent initial conditions
($x(0), y(0), \dot{x}(0), \dot{y}(0)$) around  the equilibrium point $P_0$.
This effectively allows us to study the dynamics of the system on the chosen
Poincar$\acute{e}$ sections. We consider four different initial conditions
for a given $\beta$ as follows:
(i) ($ 0.01, 0.02, 0.03, 0.04$), (ii) $( 0.01, 0.03, 0.015, 0.04$),
(iii) ($0.01, 0.04, 0.025, 0.03$), and (iv) ($0.01, 0.025, 0.03, 0.02$).
The corresponding surface of sections change differently for different
values of $\beta$. It is seen from Fig. 5 that all four orbits on the
chosen surface are closed for $\beta=0.5$. The total number of
closed orbits gradually reduce to 1 around $\beta=0.95$, and finally no closed
orbits are seen as the critical value $\beta_c=1.05$ is approached. The
breaking of closed orbits on the surface of section correspond to breaking
of tori in the original four dimensional state-velocity space. There are
only isolated points in the chaotic region, i.e. beyond $\beta > \beta_c$.

\section{Non-${\cal{PT}}$-symmetric non-Hamiltonian system}

In this section, we consider a system by modifying the nonlinear interaction $3 g x^2 y$ in Eq. (\ref{vdpd-eqn})
to $\tilde{g} y^3$. The equations of motion for the system in terms of the dimensionless variables introduced
in Eqs. (\ref{scale}, \ref{scale-1}) are,
\bea
&& \ddot{x} + 2\Gamma \left(1-a x^{2}-b y^{2}\right)\dot{x}+x+\beta y+ \alpha x^{3}=0,\nonumber \\
&& \ddot{y} - 2\Gamma \left(1-a x^{2}-b y^{2}\right)\dot{y}+y+\beta x+ \tilde{\alpha} y^{3}=0,
\label{vdp-duff-eqnn}
\eea
\noindent where we have chosen $\sgn(\beta_1)=\sgn(\beta_2)$ for simplicity and
$\tilde{\alpha}=\frac{\tilde{g}}{{\vert \beta_1 \vert} \omega^2}$. The system described by Eq. (\ref{vdp-duff-eqnn})
is a generalization of the models considered in Refs. \cite{pkg-review,khare-0}. The model of \cite{khare-0} is
obtained for $\alpha =\tilde{\alpha}, a=0=b$, while the system described in Ref. \cite{pkg-review} is obtained
for $a=0=b$ and without any constraint on $\alpha, \tilde{\alpha}$. The generalized system describes
two coupled VdPD oscillators with balanced loss and gain. The $x$-degree of freedom
describes the standard VdPD oscillator for $\beta=0=b$, while the $y$ degree of freedom describes
a damped Duffing oscillator with the linear restoring force depending on $x$. Similarly, for $\beta=0=a$,
there is a role reversal for the $x$ and $y$ degrees of freedom. The space-dependence of the gain and
loss terms should be identical so that the flow preserves the volume in the position-velocity state
space\cite{pkg-1}. We choose the space-dependence of the loss-gain terms as $1-ax^2-by^2$ so that the forms
$1-ax^2$ and  $1-b y^2$ may be considered for $b=0$ and $a=0$, respectively. It is worth mentioning here that
a similar system with additional velocity mediated coupling terms and unbalanced loss-gain terms, where the
damping/anti-damping term varies linearly with space-coordinates, has been reported to admit amplitude death\cite{iiser}.
We consider a balanced loss-gain system which exhibits periodic solutions in the regular regime.

It stems from the linear stability
analysis of the system that the  coupling between the two systems via non-vanishing $\beta$ is required
for the existence of periodic solutions. The stability criteria for the systems defined by Eqs. (\ref{vdpd-eqn-1})
and (\ref{vdp-duff-eqnn}) are identical. The system is ${\cal{PT}}$-symmetric for $a = b$ and
$\alpha = \tilde{\alpha}$, and non-${\cal{PT}}$-symmetric if any of these conditions are violated.
It will be seen that the dynamical behaviour of the system is quite rich, including existence of
periodic solutions, for the ${\cal{PT}}$-symmetric as well as non-${\cal{PT}}$-symmetric regime.
The Hamiltonian  for the system can not be constructed by using the techniques outlined in
Ref. \cite{pkg-ds, p6-deb} for $\alpha \neq 0 \neq \tilde{\alpha}$.

\begin{figure}[ht!]
\begin{subfigure}{.5\textwidth}
\centering
\includegraphics[width=.8\linewidth]{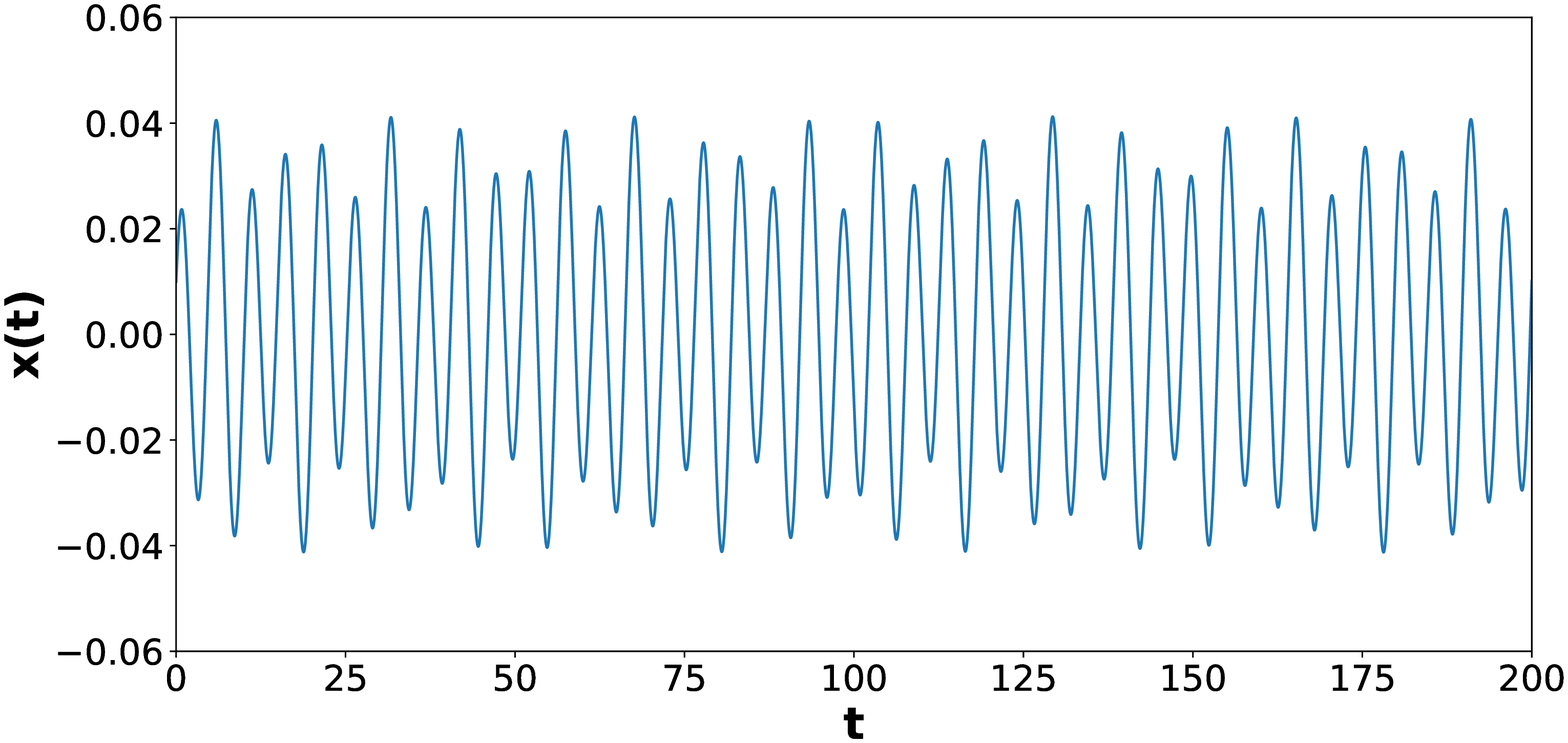} 
\caption{ $\alpha=\tilde{\alpha}=0.5, \beta=0.5, \Gamma=0.03, a=1.0, b=1.0$}
\label{xtimeseries_nh}
\end{subfigure}%
\begin{subfigure}{.5\textwidth}
\centering
\includegraphics[width=.8\linewidth]{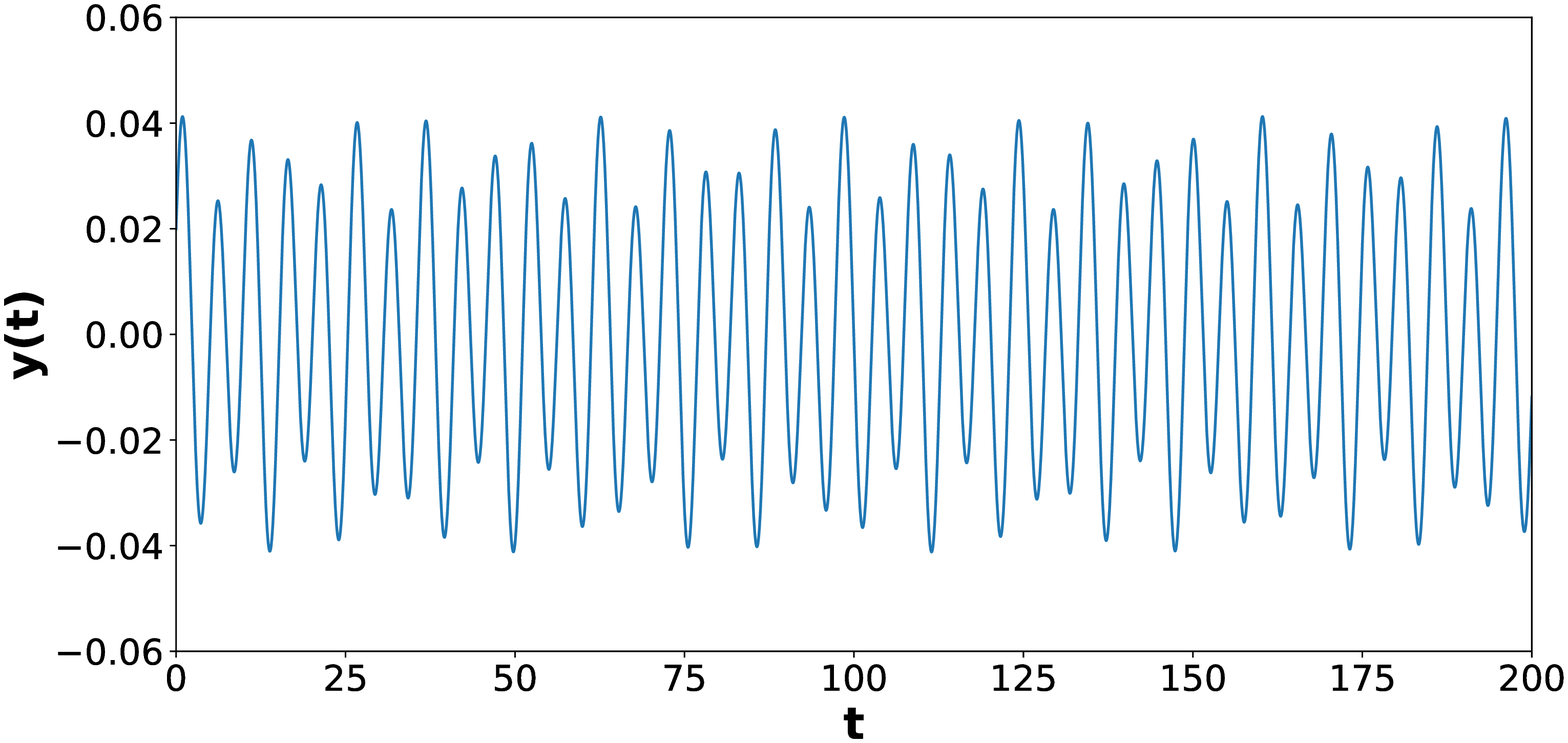}  
\caption{$\alpha=\tilde{\alpha}=0.5, \beta=0.5, \Gamma=0.03, a=1.0 ,b=1.0$}
\label{ytimeseries_nh}
\end{subfigure}%
\newline\begin{subfigure}{.5\textwidth}
\centering
\includegraphics[width=.8\linewidth]{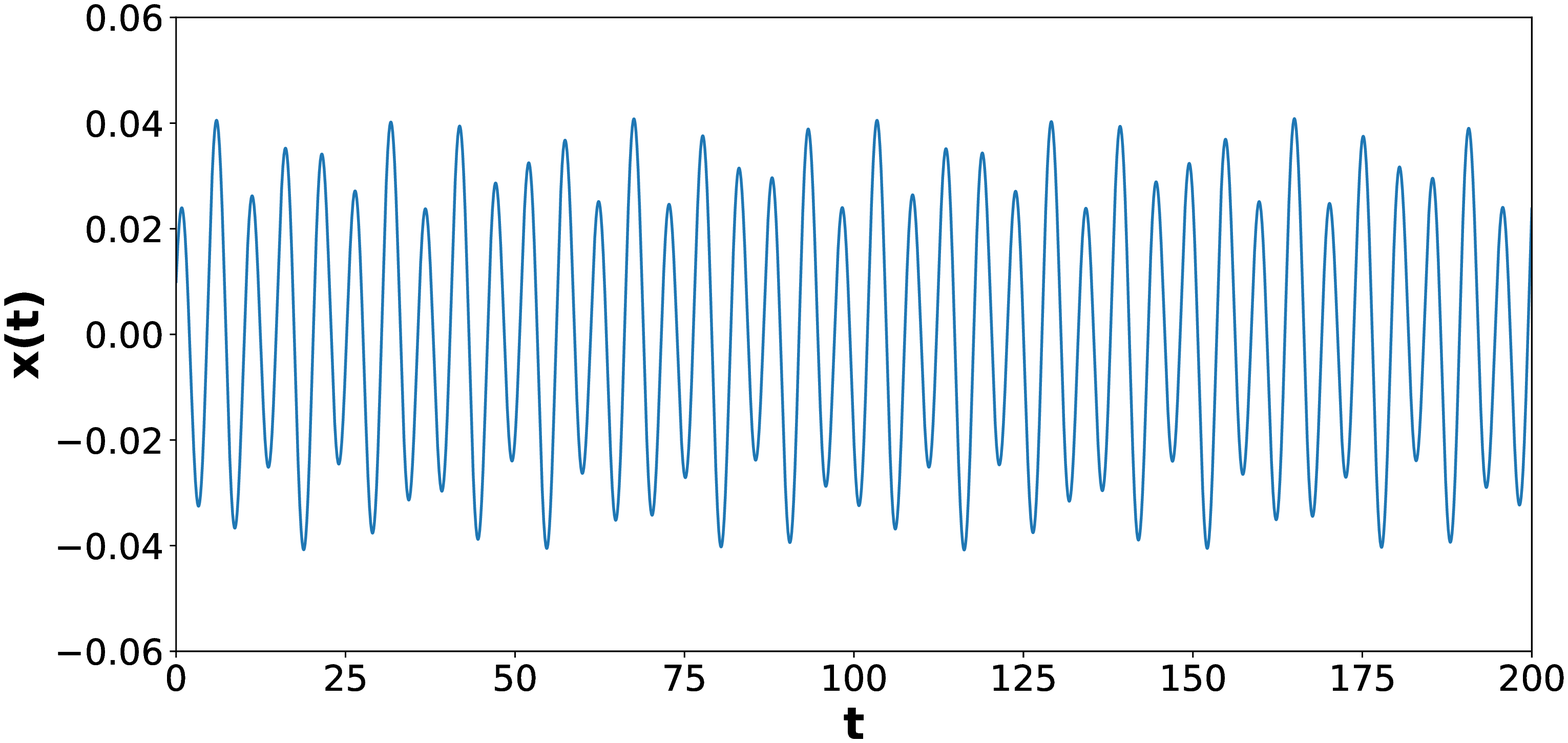} 
\caption{ $\alpha=1.0, \tilde{\alpha}=3.0, \beta=0.5, \Gamma=0.01, a=1.0, b=2.0$}
\label{xtimeseries_abneqnh}
\end{subfigure}%
\begin{subfigure}{.5\textwidth}
\centering
\includegraphics[width=.8\linewidth]{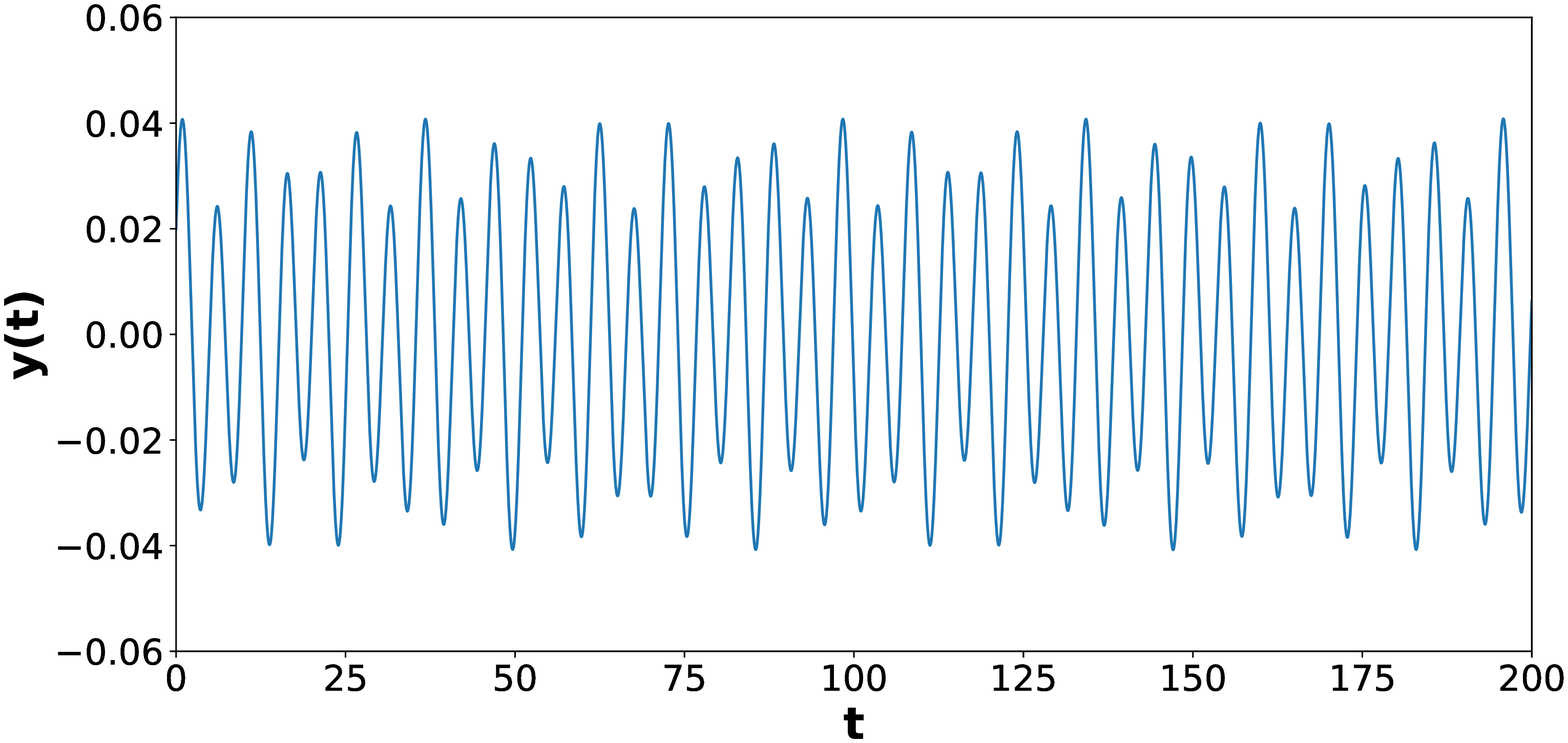}  
\caption{$\alpha=1.0, \tilde{\alpha}=3.0, \beta=0.5, \Gamma=0.01, a=1.0, b=2.0$}
\label{ytimeseries_abneqnh}
\end{subfigure}%
\caption{(Color online) \  Solutions of Eq.({\ref{vdp-duff-eqnn}}) in the vicinity of the point
$P_0$ with the initial conditions $x(0)=0.01, y(0)=0.02, \dot{x}(0)=0.03$ and $\dot{y}(0)=0.04$.
The results for ${\cal{PT}}$-symmetric and non-${\cal{PT}}$-symmetric regions are described
in the first and the second rows, respectively.}
\label{time-series-P0_nh}
\end{figure}
\begin{figure}[ht!]
\begin{subfigure}{.5\textwidth}
\centering
\includegraphics[width=0.8\linewidth]{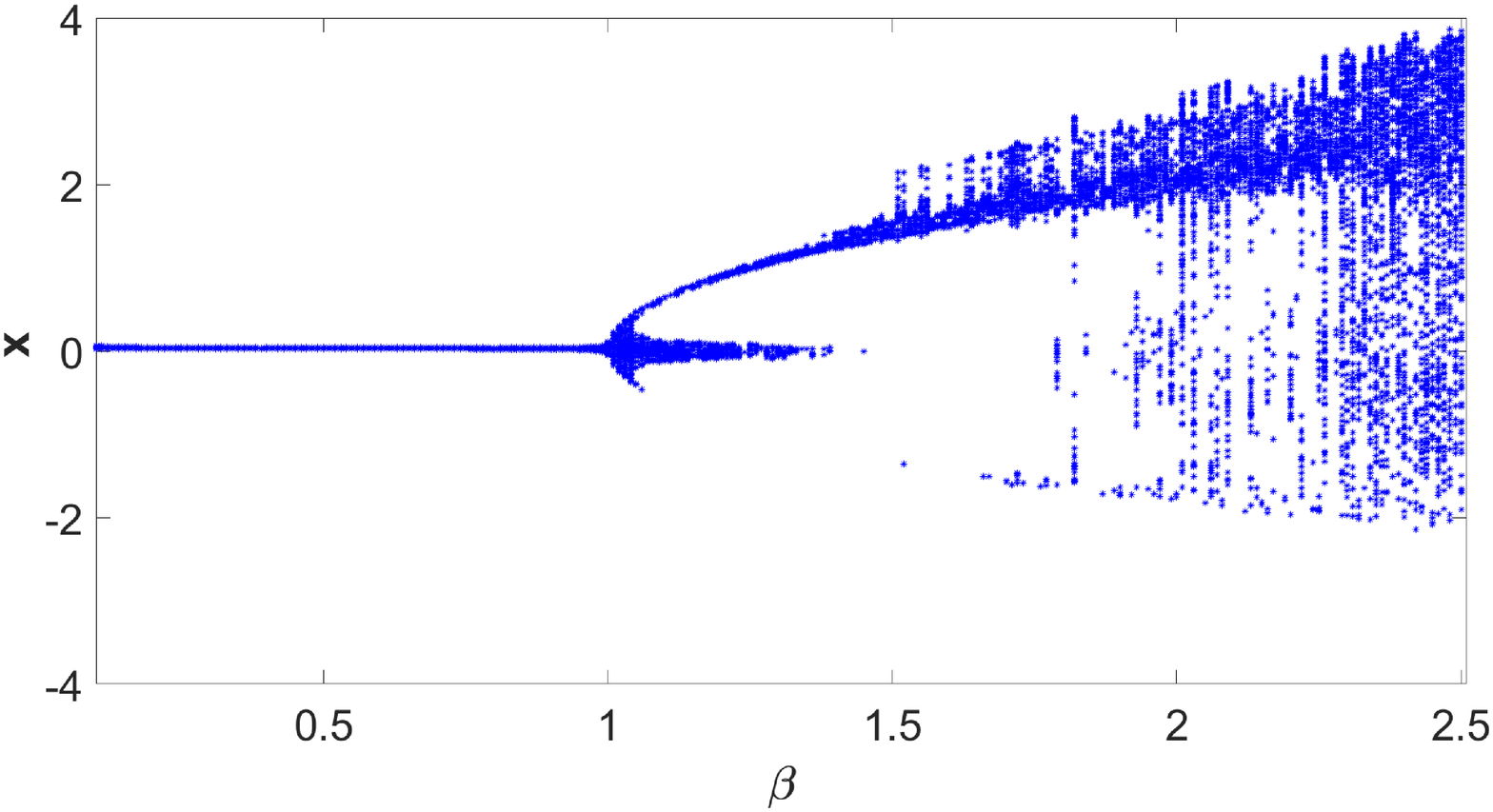} 
\caption{ $\alpha=\tilde{\alpha}=0.5,\Gamma=0.03, a=1.0, b=1.0$}
\label{bifurcation-xvsbeta-nh}
\end{subfigure}%
\begin{subfigure}{.5\textwidth}
\centering
\includegraphics[width=0.8\linewidth]{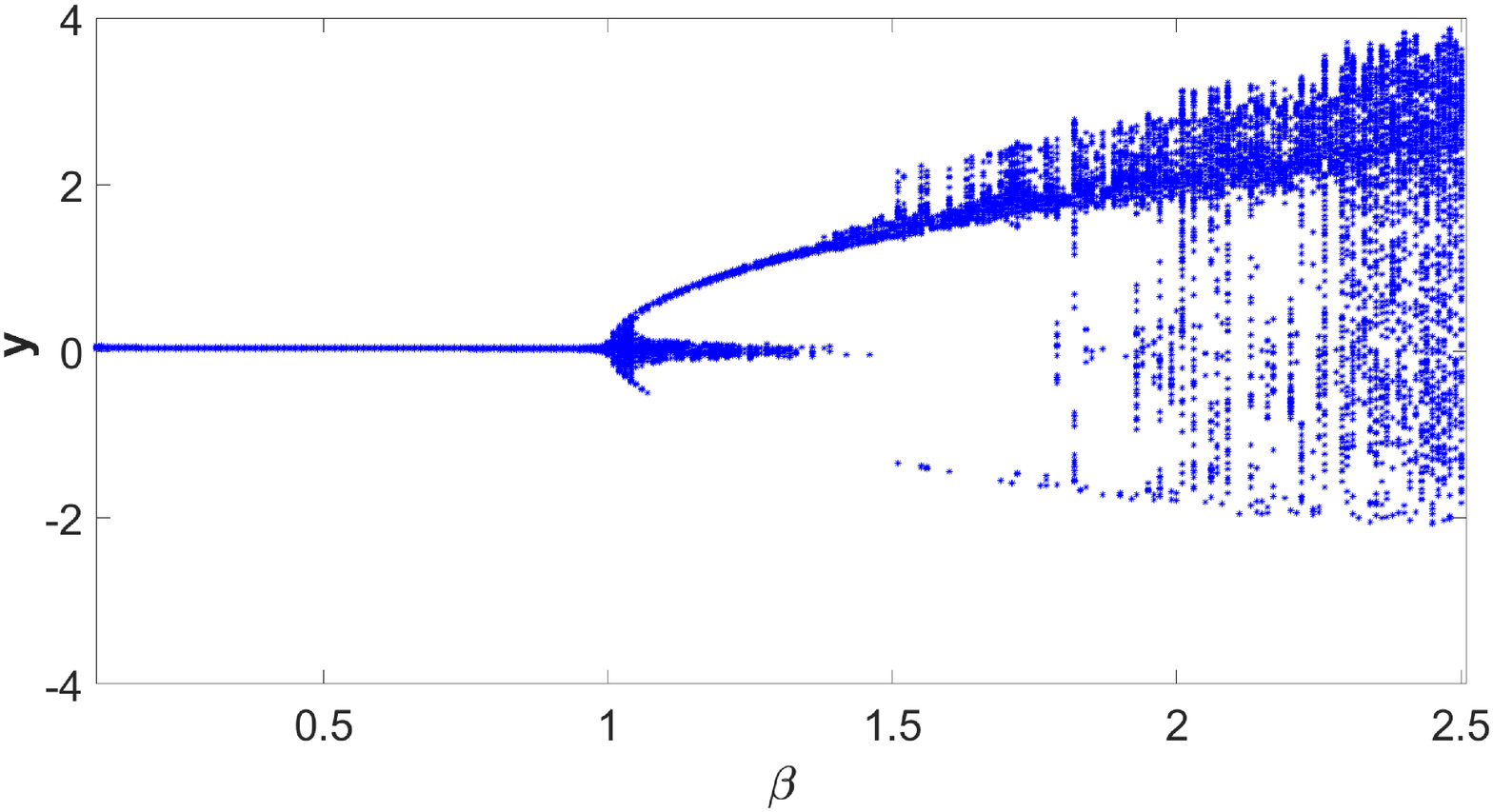}  
\caption{$\alpha=\tilde{\alpha}=0.5,\Gamma=0.03, a=1.0, b=1.0$}
\label{bifurcation-yvsbeta-nh}
\end{subfigure}%
\newline
\begin{subfigure}{.5\textwidth}
\centering
\includegraphics[width=0.8\linewidth]{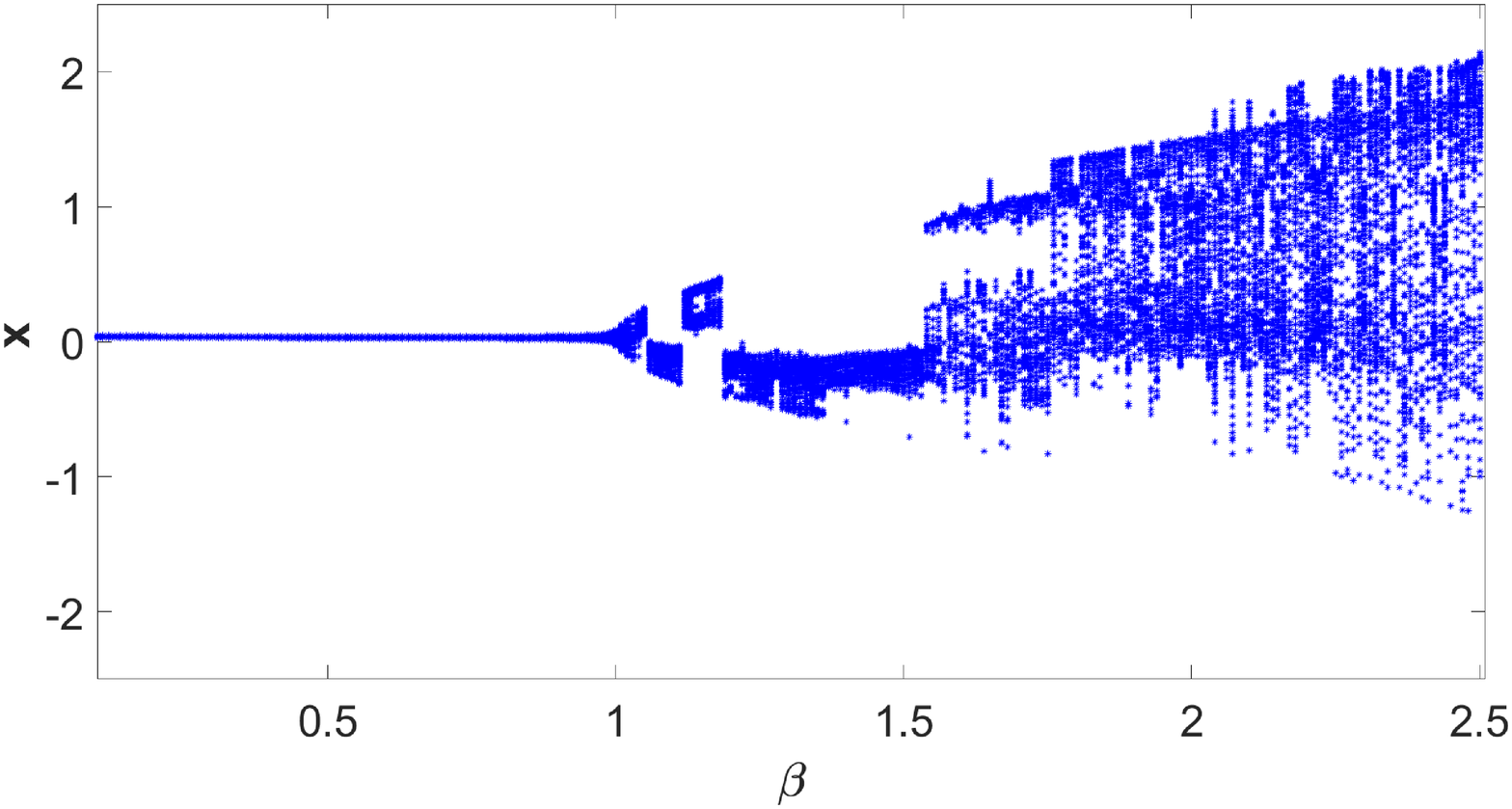} 
\caption{ $\alpha=1.0,\tilde{\alpha}=3.0,\Gamma=0.01, a=1.0, b=2.0$}
\label{bifurcation-xvsbeta-abneqnh}
\end{subfigure}%
\begin{subfigure}{.5\textwidth}
\centering
\includegraphics[width=0.8\linewidth]{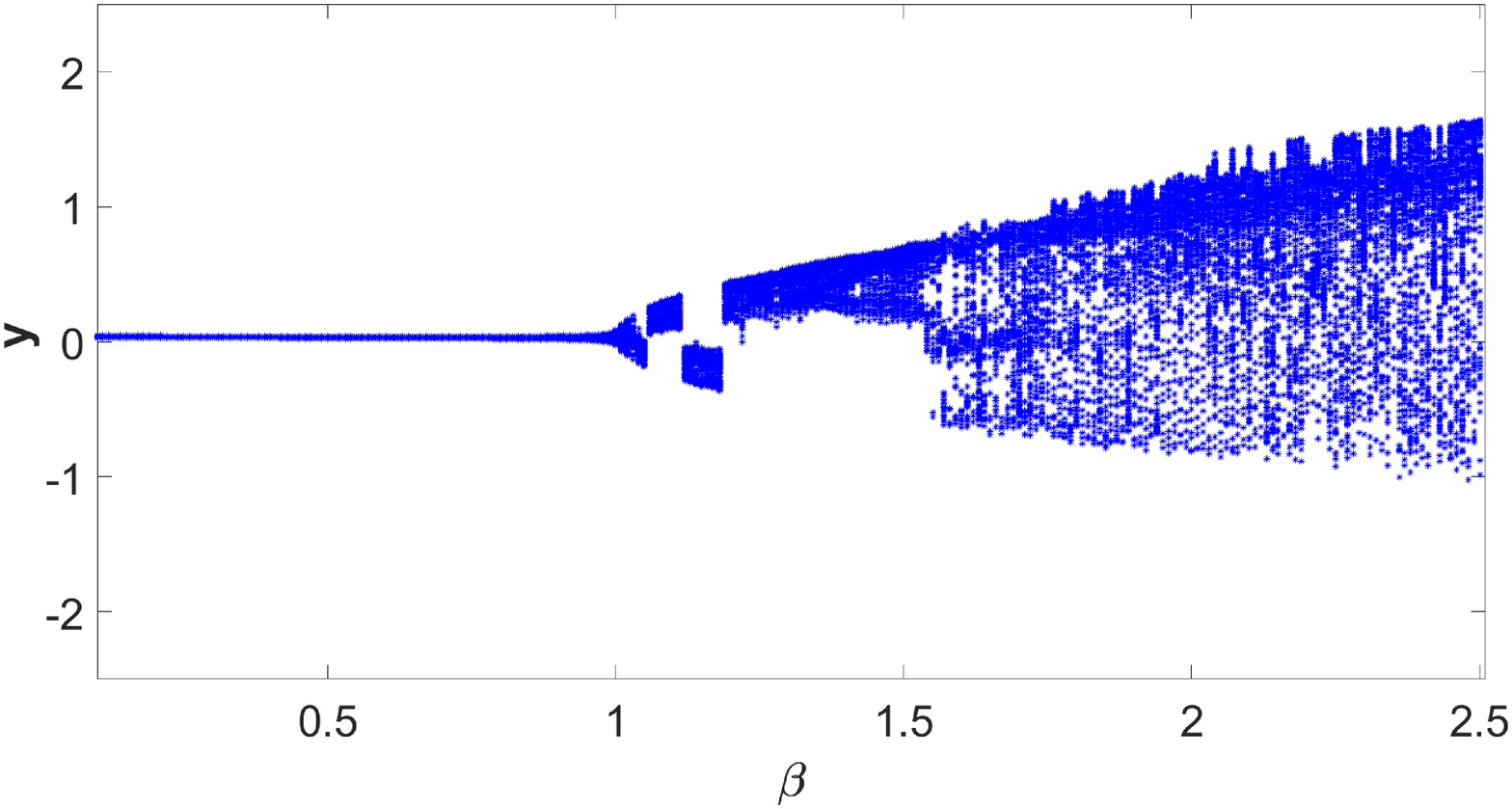}  
\caption{$\alpha=1.0,\tilde{\alpha}=3.0,\Gamma=0.01, a=1.0, b=2.0$}
\label{bifurcation-yvsbeta-abneqnh}
\end{subfigure}%
\caption{(Color online) \ Bifurcation diagrams of Eq. (\ref{vdp-duff-eqnn}) in the vicinity of the point
$P_0$ with the initial conditions $x(0)=0.01,y(0 )=0.02,\dot{x}(0)=0.03$ and $\dot{y}(0)=0.04$. The
results for the ${\cal{PT}}$-symmetric and non-${\cal{PT}}$-symmetric regions are given in the first
and the second rows, respectively.}
\label{bifurcation-nh-npt}
\end{figure}

\begin{figure}[ht!]
\begin{subfigure}{.5\textwidth}
\centering
\includegraphics[width=.8\linewidth]{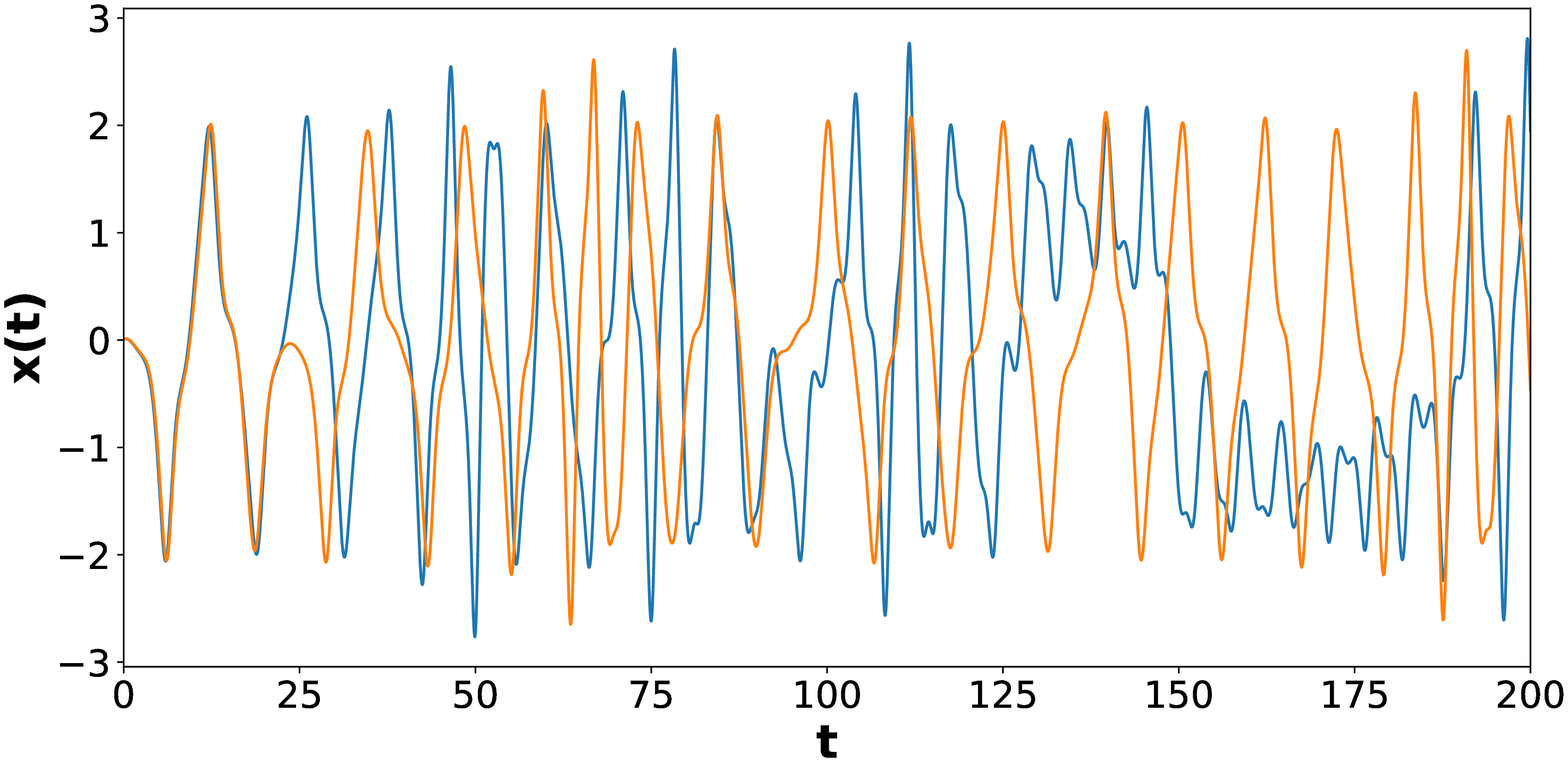} 
\caption{ $\Gamma=0.03,\beta=2.0,\alpha=\tilde{\alpha}=0.5, a=1.0, b=1.0$}
\label{chaotic1_nh}
\end{subfigure}%
\begin{subfigure}{.5\textwidth}
\centering
\includegraphics[width=.8\linewidth]{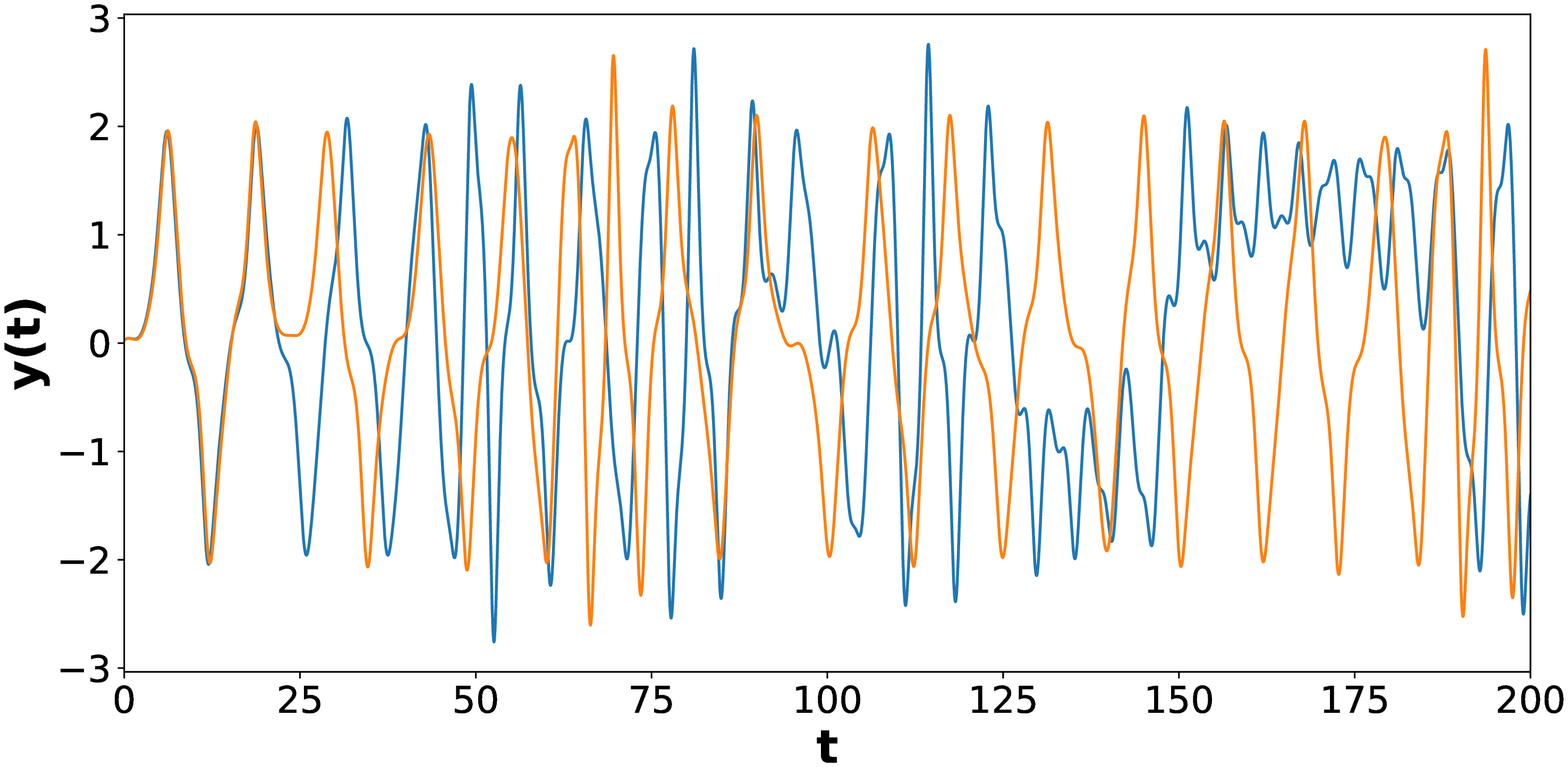}  
\caption{$\Gamma=0.03,\beta=2.0,\alpha=\tilde{\alpha}=0.5, a=1.0, b=1.0$}
\label{chaotic2_nh}
\end{subfigure}%
\newline 
\begin{subfigure}{.5\textwidth}
\centering
\includegraphics[width=.8\linewidth]{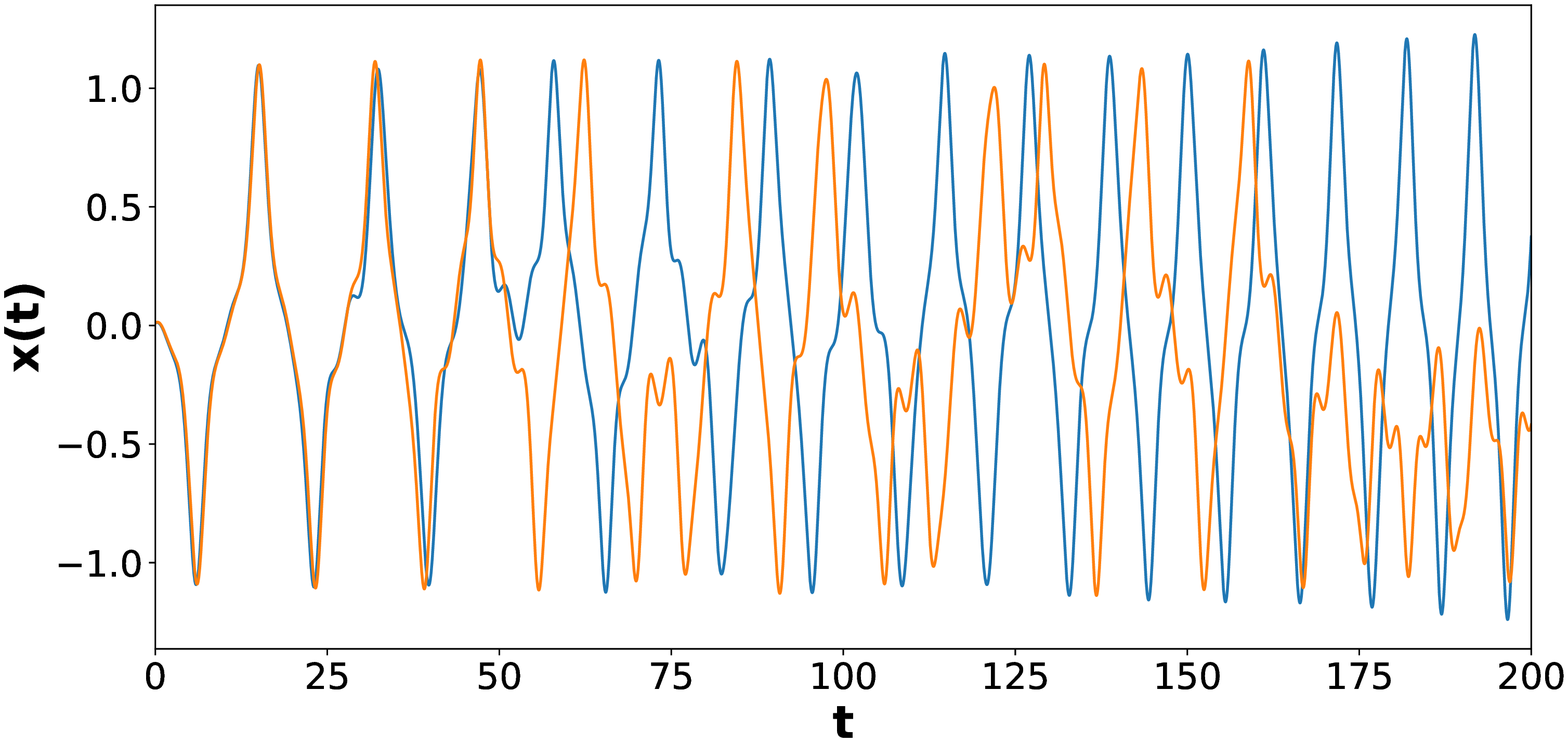} 
\caption{$\Gamma=0.01,\beta=1.8,\alpha=1.0,\tilde{\alpha}=3.0,a=1.0,
\newline b=2.0$}
\label{chaotic1_abneqnh}
\end{subfigure}%
\begin{subfigure}{.5\textwidth}
\centering
\includegraphics[width=.8\linewidth]{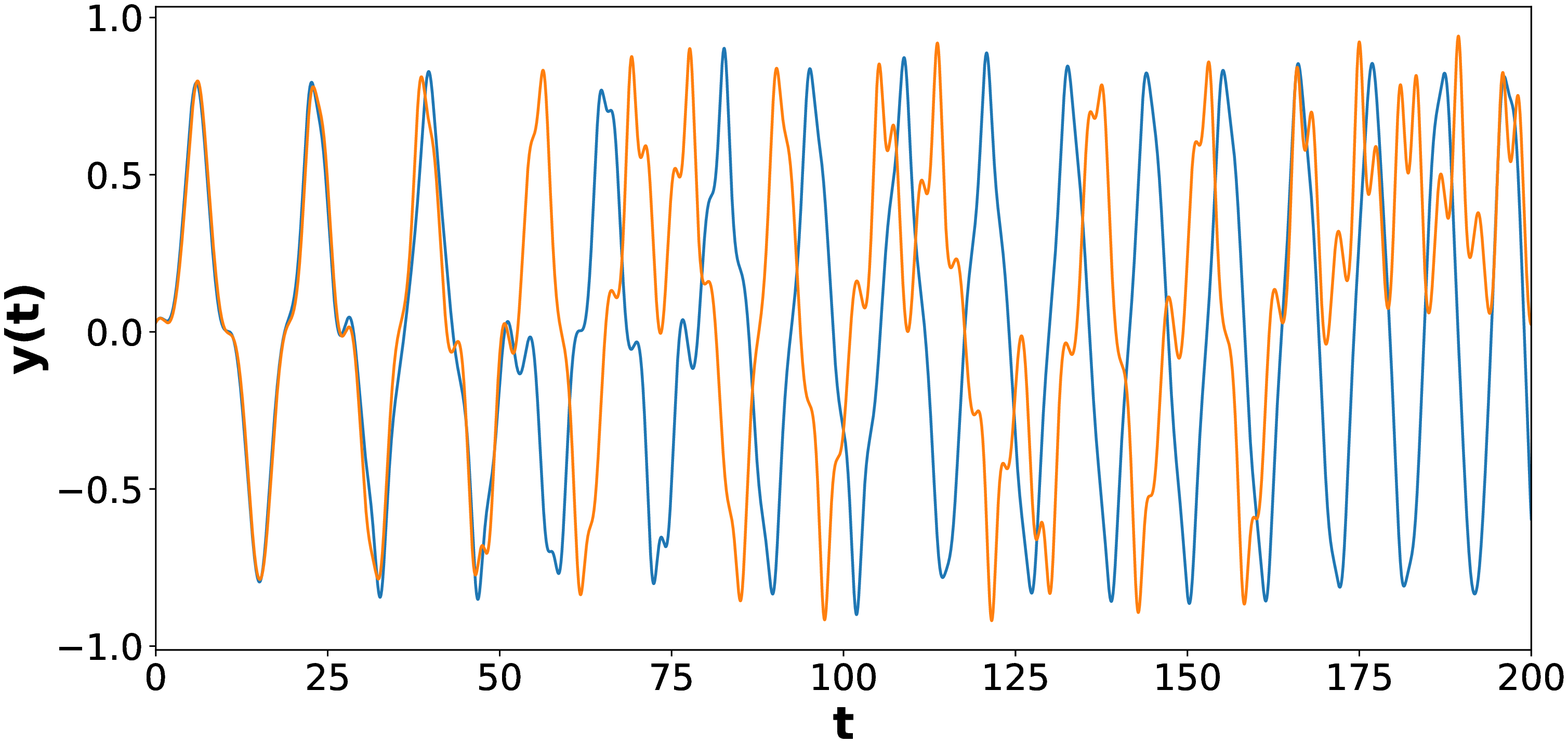}  
\caption{$\Gamma=0.01,\beta=1.8,\alpha=1.0,\tilde{\alpha}=3.0,a=1.0,\newline b=2.0$}
\label{chaotic2_abneqnh}
\end{subfigure}%
 
\caption{(Color online) Chaotic solutions of Eq. ({\ref{vdp-duff-eqnn}}) with two sets of initial
conditions (a) $x(0)=0.01,y(0)=0.02,\dot{x}(0)=0.03,\dot{y}(0)=0.04$ (blue color) and 
(b) $x(0)=0.01,y(0)=0.025,\dot{x}(0)=0.03,\dot{y}(0)=0.04$ (orange color). The first row describes
results in the ${\cal{PT}}$-symmetric region, while the figures in the second row correspond to
non-${\cal{PT}}$-symmetric regime.} 
\label{chaotic_nonhamiltonian}
\end{figure}
\begin{figure}[ht!]
\begin{subfigure}{.5\textwidth}
\centering
\includegraphics[width=.8\linewidth]{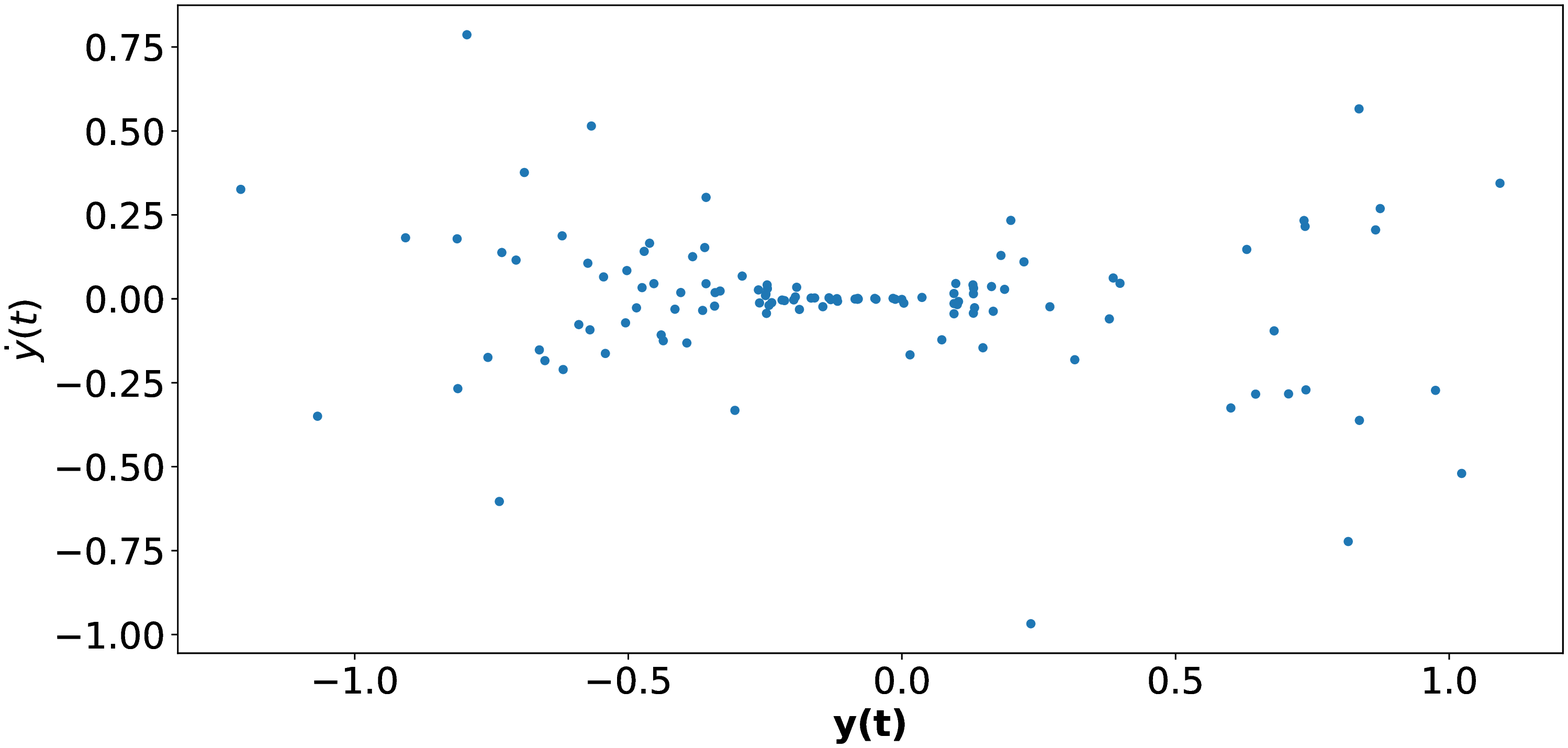}
\caption{Poincar$\acute{e}$ section: $\dot{y}(t)$ VS. $y(t)$ plot}
\label{vdpd-poincare}
\end{subfigure}%
\begin{subfigure}{.5\textwidth}
\centering
\includegraphics[width=.8\linewidth]{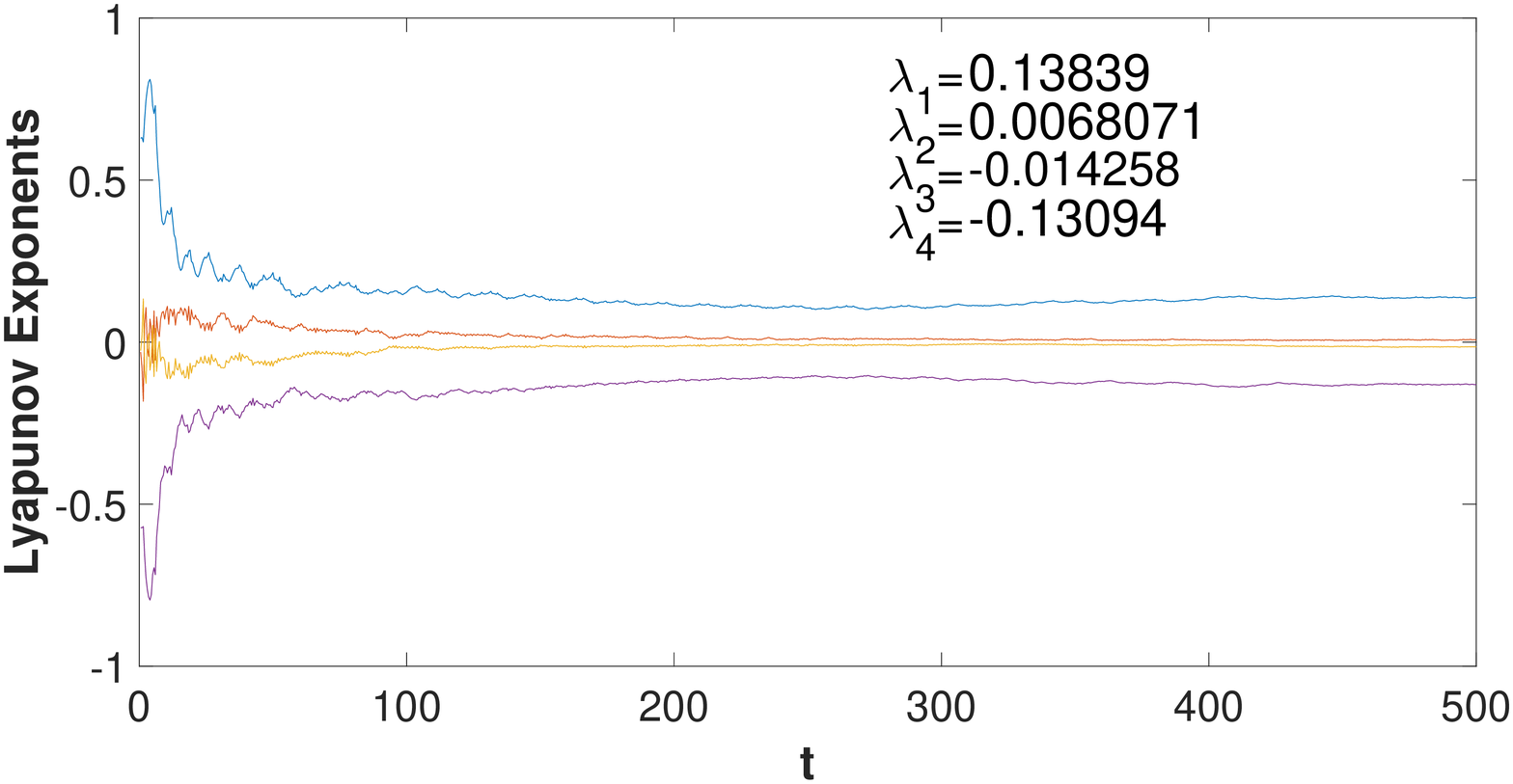} 
\caption{ lyapunov Exponents}
\label{vdpd-lyapunov}
\end{subfigure}%
\newline
\begin{subfigure}{.5\textwidth}
\centering
\includegraphics[width=.8\linewidth]{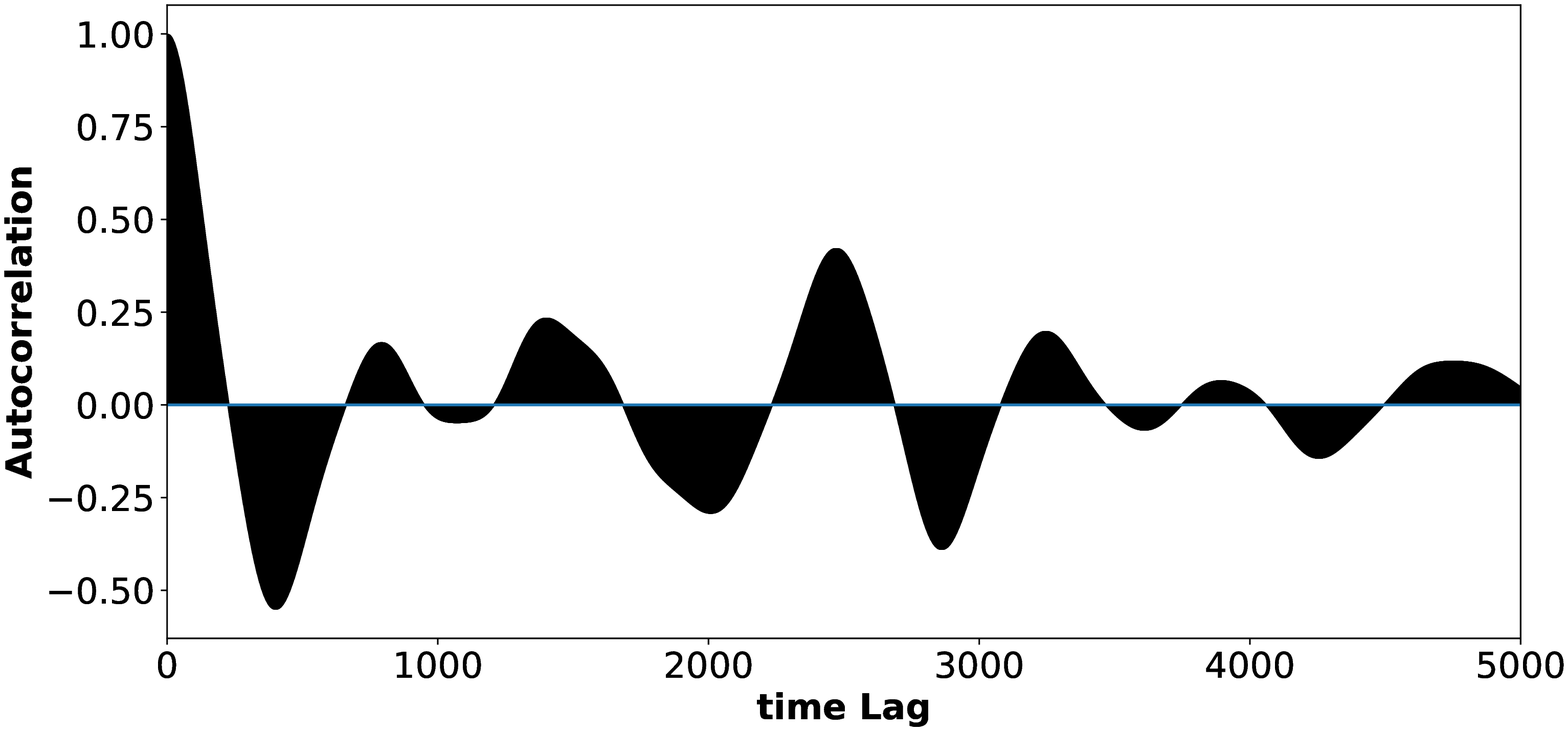}
\caption{Autocorrelation function of $x(t)$}
\label{vdpd-xauto}
\end{subfigure}%
\begin{subfigure}{.5\textwidth}
\centering
\includegraphics[width=.8\linewidth]{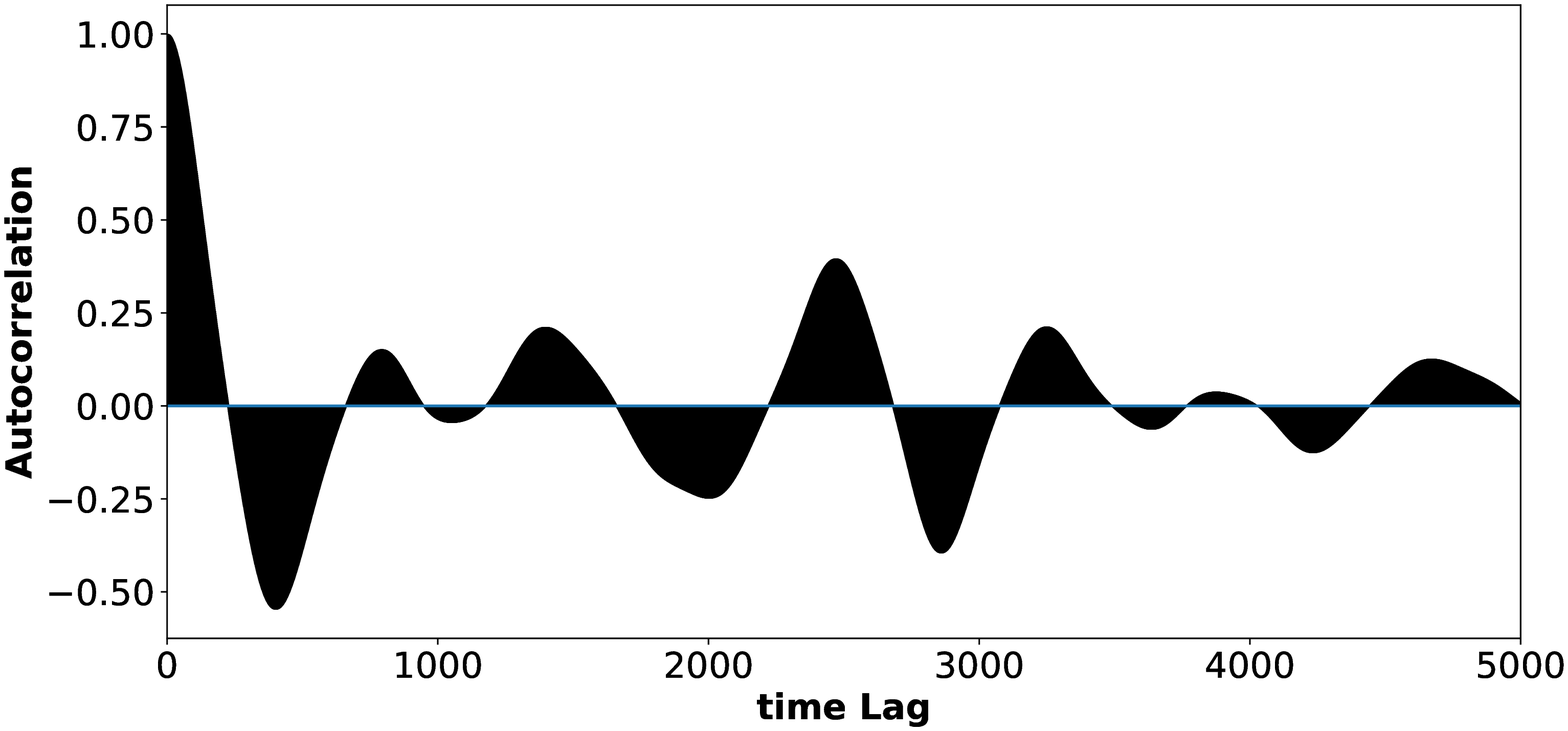}
\caption{Autocorrelation function of $y(t)$}
\label{vdpd-yaoto}
\end{subfigure}%
\newline
\begin{subfigure}{.5\textwidth}
\centering
\includegraphics[width=.8\linewidth]{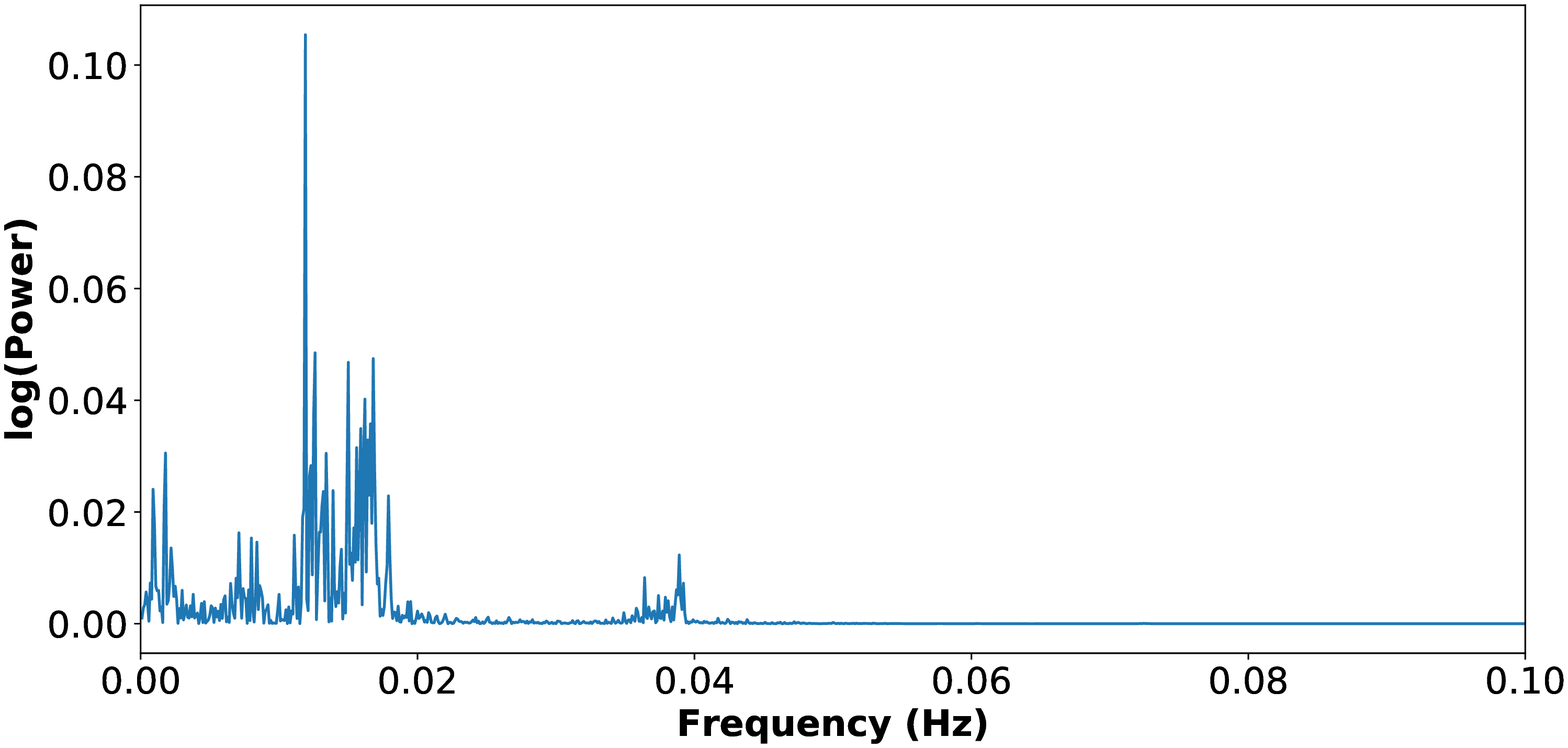}
\caption{Powerspectra of $x(t)$}
\label{vdpd-xpower}
\end{subfigure}%
\begin{subfigure}{.5\textwidth}
\centering
\includegraphics[width=.8\linewidth]{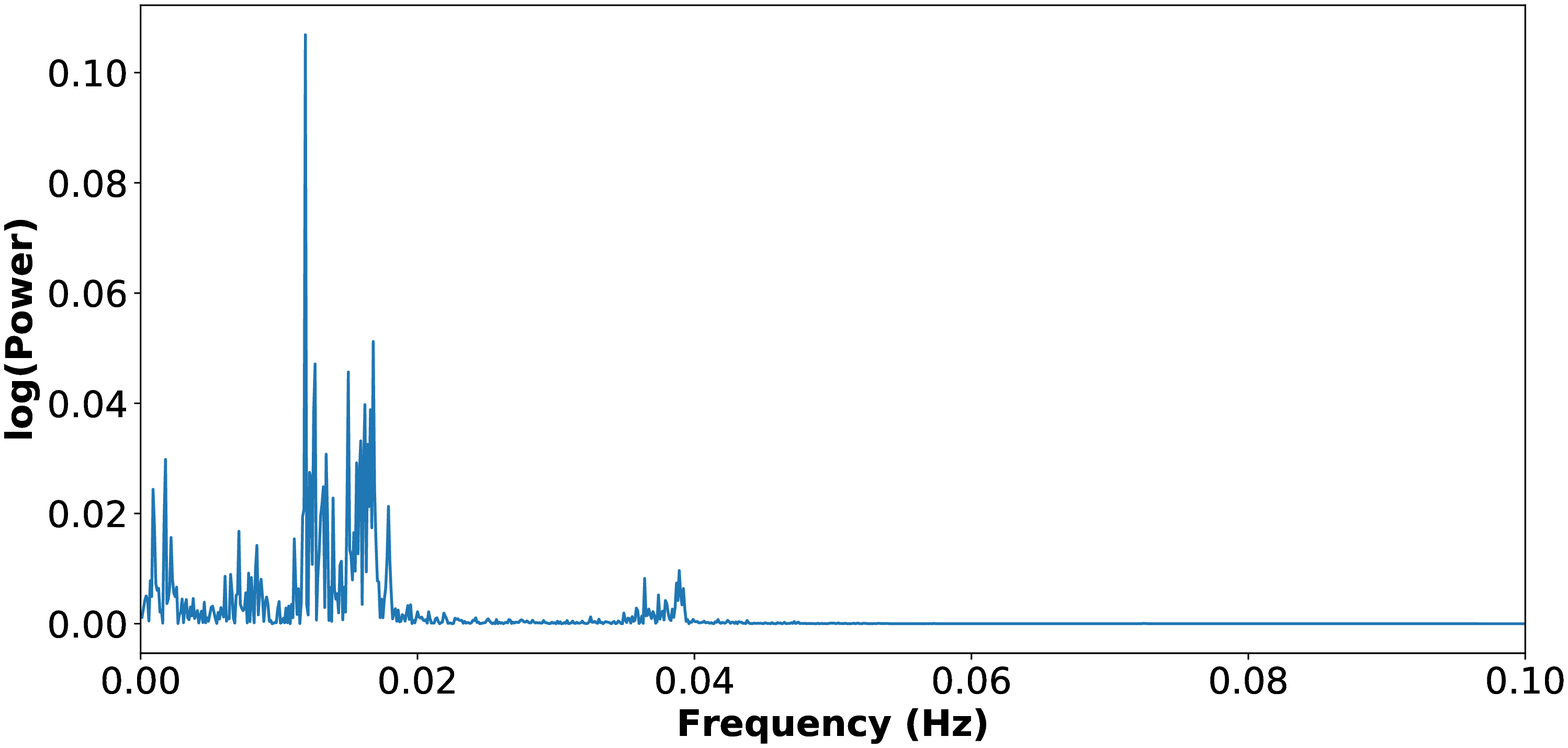}
\caption{Powerspectra of $y(t)$}
\label{vdpd-ypower}
\end{subfigure}%
\caption{(Color online)  Poincar$\acute{e}$ section, Lyapunov exponents, autocorrelation function
and power spectra for $\Gamma=0.03,\beta=2.0,\alpha=\tilde{\alpha}=0.5, a=1.0, b=1.0 $ with the initial conditions
$x(0)=0.01,y(0)=0.02,\dot{x}(0)=0.03,\dot{y}(0)=0.04$ }
\label{multi_nh}
\end{figure}

\begin{figure}[ht!]
\begin{subfigure}{.5\textwidth}
\centering
\includegraphics[width=.8\linewidth]{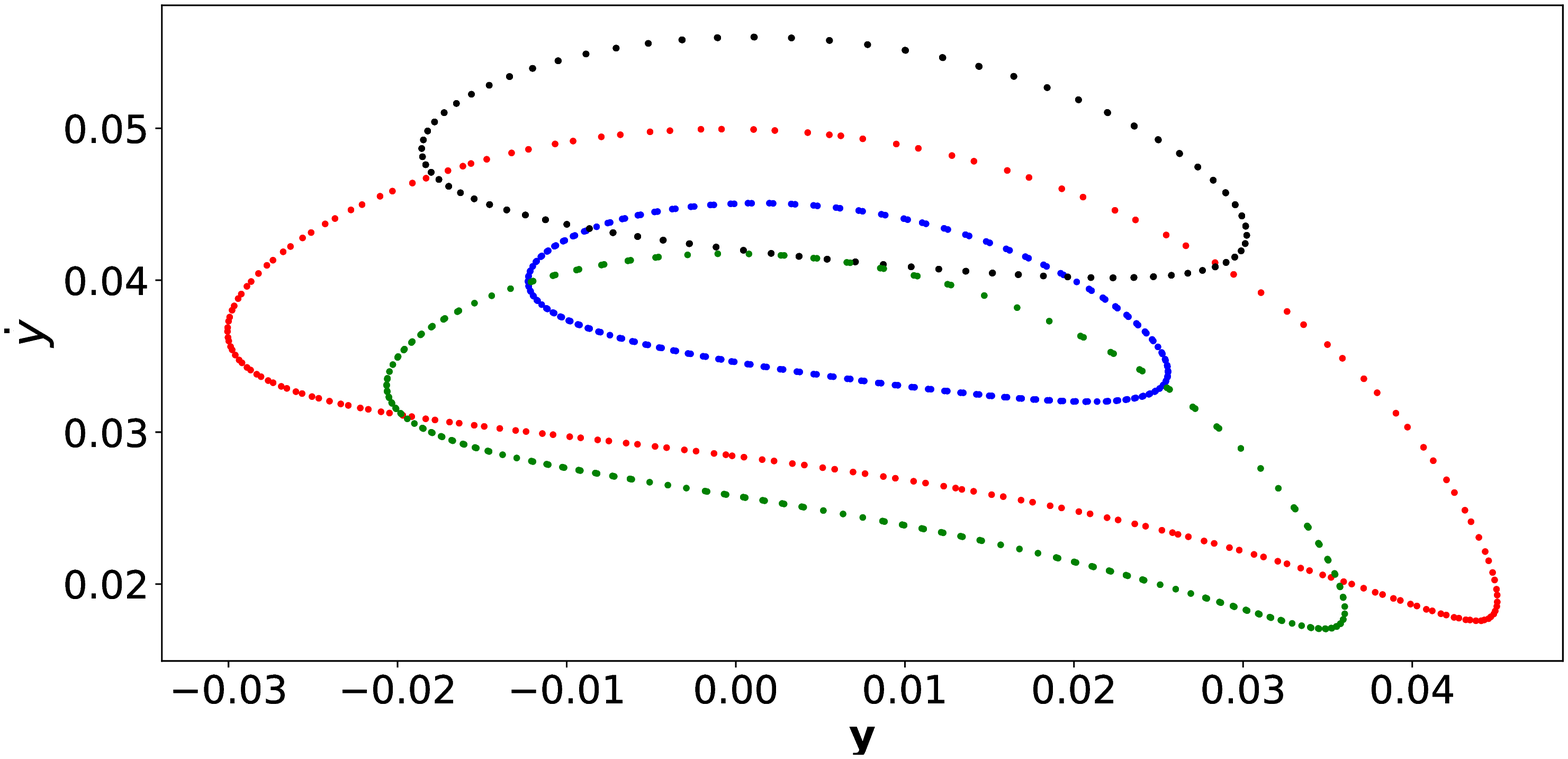} 
\caption{ $\alpha=\tilde{\alpha}=0.5, \beta=0.65, \Gamma=0.03,a=1.0,b=1.0$}
\label{poincare1nh}
\end{subfigure}%
\begin{subfigure}{.5\textwidth}
\centering
\includegraphics[width=.8\linewidth]{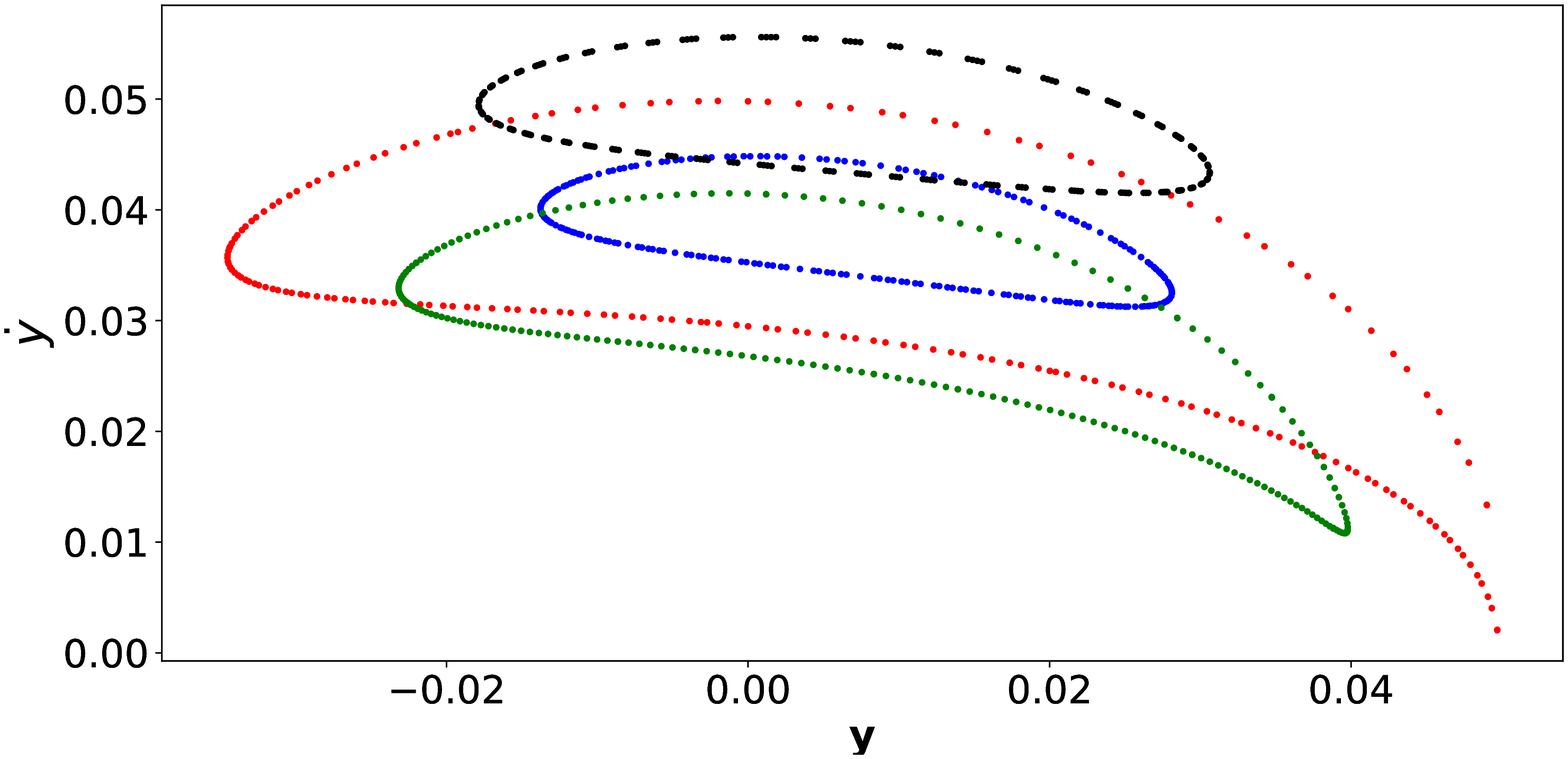}  
\caption{$\alpha=\tilde{\alpha}=0.5, \beta=0.76, \Gamma=0.03,a=1.0,b=1.0$}
\label{poincare2nh}
\end{subfigure}%
\newline
\begin{subfigure}{.5\textwidth}
\centering
\includegraphics[width=.8\linewidth]{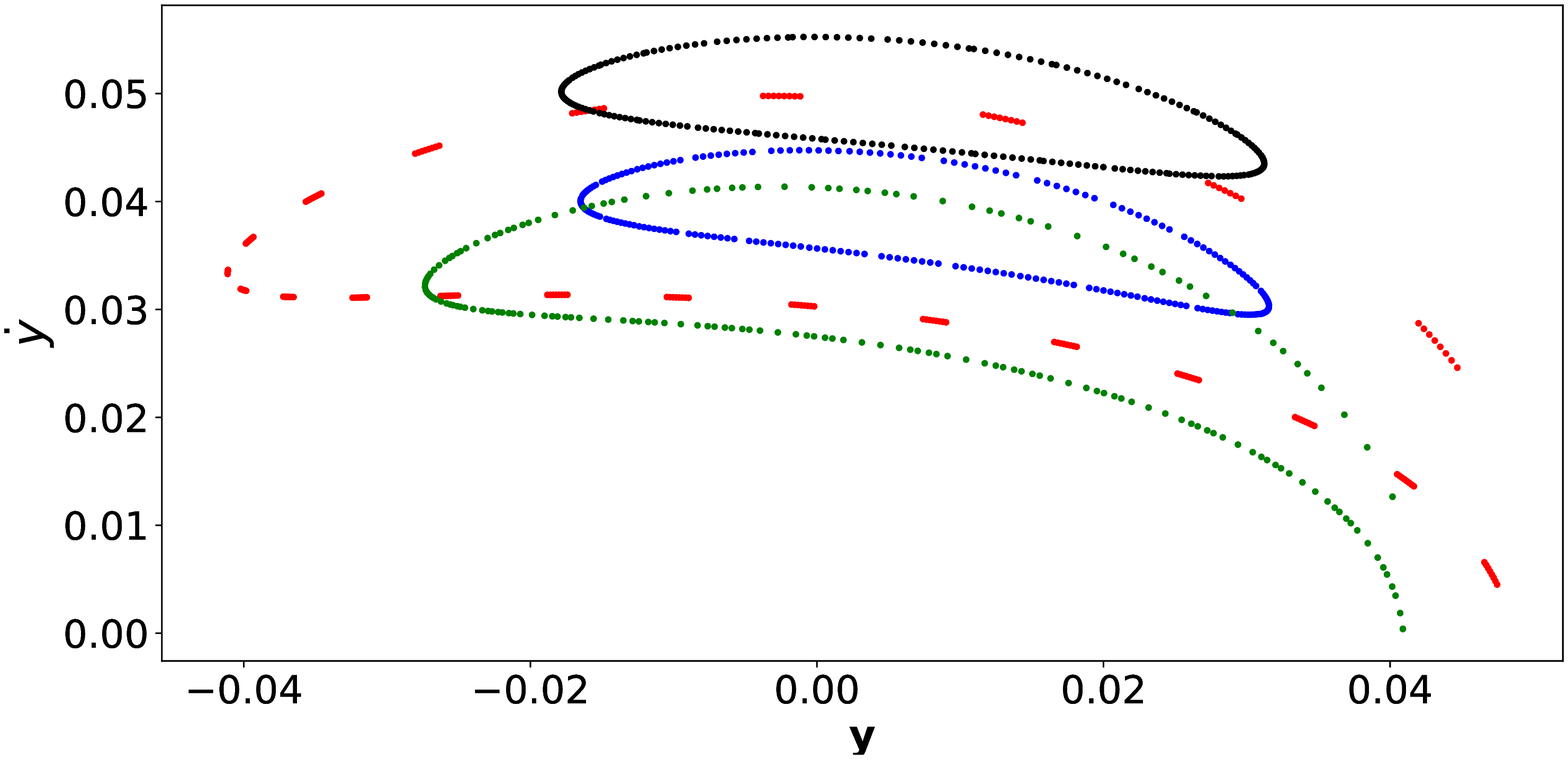}
\caption{$\alpha=\tilde{\alpha}=0.5, \beta=0.85, \Gamma=0.03,a=1.0,b=1.0$}
\label{poincare3nh}
\end{subfigure}%
\begin{subfigure}{.5\textwidth}
\centering
\includegraphics[width=.8\linewidth]{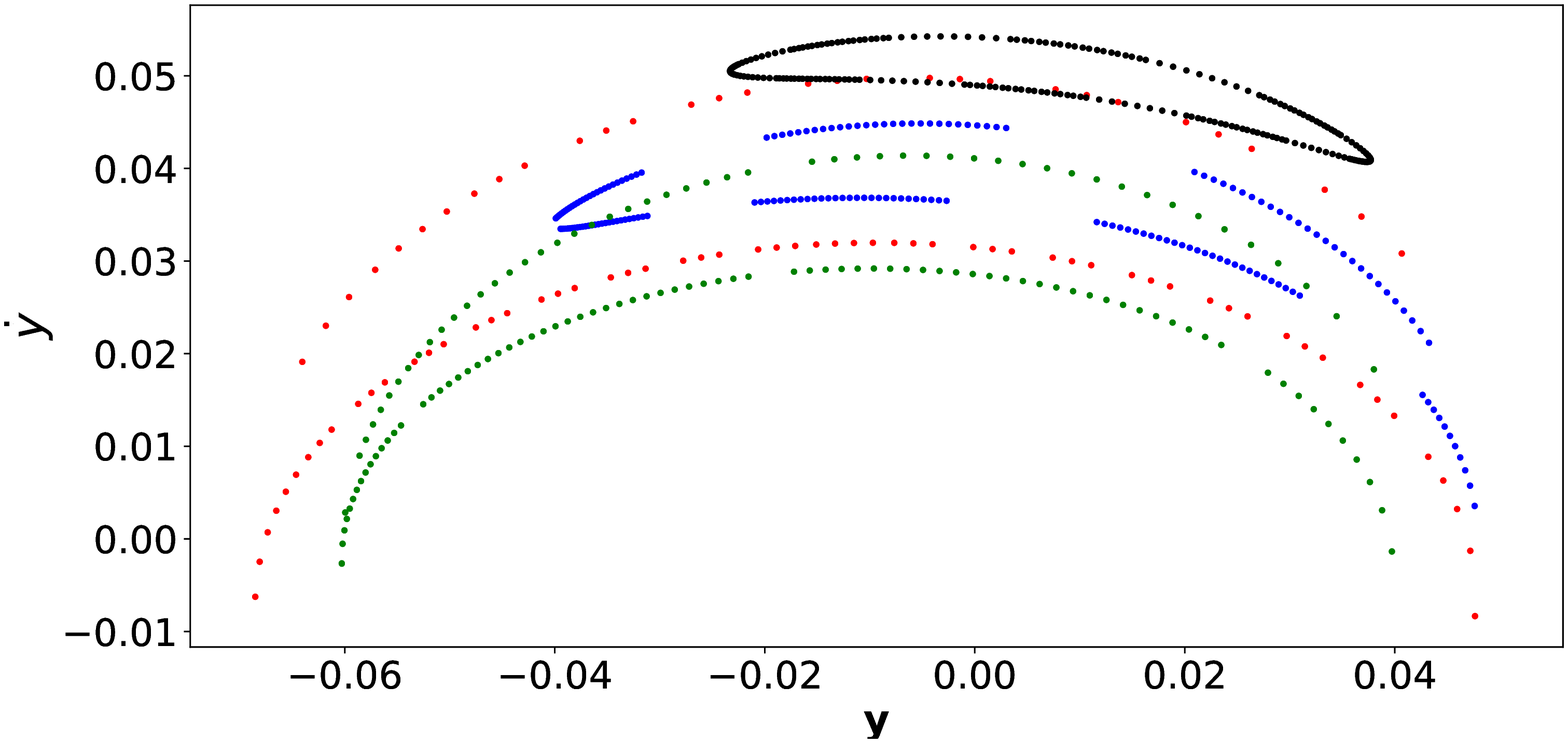}
\caption{$\alpha=\tilde{\alpha}=0.5,\beta=0.97,\Gamma=0.03,a=1.0,b=1.0$}
\label{poincare4nh}
\end{subfigure}%
\newline
\begin{subfigure}{.5\textwidth}
\centering
\includegraphics[width=.8\linewidth]{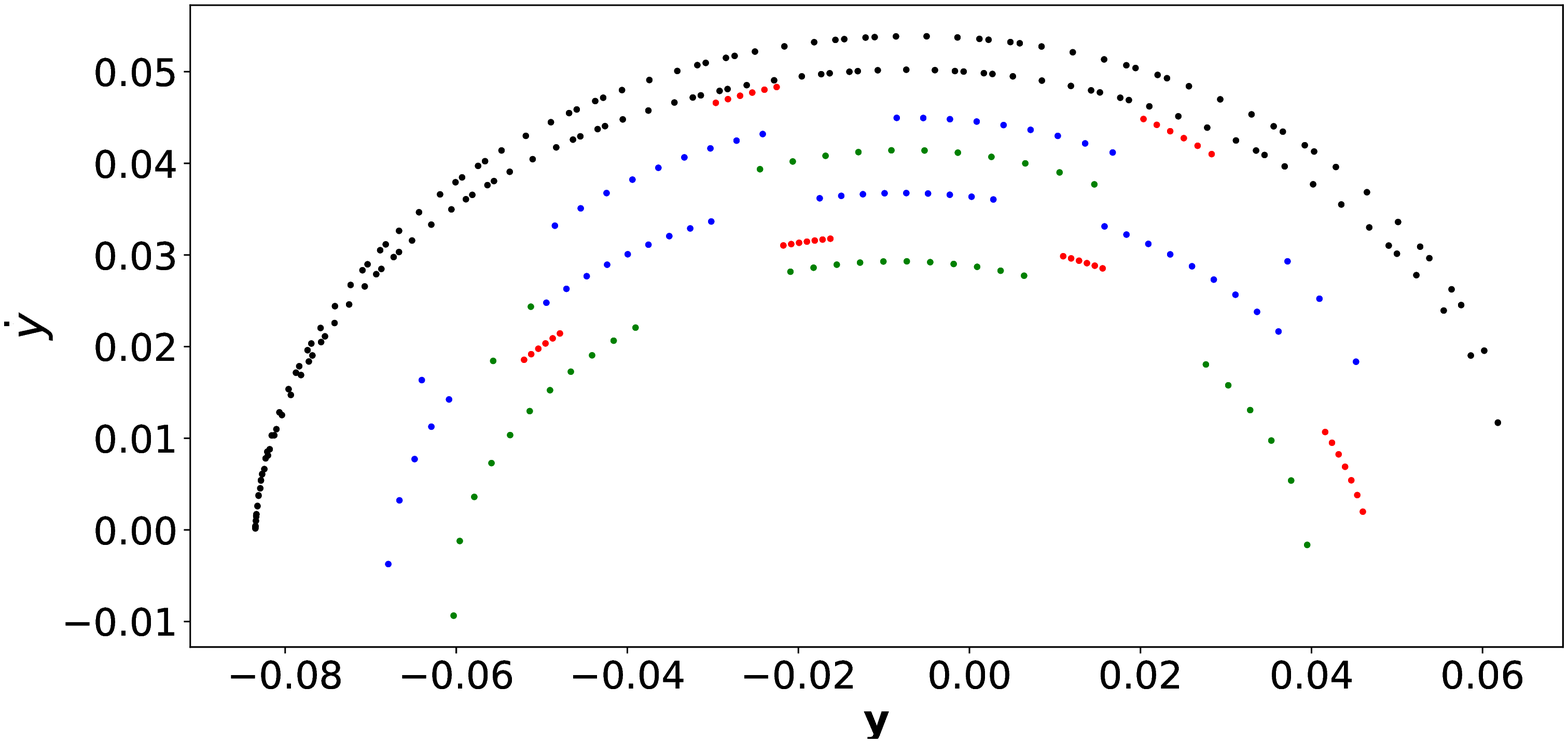}
\caption{$\alpha=\tilde{\alpha}=0.5,\beta=1.0,\Gamma=0.03,a=1.0,b=1.0$}
\label{poincare5nh}
\end{subfigure}%
\begin{subfigure}{.5\textwidth}
\centering
\includegraphics[width=.8\linewidth]{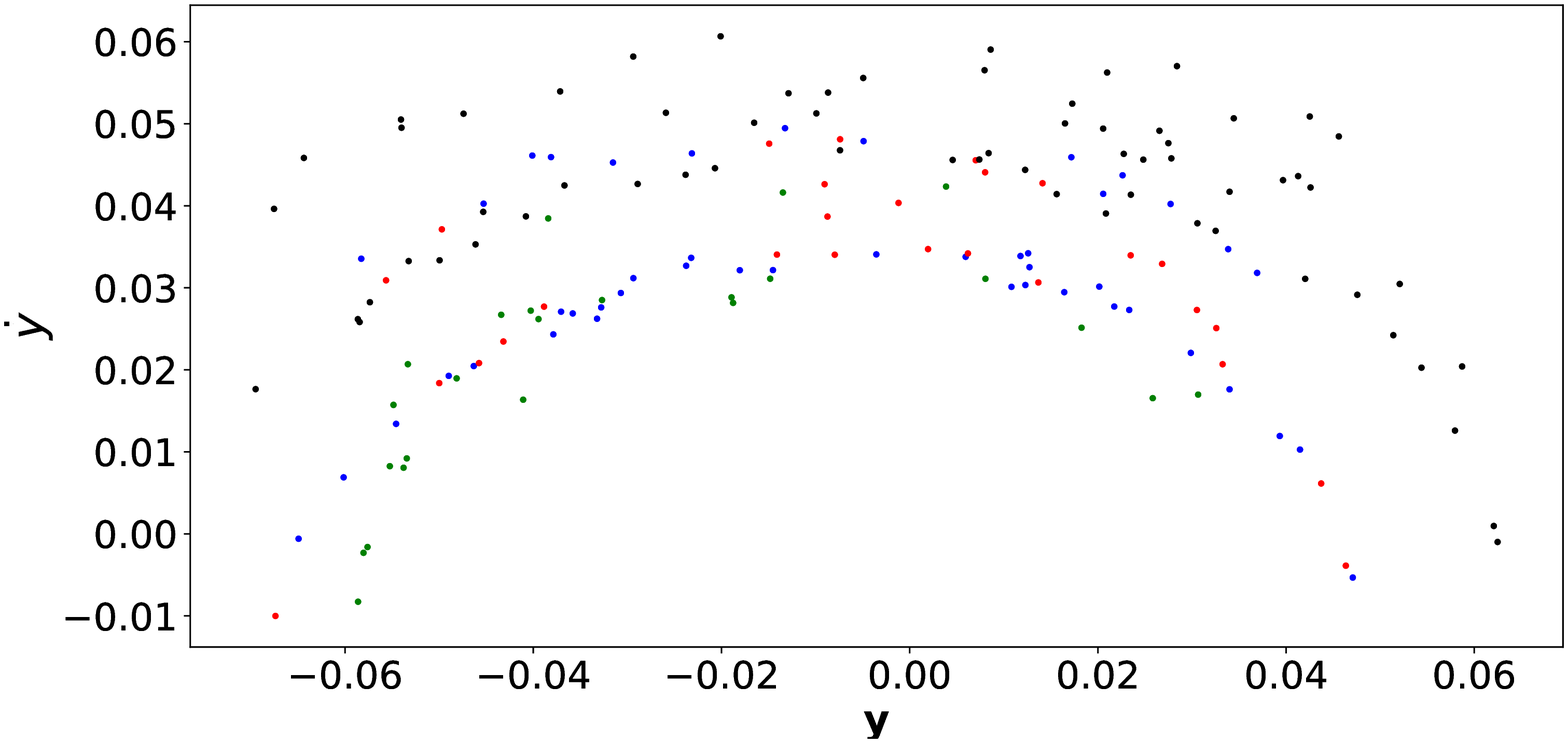}
\caption{$\alpha=\tilde{\alpha}=0.5,\beta=1.08,\Gamma=0.03,a=1.0,b=1.0$}
\label{poincare6nh}
\end{subfigure}%

\caption{(Color online) Poincar$\acute{e}$ section of Eq. ({\ref{vdp-duff-eqnn}}) with four sets of initial
conditions (a) $x(0)=0.01,y(0)=0.02,\dot{x}(0)=0.03,\dot{y}(0)=0.04$ (blue color) ,
(b) $x(0)=0.01,y(0)=0.03,\dot{x}(0)=0.02,\dot{y}(0)=0.04$ (red color),(c) $x(0)=0.02,y(0)=0.04,\dot{x}(0)=0.03,\dot{y}(0)=0.03$ (black color) and (d) $x(0)=0.01,y(0)=0.025,\dot{x}(0)=0.03,\dot{y}(0)=0.02$ (green color)
}
\label{poincarenh}
\end{figure}
The MSA and the RG techniques can be carried forward in a straightforward way and the slow
variation of the amplitude is governed by the equation,
\bea
2i \frac{\partial{\cal{A}}}{\partial T_1} + \beta_0 \sigma_1 {\cal{A}}
+ 2 i \Gamma_0 \sigma_3 \bp A (1-a {\vert A \vert}^2 - 2 b {\vert B \vert}^2 ) + b B^{2}A^{*}\\
B (1-b {\vert B \vert}^{2} - 2 a {\vert A \vert}^2 ) +a A^{2} B^{*} \ep 
+  3 \alpha_0 \bp {\vert A \vert}^2 A\\ {\vert B \vert}^2 B \ep = 0
\label{amp-eqn-nh}
\eea
\noindent Note that Eq. (\ref{amp-eqn-nh}) differs from the corresponding Eq. (\ref{polar-form2})
for the Hamiltonian system in the last term. In the limit of small $\theta_A, \theta_B$, the
amplitudes $A_0, B_0$ and the phase $\theta_A$ are determined by the Eqs. (\ref{flow-1},
\ref{flow-2}) and the first equation of (\ref{flow-3}). The equation for $\theta_B$ is different and
given by,
\bea
\frac{\partial \theta_B}{\partial t} = \frac{\beta A_0}{2 B_0} + \frac{3 \alpha B_0^2}{2}
\Rightarrow \theta_B = \frac{\beta}{2 K} p + \frac{3 \alpha K}{2p} + \theta_{B0}
\eea
\noindent where we have used $B_0=\frac{K}{A_0}$ in the last step.
Thus, in the leading order of the perturbation, only $\theta_B$ is different for the systems
defined by Eqs. (\ref{vdpd-eqn-1}) and (\ref{vdp-duff-eqnn}).
The solution for $p$ from Eqs (\ref{flow-A}) may be substituted to get the expression for $\theta_B$.
\begin{figure}[ht!]
\begin{subfigure}{.5\textwidth}
\centering
\includegraphics[width=.8\linewidth]{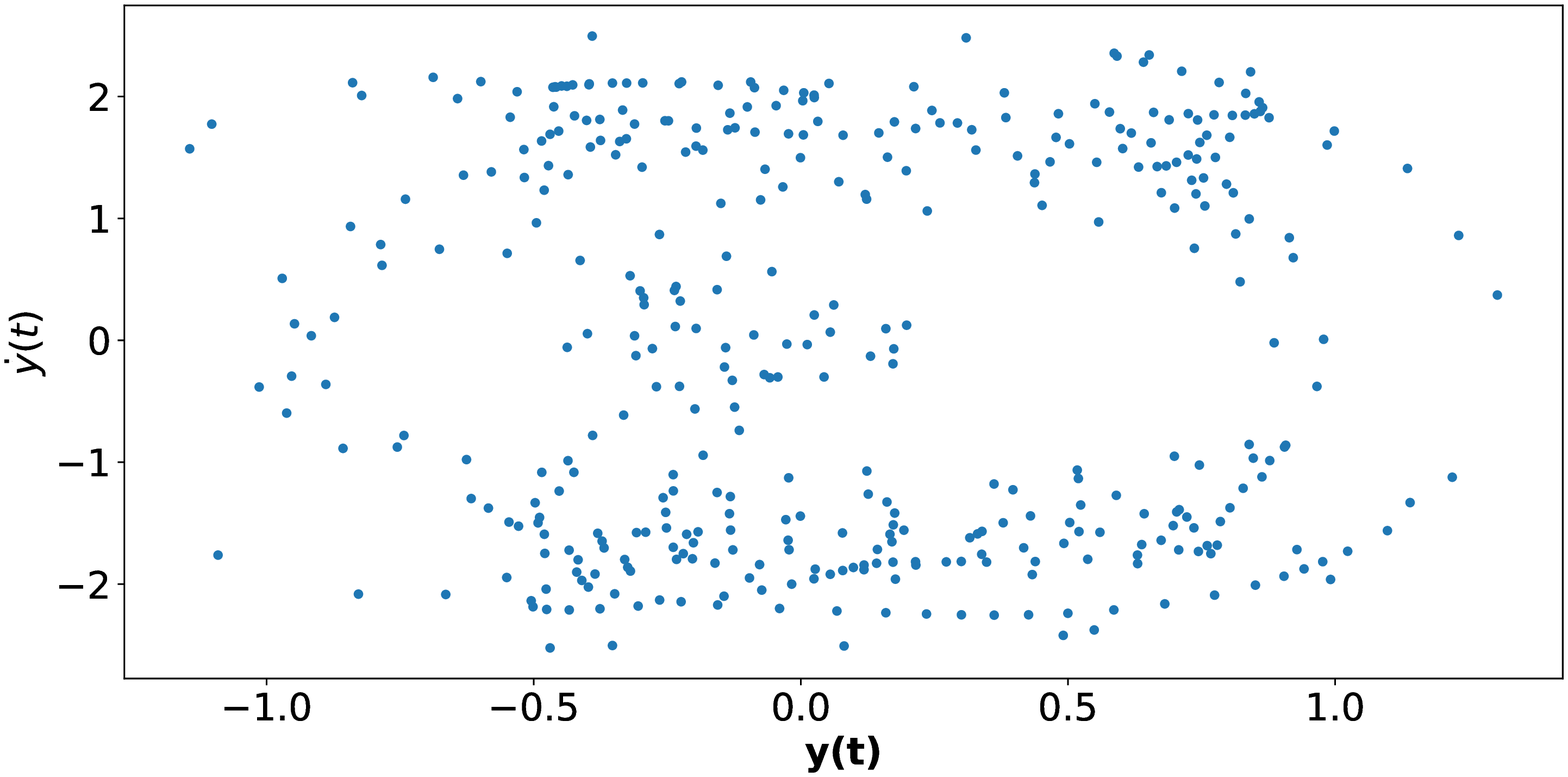}
\caption{Poincar$\acute{e}$ section: $\dot{y}(t)$ VS. $y(t)$ plot}
\label{vdpd-poincare_nhabneq}
\end{subfigure}%
\begin{subfigure}{.5\textwidth}
\centering
\includegraphics[width=.8\linewidth]{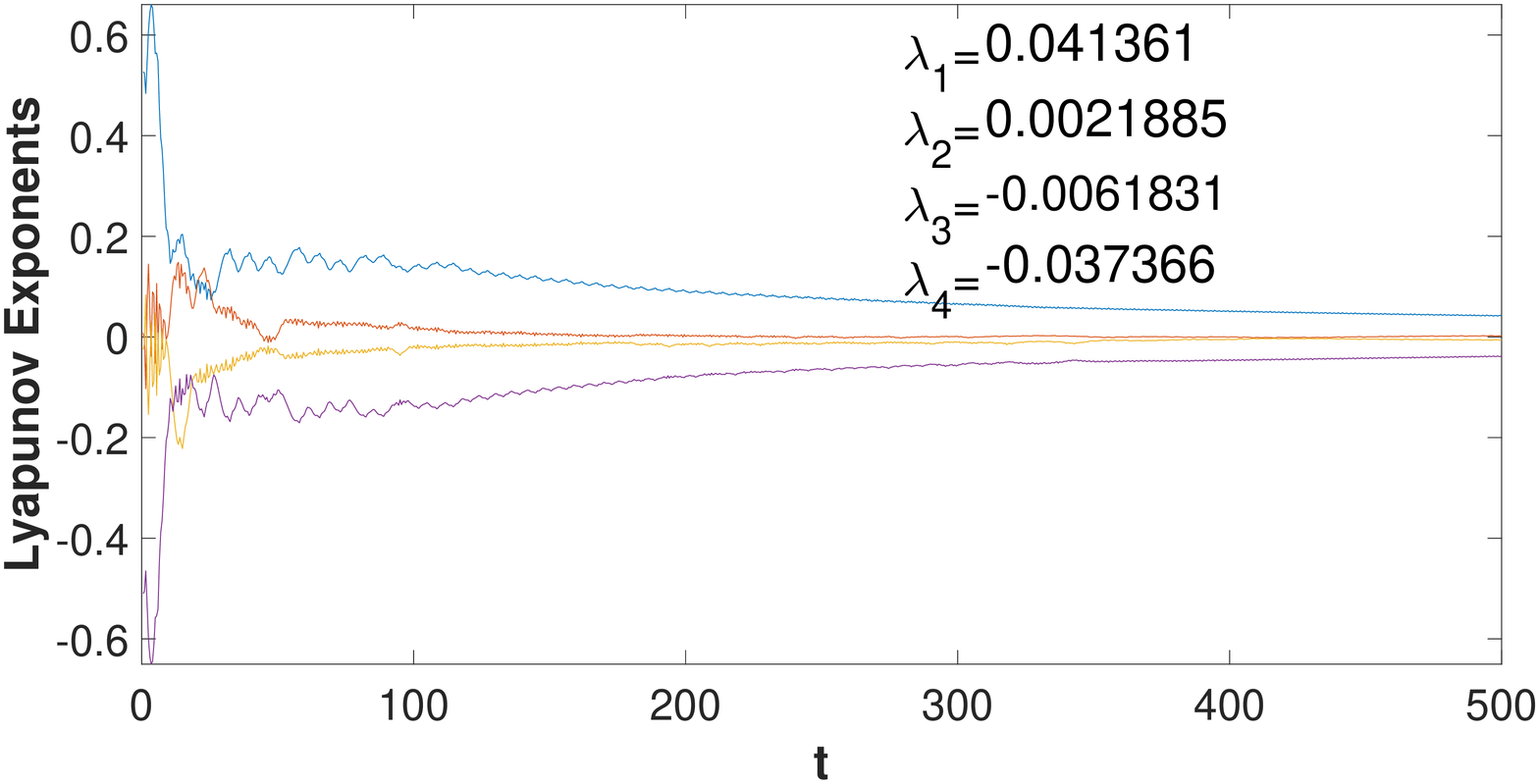} 
\caption{ lyapunov Exponents}
\label{vdpd-lyapunov_nhabneq}
\end{subfigure}%
\newline
\begin{subfigure}{.5\textwidth}
\centering
\includegraphics[width=.8\linewidth]{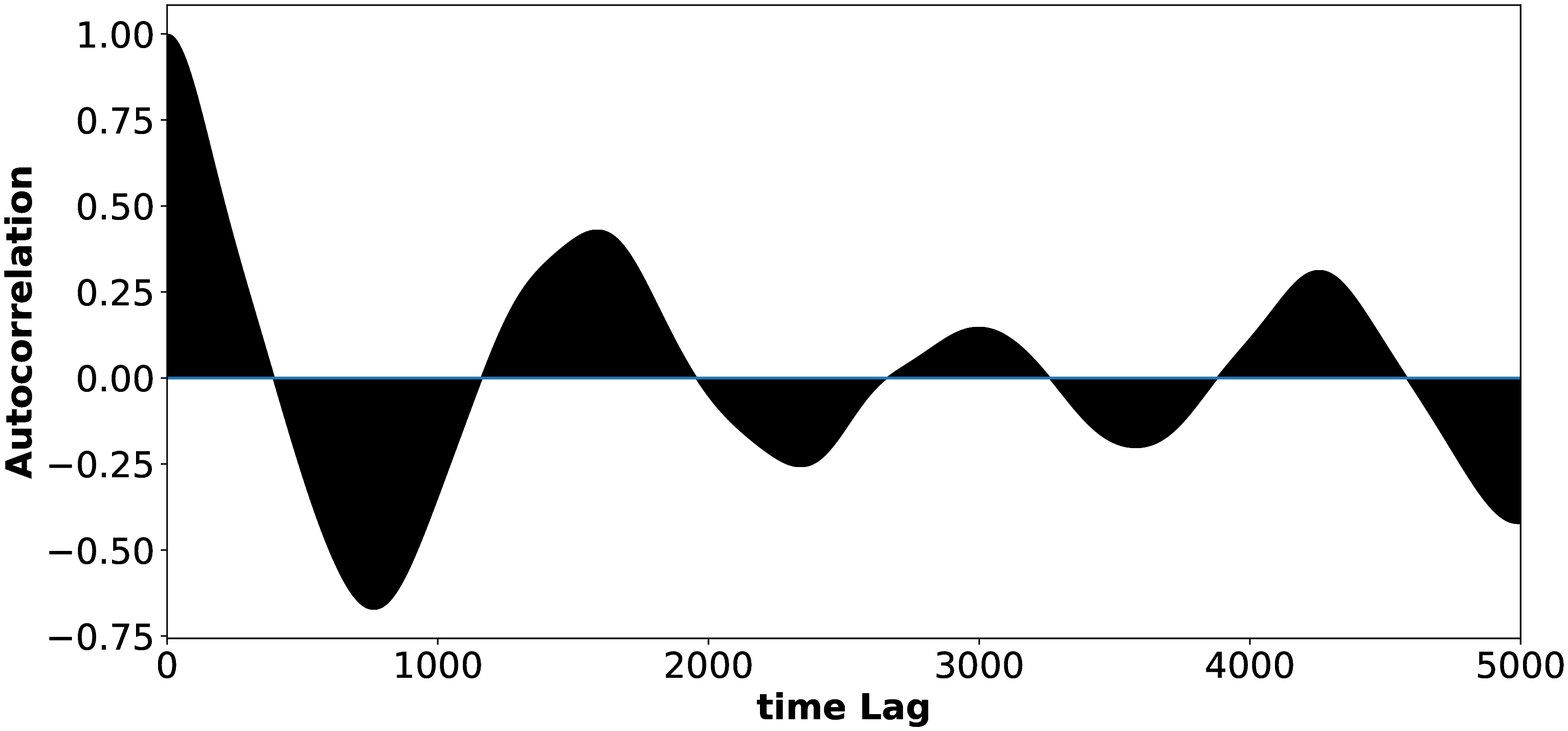}
\caption{Autocorrelation function of $x(t)$}
\label{vdpd-xauto_nhabneq}
\end{subfigure}%
\begin{subfigure}{.5\textwidth}
\centering
\includegraphics[width=.8\linewidth]{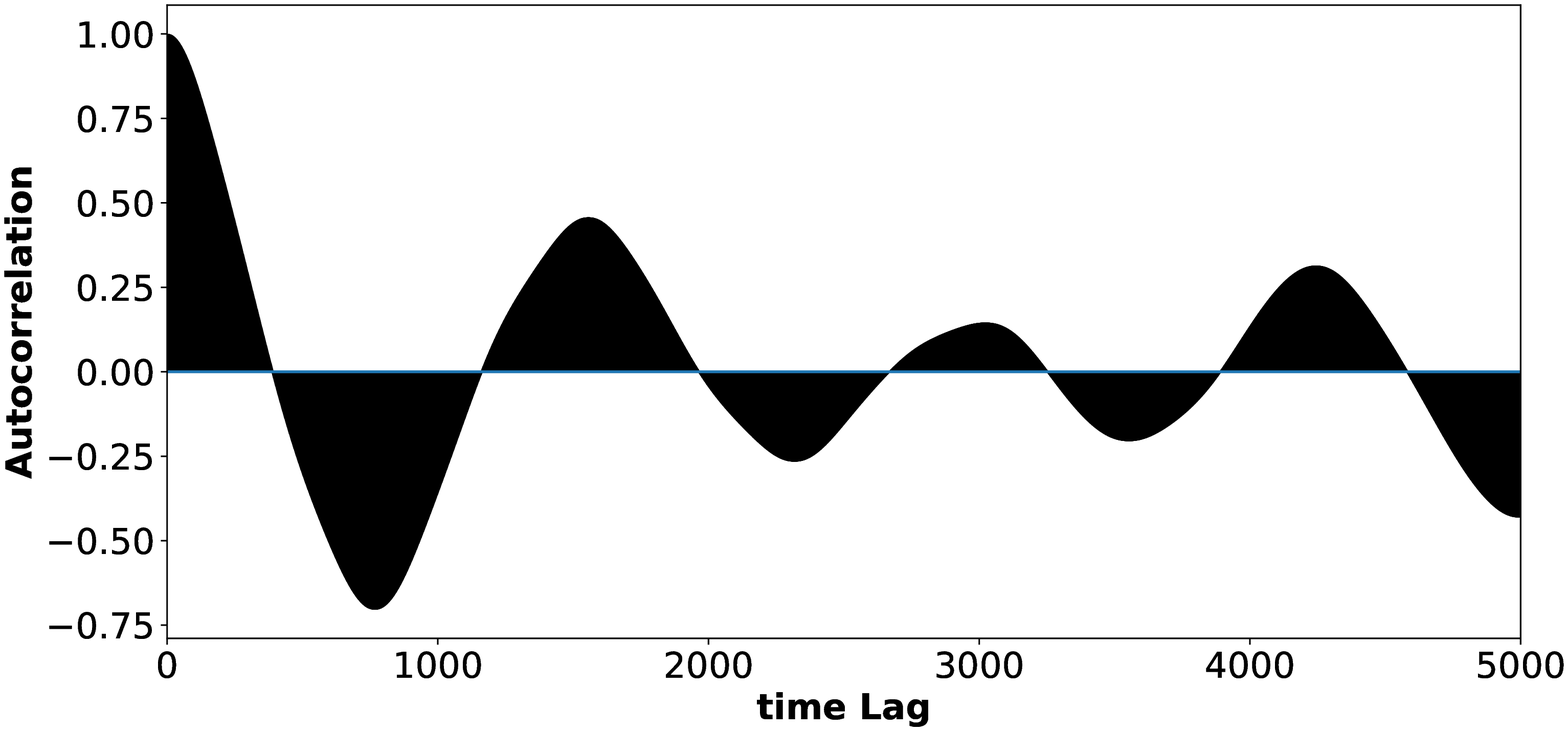}
\caption{Autocorrelation function of $y(t)$}
\label{vdpd-yaoto_nhabneq}
\end{subfigure}%
\newline
\begin{subfigure}{.5\textwidth}
\centering
\includegraphics[width=.8\linewidth]{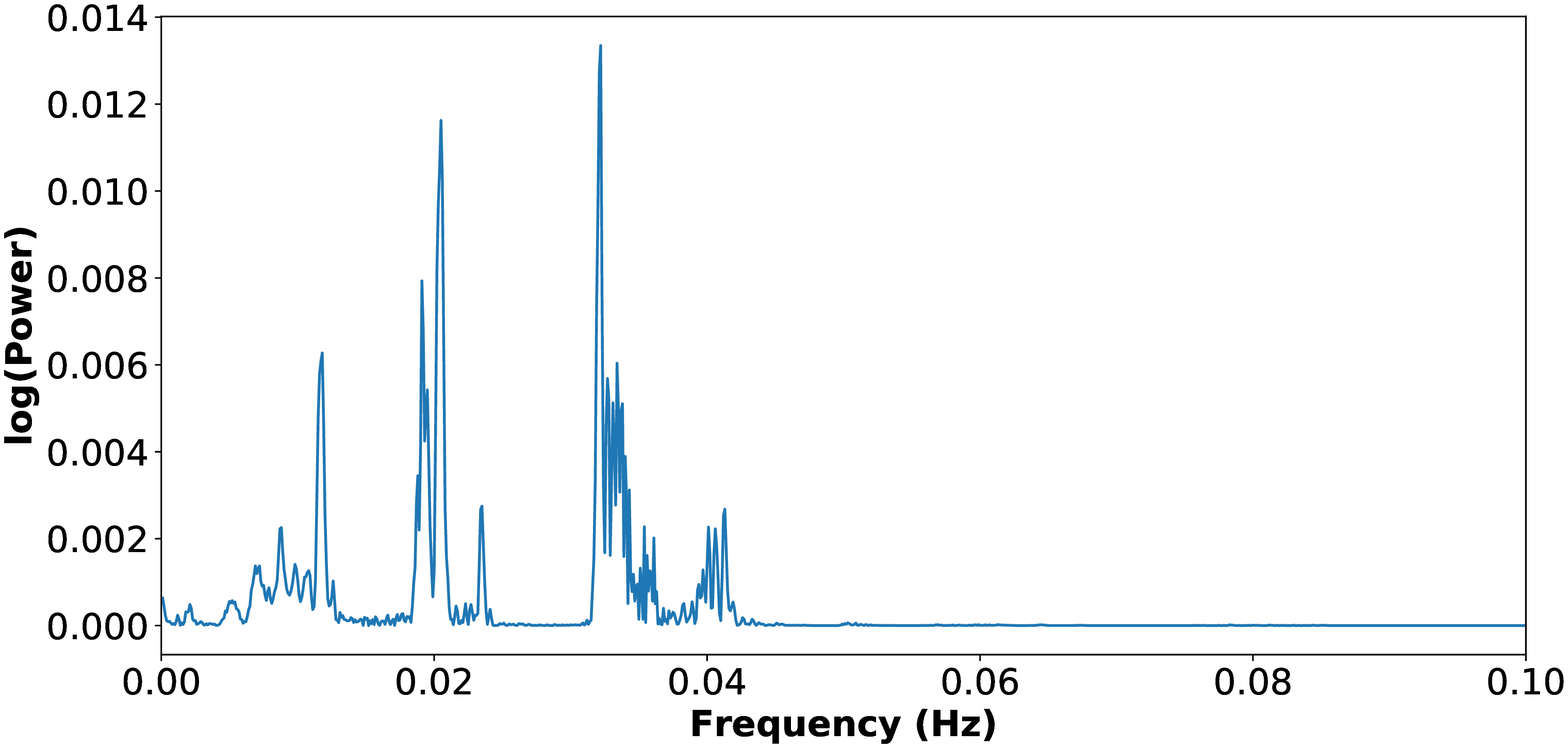}
\caption{Powerspectra of $x(t)$}
\label{vdpd-xpower_nhabneq}
\end{subfigure}%
\begin{subfigure}{.5\textwidth}
\centering
\includegraphics[width=.8\linewidth]{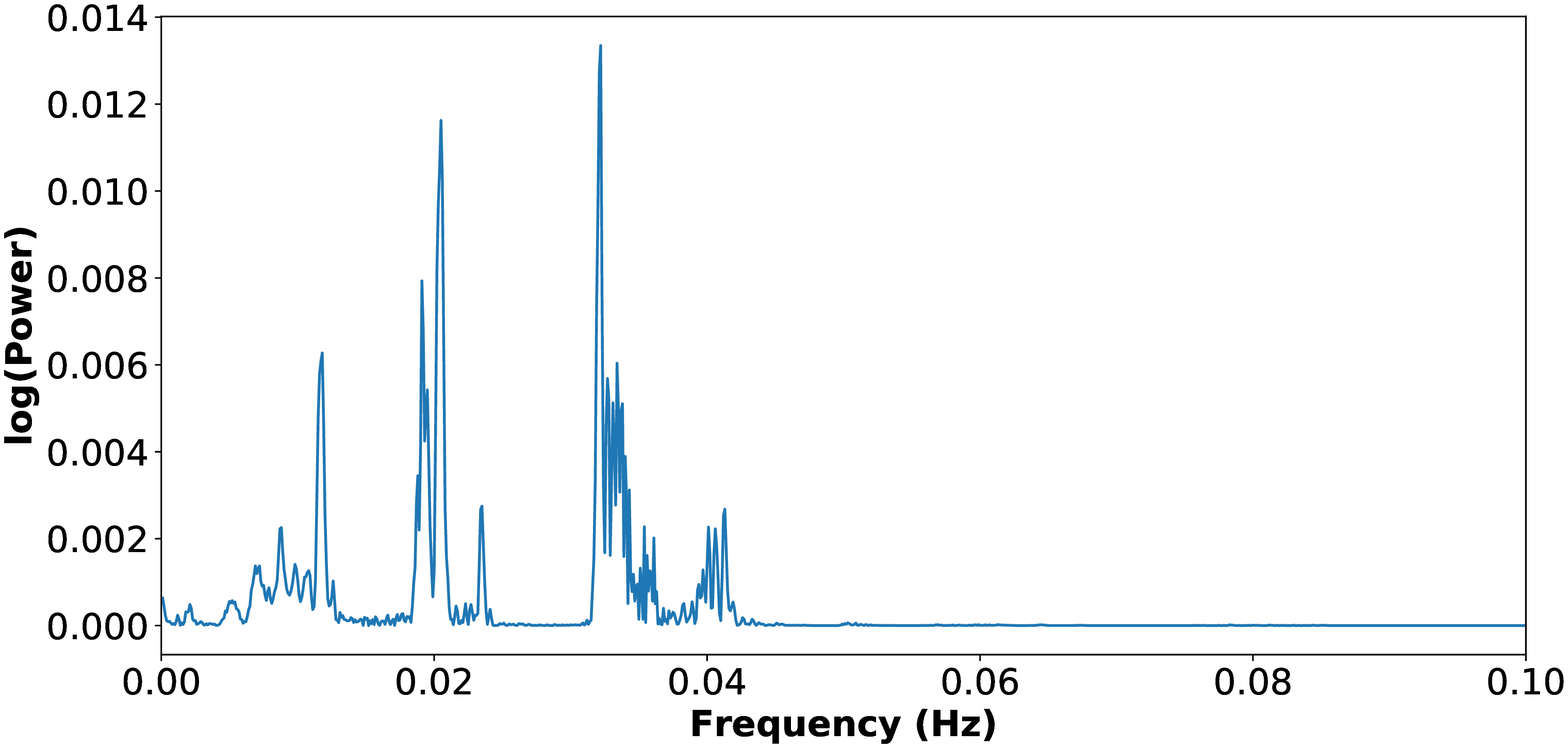}
\caption{Powerspectra of $y(t)$}
\label{vdpd-ypower_nhabneq}
\end{subfigure}%
\caption{(Color online)  Poincar$\acute{e}$ section, Lyapunov exponents, autocorrelation function
and power spectra for $\Gamma=0.01,\beta=1.8,\alpha=1.0,\tilde{\alpha}=3.0, a=1.0, b=2.0$ with the initial condition
$x(0)=0.01, y(0)=0.02, \dot{x}(0)=0.03, \dot{y}(0)=0.04$.}
\label{multi_abneqnh}
\end{figure}

\begin{figure}[ht!]
\begin{subfigure}{.5\textwidth}
\centering
\includegraphics[width=.8\linewidth]{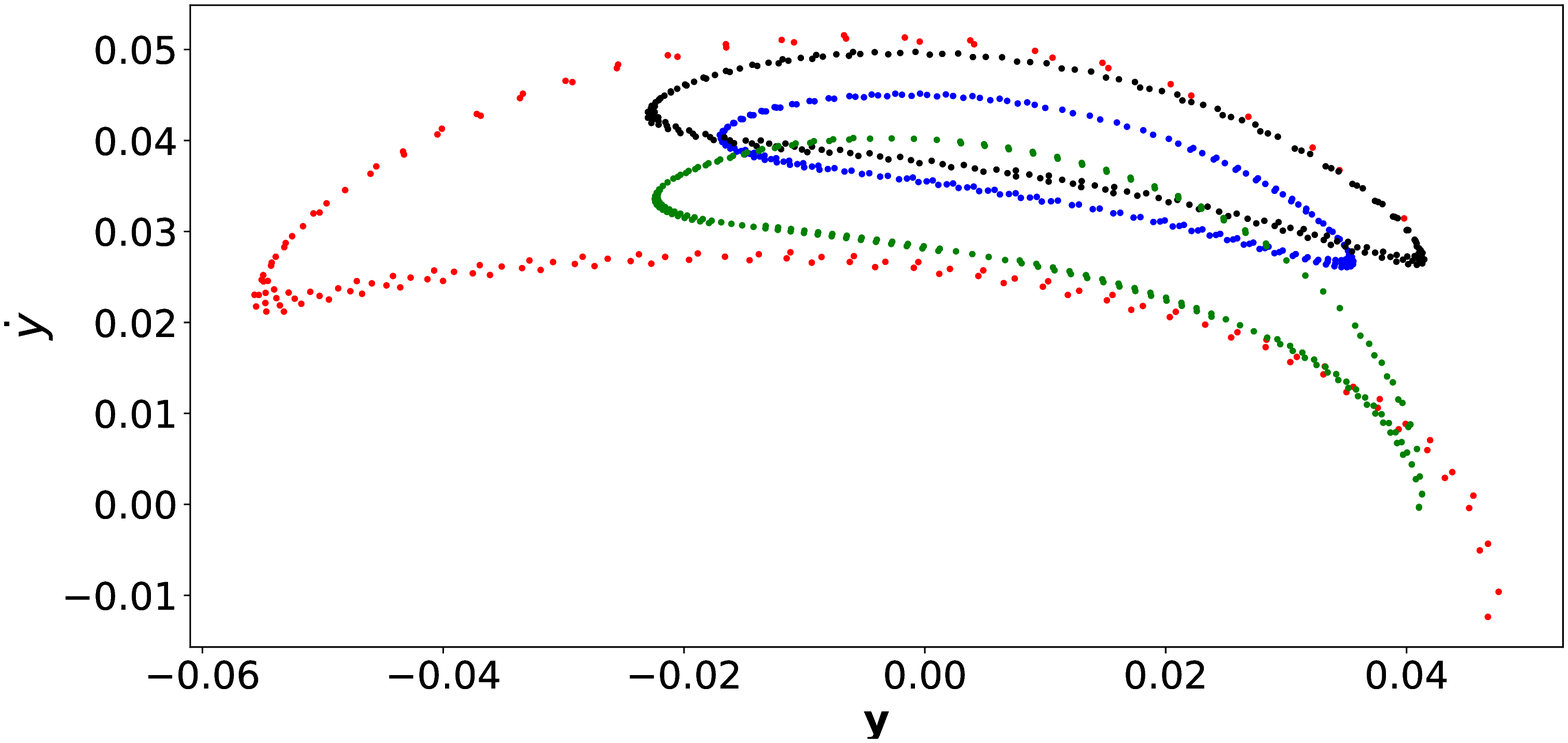}
\caption{ $\alpha=1.0,\tilde{\alpha}=3.0,\beta=0.85,\Gamma=0.01,a=1.0,\newline b=2.0$}
\label{poincareabneqnon1}
\end{subfigure}%
\begin{subfigure}{.5\textwidth}
\centering
\includegraphics[width=.8\linewidth]{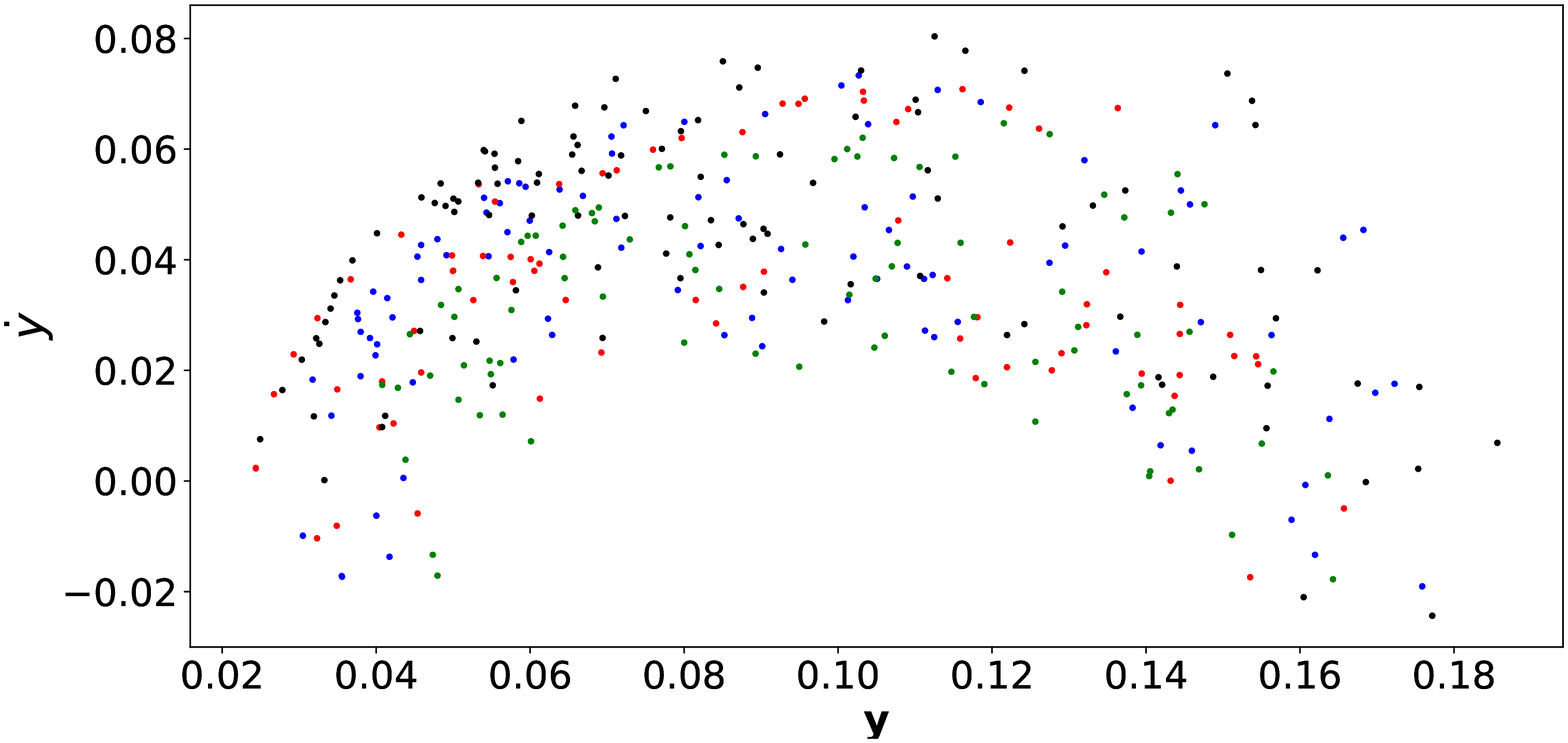}
\caption{$\alpha=1.0,\tilde{\alpha}=3.0,\beta=1.05,\Gamma=0.01,a=1.0,\newline b=2.0$}
\label{poincareabneqnon2}
\end{subfigure}%

\caption{(Color online) Poincar$\acute{e}$ section of Eq. ({\ref{vdp-duff-eqnn}}) with four sets of initial
conditions (a) $x(0)=0.01, y(0)=0.02, \dot{x}(0)=0.03, \dot{y}(0)=0.04$ (blue color) ,(b) $x(0)=0.01, y(0)=0.03,
\dot{x}(0)=0.015, \dot{y}(0)=0.04$ (red color),(c) $x(0)=0.01, y(0)=0.04, \dot{x}(0)=0.025, \dot{y}(0)=0.03$
(black color) and (d) $x(0)=0.01, y(0)=0.025, \dot{x}(0)=0.03, \dot{y}(0)=0.02$ (green color).
The parameters  $\alpha=1.0,\tilde{\alpha}=3.0, \Gamma=0.01,a=1.0,b=2.0$ are same for both the plots.
The parameter $\beta=0.85$ for the first plot, while $\beta=1.05$ for the second plot.}
\end{figure}

\subsection{Numerical Solution}

The system is analyzed by numerical analysis which reveals a rich complex dynamical behaviour. The regular
dynamics admitting periodic solutions is consistent with perturbative analysis. The results of
numerical investigations are presented in the ${\cal{PT}}$-symmetric regime ($a=b, \alpha =\tilde{\alpha}$) as
well as in the non-${\cal{PT}}$-symmetric regime ($a \neq b, \alpha \neq \tilde{\alpha}$) for different values
of $\Gamma$ and $\beta$ for which $P_0$ is a center. We choose $a=b=1.0, \alpha=\tilde{\alpha}=0.5$ for
numerical results in the ${\cal{PT}}$-symmetric region. The results in the non-${\cal{PT}}$-symmetric region
are presented with $a=1.0, b=2.0, \alpha=1.0, \tilde{\alpha}=3.0$. The values of $\Gamma$ and $\beta$ are chosen
depending on the particular object of physical interest. The initial conditions for all the numerical investigations
are chosen as $x(0)=0.01, y(0)=0.02, \dot{x}(0)=0.03$ and $\dot{y}(0)=0.04$, unless specified otherwise,
which correspond to small fluctuations around the equilibrium point $P_0$. The time-series of $x$ and $y$
in the ${\cal{PT}}$-symmetric regime is presented in the first row of Fig. \ref{time-series-P0_nh} for
$\beta = 0.5, \Gamma= 0.03$.
The periodic solution in the non-${\cal{PT}}$-symmetric region for $\Gamma=0.01$ and $\beta=0.5$ is presented
in the second row of Fig. \ref{time-series-P0_nh}. The periodic solution of non-Hamiltonian, non-${\cal{PT}}$-symmetric
VdPD oscillator is yet another example where the existence of bounded solution is not attributed to unbroken
${\cal{PT}}$-symmetry. The bifurcation diagrams in Fig . \ref{bifurcation-nh-npt} show that the system 
is chaotic for $\beta > \beta_c$ both in the ${\cal{PT}}$-symmetric as well as non-${\cal{PT}}$-symmetric
regimes. The value of $\beta_c$ in the ${\cal{PT}}$-symmetric regime is $1.07$, while it is $1.05$ in the
non-${\cal{PT}}$-symmetric region. The values of the parameters for these two cases are different and a
direct comparison of the values of $\beta_c$ is not going to give any information.
The sensitivity of the dynamical variables to the initial conditions are studied in different regions of
the parameter-space by considering two sets of initial conditions: (a) $x(0)=0.01, y(0)=0.02, \dot{x}(0)=0.03,
\dot{y}(0)=0.04$ and (b) $x(0)=0.01, y(0)=0.02, \dot{x}(0)=0.03, \dot{y}(0)=0.025$.  These two initial conditions
are identical except for the values of $\dot{y}(0)$ which differ by $0.015$. The time series of the dynamical
variables in the chaotic regime is presented in Fig. \ref{chaotic_nonhamiltonian} \textemdash
the first row corresponds to ${\cal{PT}}$-symmetric region with $\beta=2.0, \Gamma=0.03$ and the second-row for
the non-${\cal{PT}}$-symmetric region for $\beta=1.8, \Gamma=0.01$. The chaotic behaviour in the model has been
confirmed by other independent methods also. In this regard, the auto-correlation function, Lyapunov exponent,
Poincar$\acute{e}$ section and power spectra are plotted in Figs. \ref{multi_nh} and \ref{multi_abneqnh}
for ${\cal{PT}}$-symmetric and non-${\cal{PT}}$-symmetric regimes, respectively. The Lyapunov exponents in
the ${\cal{PT}}$-symmetric region is computed up to six decimal places $(0.13839,0.068071,-0.014258,-0.13094)$,
while it is computed up to seven decimal places $(0.041361 , 0.0021885 , -0.0061831, -0.037366)$ for the
non-${\cal{PT}}$-symmetric case.

The route to chaos for the non-Hamiltonian system with or without ${\cal{PT}}$-symmetry may be studied
in terms of the qualitative changes in a set of Poincar$\acute{e}$ sections of the system as the parameter $\beta$ is
varied from the non-chaotic to the chaotic region. The qualitative features are similar to that of the Hamiltonian
system. It is seen from Fig. 10 corresponding to the ${\cal{PT}}$-symmetric non-Hamiltonian system that all four orbits
on the chosen surface are closed for $\beta=0.65$ and only one orbit is closed around $\beta=0.97$, and finally no closed
orbits are seen as the critical value $\beta_c=1.05$ is approached.  There are only isolated points in the chaotic region,
i.e. beyond $\beta > \beta_c$. The scenario for non-${\cal{PT}}$-symmetric non-Hamiltonian system is similar
to the ${\cal{PT}}$-symmetric case, and depicted in Fig. 12 with only two plots corresponding to non-chaotic and
chaotic regions in order to avoid repetition.

\section{Conclusions $\&$ Discussions}

We have investigated two models of coupled VdPD oscillators with balanced loss and gain.
The first model is a Hamiltonian system describing a Van der Pol oscillator coupled to another VdPD
oscillator through linear as well as nonlinear coupling. The nonlinear coupling is introduced
via balanced loss-gain terms. Further, the strength of the linear restoring force for
the Van der Pol oscillator depends on the degree of freedom describing the VdPD oscillator.
The space-dependent linear restoring force also introduces nonlinearity in the differential equation.
The model is a generalization of coupled Duffing oscillator system with balanced loss and
gain\cite{pkg-pr} by allowing the loss-gain terms to be space-dependent. In general, the system
is non-${\cal{PT}}$-symmetric. The system admits five equilibrium points out of which only three
are stable. The co-ordinates of the equilibrium points in the phase-space are different from
the coupled Duffing oscillator model\cite{pkg-pr}. However the stability criteria are same
for both the models. The equations of motion are studied by using two different perturbation
techniques, namely MSA and RG methods. The approximate solutions in the leading order of the
perturbation are identical for both the methods. These solutions are
periodic in time with the time-dependence of the amplitudes and phases determined from the RG flow
equation. The RG flow equation defines a dimer model and is exactly solvable for specific initial conditions.

We have investigated the system numerically which confirms the findings of the linear stability
analysis and perturbation methods. The system admits regular periodic solution in the
${\cal{PT}}$-symmetric as well as non-${\cal{PT}}$-symmetric regimes in the parameter-space. The
existence of bounded and unbounded solutions in different regimes of the parameter-space is to be
understood in terms of the standard techniques of dynamical systems, instead of broken and unbroken
phases of ${\cal{PT}}$ symmetry. The bifurcation diagram exhibits chaotic
behaviour in the system beyond a critical value of the linear coupling. There is no external
driving force and coupling to the Van der Pol oscillator acts as a source of energy. The model
provides yet another example of a Hamiltonian chaos within the ambit of systems with balanced
loss and gain. The chaotic behaviour is studied in detail through sensitivity of the solution
to the initial conditions, Poincar${\acute{e}}$-sections, auto-correlation functions and powerspectra.
The Lyapunov exponents are computed numerically.

The second model of coupled VdPD oscillators is obtained by modifying the nonlinear
interaction in the first model which is a generalization of the models considered
in Refs. \cite{pkg-review, khare-0}. The model is non-Hamiltonian for the non-vanishing
nonlinear interaction. It describes two  damped Duffing oscillators coupled via
linear as well as nonlinear coupling. The nonlinear coupling is achieved through
the balanced loss-gain terms. The system has ${\cal{PT}}$ symmetric as well as
non-${\cal{PT}}$-symmetric regions in the parameter space. We have obtained
regular periodic solutions in the  ${\cal{PT}}$ symmetric as well as 
non-${\cal{PT}}$-symmetric regions. The system is chaotic which is investigated
through  time-series, Poincar$\acute{e}$-section, power-spectra, autocorrelation
function. The Lyapunov exponents are computed numerically.

The Duffing oscillator, Van der Pol oscillator and their various generalizations have
been studied  extensively in the literature over several decades with interesting
theoretical as well as experimental results. The generalization of these systems
by including balanced loss and gain is a recent phenomenon and reveal rich dynamical
behaviour. It is expected that further investigations of these new class of systems
from different perspectives may shed new insights into hitherto unexplored areas
of dynamical systems. A few particular directions of physical interests may be 
to investigate the quantized versions of these system along with studies on quantum
chaos, quantum criticality, quantum synchronizations etc. Some of these issues
will be addressed in future.
 
\section{Acknowledgements}

This work of PKG is supported by a grant ({\bf SERB Ref. No. MTR/2018/001036})
from the Science \& Engineering Research Board(SERB), Department of Science
\& Technology, Govt. of India under the {\bf MATRICS} scheme. The work of PR
is supported by CSIR-NET fellowship({\bf CSIR File No.:  09/202(0072)/2017-EMR-I})
of Govt. of India.

\end{document}